\documentclass[pdflatex,sn-basic]{sn-jnl}

\usepackage{graphicx}%
\usepackage{multirow}%
\usepackage{amsmath,amssymb,amsfonts}%
\usepackage{amsthm}%
\usepackage{mathrsfs}%
\usepackage[title]{appendix}%
\usepackage{xcolor}%
\usepackage{textcomp}%
\usepackage{manyfoot}%
\usepackage{booktabs}%
\usepackage{algorithm}%
\usepackage{algorithmicx}%
\usepackage{algpseudocode}%
\usepackage{listings}%

\usepackage[normalem]{ulem}
\usepackage{anyfontsize}
\usepackage{multirow}

\newcommand{\Msun}{\mathrm{M}_{\odot}} 
\newcommand{\Tsun}{T_{\odot}}

\graphicspath{ {./} }

\begin{document}

\title[Gravity experiments with radio pulsars]{Gravity experiments with radio pulsars}

\author[1]{\fnm{Paulo C. C.}\sur{Freire}}\email{pfreire@mpifr-bonn.mpg.de}
\author*[1]{\fnm{Norbert}\sur{Wex}}\email{wex@mpifr-bonn.mpg.de}

\affil*[1]{\orgname{Max-Planck-Institut f\"ur Radioastronomie}, \orgaddress{\street{Auf dem H\"ugel 69}, \city{Bonn}, \postcode{53121}, \country{Germany}}}


\abstract{
The discovery of the first pulsar in a binary star system, the Hulse--Taylor pulsar, 50 years ago opened up an entirely new field of experimental gravity. For the first time it was possible to investigate strong-field and radiative aspects of the gravitational interaction. Continued observations of the Hulse--Taylor pulsar eventually led, among other confirmations of the predictions of general relativity (GR), to the first evidence for the reality of gravitational waves.
In the meantime, many more radio pulsars have been discovered that are suitable for testing GR and its alternatives. One particularly remarkable binary system is the Double Pulsar, which has far surpassed the Hulse--Taylor pulsar in several respects. In addition, binary pulsar-white dwarf systems have been shown to be particularly suitable for testing alternative gravitational theories, as they often predict strong dipolar gravitational radiation for such asymmetric systems. A rather unique pulsar laboratory is the pulsar in a hierarchical stellar triple, that led to by far the most precise confirmation of the strong-field version of the universality of free fall. Using radio pulsars, it could be shown that additional aspects of the Strong Equivalence Principle apply to the dynamics of strongly self-gravitating bodies, like the local position and local Lorentz invariance of the gravitational interaction.
So far, GR has passed all pulsar tests with flying colours, while at the same time many alternative gravity theories have either been strongly constrained or even falsified. New telescopes, instrumentation, timing and search algorithms promise a significant improvement of the existing tests and the discovery of (qualitatively) new, more relativistic binary systems. 
}




\maketitle


\setcounter{tocdepth}{3} 
\tableofcontents


\newpage
\section*{Abbreviations}

The following list is a compilation of all (frequently) used abbreviations in this review: \\

\noindent
\begin{tabular}{ll}
    BH     & black hole \\
    CM     & centre of mass \\
    CMB    & cosmic microwave background \\
    DEF    & Damour--Esposito-Far{\`e}se \\
    DNS    & double neutron star \\
    EEP    & Einstein equivalence principle \\
    EM     & electromagnetic \\
    EoS    & equation of state \\
    GR     & general relativity \\
    GW     & gravitational wave \\
    ISM    & interstellar medium \\
    JFBD   & Jordan--Fierz--Brans--Dicke \\
    LIGO   & Laser Interferometer Gravitational-wave Observatory\\
    LLR    & Lunar Laser Ranging \\
    LT     & Lense--Thirring \\
    MoI    & moment of inertia \\
    MOND   & modified Newtonian dynamics \\
    MS     & main sequence star \\
    MSP    & millisecond pulsar \\
    NS     & neutron star \\
    PA     & position angle \\
    PK     & post-Keplerian \\
    PN     & post-Newtonian \\
    PPN    & parameterised post-Newtonian \\
    PTA    & Pulsar Timing Array \\
    SEP    & strong equivalence principle \\
    S/N    & signal-to-noise ratio \\
    SN     & supernova \\
    SSB    & Solar System barycentre \\
    TCB    & Temps-coordonn{\'e}e barycentrique (Barycentric Coordinate Time) \\
    ToA    & time of arrival \\
    UFF    & universality of free fall \\
    VLBI   & very long baseline interferometry \\
    WD     & white dwarf \\
    WEP    & weak equivalence principle 
\end{tabular}

\newpage
\section{Introduction}
\label{secion:introduction}

For many years, Einstein's general theory of relativity (GR), which was finalised in 1915 \citep{Einstein:1915ca}, could only be tested in the weak-field slow-motion regime of the Solar System\footnote{Attempts to measure the gravitation redshift with the help of white dwarfs remained inconclusive for quite a while due to observational and astrophysical challenges, never superseding Solar System tests (see e.g. \citealt{Holberg:2010}).}, and testing GR and its alternatives beyond their first post-Newtonian (PN) approximation was way out of reach. Then the discovery of the first binary pulsar by Russell Hulse and Joseph Taylor in the summer of 1974 \citep{ht75a} provided the physics community with a completely new testbed for investigating our understanding of gravity, space and time. With the discovery and continued observation of the Hulse--Taylor pulsar, aspects of the gravitational interaction of strongly self-gravitating bodies---specifically, two neutron stars (NSs)---could be investigated for the first time. Furthermore, at a quite early stage it was clear that the Hulse--Taylor pulsar provides a unique opportunity to test for the existence of gravitational waves (GWs) emitted by accelerated masses, which was confirmed with high precision in the following decade \citep{tw82,tw89}.

This was important for the opening of the GW window on the Universe: not only did it greatly facilitate the construction of ground-based GW observatories, but it also motivated the current attempts at detecting very low frequency GWs via pulsar timing arrays \citep{IPTA2019}, which at the time of writing seems close \citep{PTA_1,PTA_2,PTA_3}. This highlights the double use of pulsar timing, both for precise tests of gravity theories and for detecting low-frequency GWs from other sources.\footnote{There are also prospects for future tests of gravity with Pulsar Timing Arrays (PTAs). These are discussed in detail by \cite{ys13}.}

Meanwhile, many other radio pulsar systems were, and are being found that can be used for testing gravitational physics and our understanding of space and time. Depending on their orbital properties and the characteristics of their companions, these pulsars allow the study of different aspects of relativistic gravity and the derivation of the tightest constraints on some alternatives to GR; as an example, pulsar experiments with a millisecond pulsar (MSP) in a triple stellar system have provided the largest lower limit (around 150,000) for the Brans--Dicke parameter $\omega_\mathrm{BD}$ (see \citealt{vcf+20} and Sect.~\ref{sec:Nordtvedt}). Besides tests of specific gravity theories, there are pulsars that allow even for quite generic constraints on potential deviations from GR in the quasi-stationary strong-field regime and the radiative aspects of gravity. So far, GR has passed all the pulsar tests with flying colours! These experiments play a crucial complementary role to other modern gravity tests, such as those conducted by GW observatories \citep{LVC_2016_GR,LVC_2021_GR,LVKC_2021_GR} or the Event Horizon Telescope \citep{EHT_2019_VI,EHT_2022_VI}, as well as modern Solar System tests and tests on cosmological scales \citep{Berti_2015}. 

Despite the growth in the number of systems that we can use, and the numbers and types of experiments, this work stands on the foundations laid after Jocelyn Bell noticed the first pulsar on her chart, and later when PSR~B1913+16 was found by Joe Taylor and collaborators. These foundations include the experimental techniques that were developed for the precise timing of pulsars and binary pulsars, but also the many theoretical developments triggered by the discovery of ``the'' binary pulsar, which provide a solid conceptual foundation for our experiments. It is important to recognize, 50 years after the discovery of PSR~B1913+16, our great debt to those who opened the path for us.

In this review, we provide a summary of the tests of gravity theories with radio pulsars. This is not meant to be exhaustive, and covers only a bare minimum of necessary historical material, focusing instead on the results of greater significance for the study of gravity. In Sect.~\ref{sec:pulsar_intro}, we provide a very brief overview of pulsar radio emission, how it is detected and its data processed for timing purposes, and some information on the pulsars as astrophysical objects, especially the formation and evolution of binary pulsars. This is important for understanding the systems themselves, but also why some systems are better laboratories than others; the reader familiar with these topics can skip this section. In Sect.~\ref{sec:timing}, we present an outline of the pulsar timing technique and describe the relativistic effects that have been seen in binary pulsars, mostly via this timing technique. We give special emphasis to the ``post-Keplerian'' (PK) parameters (which quantify these relativistic effects) and their interpretation in GR, which is the basis of most of the tests discussed in this review. In Sect.~\ref{sec:GR_tests}, we discuss the main tests of GR with binary pulsars. In Sect.~\ref{sec:beyond_GR}, we discuss pulsar tests of our general understanding of gravity and gravitational symmetries, and a search for possible deviations from GR, with a particular focus on phenomena that are predicted by alternative theories of gravity. The non-detection of these phenomena represents strong constraints on such alternative theories. As an illustrative example, we will present pulsar tests of Damour--Esposito-Far{\`e}se (DEF) gravity in some more detail. Finally, in Sect.~\ref{sec:conclusions}, we summarize the results and discuss future prospects.

As a final comment, it is important to emphasise that the vast majority of the work discussed here appeared in the years since the last \textit{Living Review in Relativity} on this topic was published \citep{sta03}.


\section{Pulsars: the neutron stars and their radio emission}
\label{sec:pulsar_intro}

\subsection{Radio pulsars}

Jocelyn Bell's discovery of radio pulsars in 1967 \citep{hbp+68} was a complete surprise. The first radio pulsar, then known as CP~1919 (now known as PSR~B1919+21) showed an extraordinarily stable pulsation period of $1.3372795 \pm 0.0000020$ seconds. This stable periodicity is the feature that makes pulsars uniquely useful for a wide range of applications in astrophysics and, as discussed below, fundamental physics.

That the signals, first detected at a radio frequency of 81.5 MHz, were of astrophysical origin was soon firmly established. The discovery itself happened because Jocelyn Bell noticed that the signal reappeared, like the distant stars, 4 minutes earlier every day within the beam of the Cambridge radio telescope. Additional evidence that the signals were of interstellar origin was provided by the detection of \emph{dispersion}: the radio pulsations at lower radio frequencies arrive with a delay relative to the same pulsations at higher frequencies, as expected from a signal traveling through diluted ionised gas for distances of hundreds of parsecs (see Sect.~\ref{sec:dispersion}). Final proof was provided by the fact that the observed periodicity varied precisely with the Doppler shift caused by the projection of Earth's velocity along the direction to the pulsar \citep{hbp+68}. 

\begin{figure}[ht]
    \centering
    \includegraphics[width=0.7\textwidth]{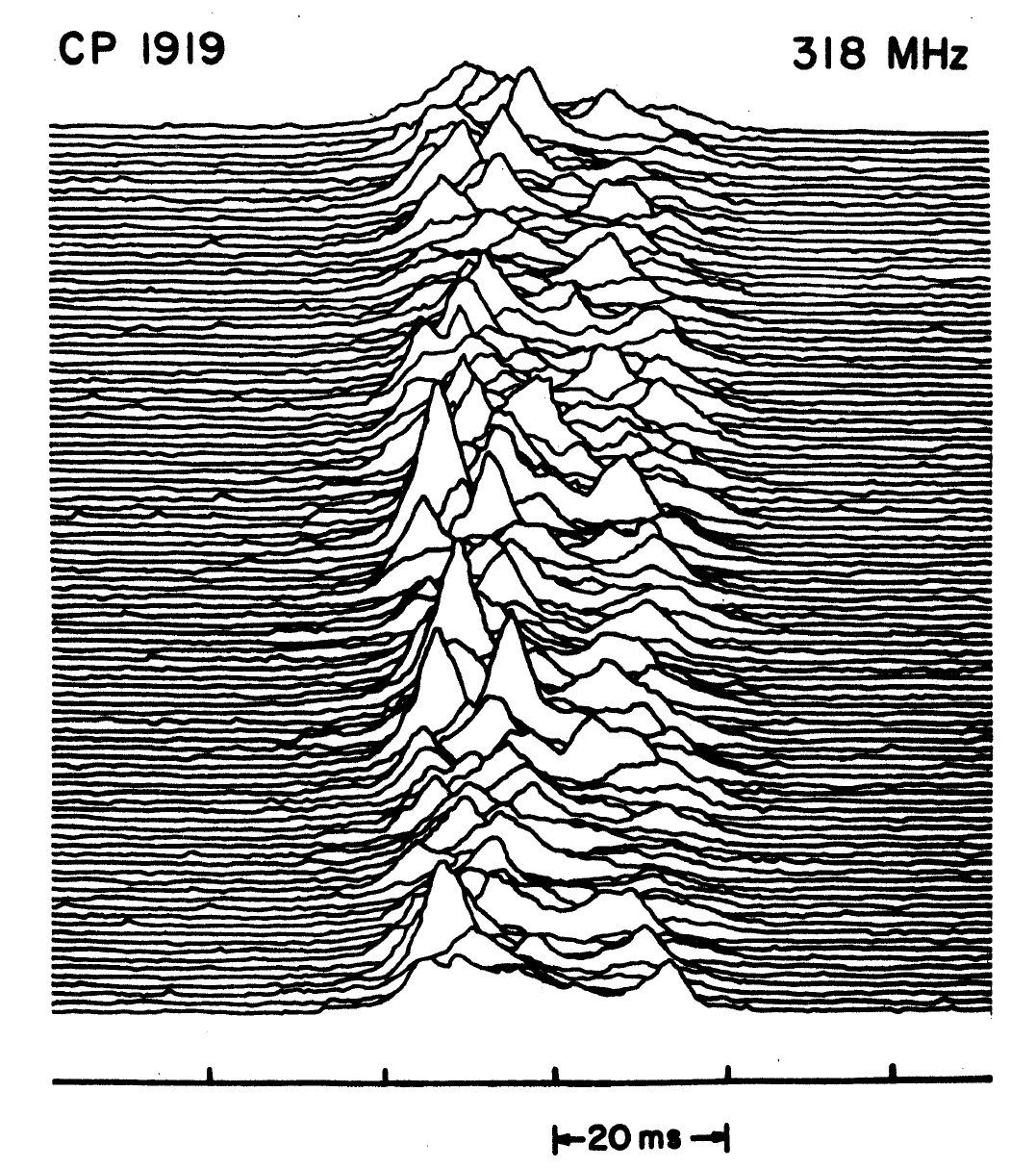}
    \caption{Sequence of radio pulses from PSR~B1919+21 observed in a radio intensity time series recorded at a radio frequency of 318 MHz with the Arecibo 305-m radio telescope. The figure shows how the intensity of the radio signal (vertical axis) changes as a function of spin phase (horizontal axis, increasing towards the right, only a short window of about 10\% of a full rotation is shown). Successive pulses are displayed with increasing vertical offsets for visibility. Note the bar, indicating a time interval of 20 ms, which shows the high time resolution of the signal. From \cite{cra70}.}
    \label{fig:CP1919}
\end{figure}

Figure~\ref{fig:CP1919} shows a well-known pulse train from this pulsar, recorded later at a radio frequency of 318~MHz \citep{cra70}. This shows clearly that the individual pulses are nearly random.
However, and most importantly, adding a large number of such pulses in phase, we arrive at a stable \emph{pulse profile}, which is characteristic of each pulsar.

\subsection{Neutron stars}

The earlier interpretations assigned the signal to the smallest sources known to exist at the time, white dwarf (WD) stars (see e.g., discussion in \citealt{hn68}). However, the discovery of the Vela and Crab pulsars \citep{lvm68,sr68}, and in particular the discovery of the very short periodicity of the latter pulsar, 33.091 ms \citep{ccl+69}, which is slowing down with time \citep{rc69}, forced the acceptance of the model that assigned the periodic pulsating signals to the rotation of a NS \citep{gol68,pac68}.

This was an extraordinary breakthrough for two main reasons:

\begin{enumerate}
\item It established beyond doubt the existence of NSs: that there is a dense, 
  stable remnant considerably more compact than WDs had been uncertain since 1934, when they were first proposed by \cite{bz34}.\footnote{Often \cite{lan32} is cited as the first to predict neutron stars. This, however, is strictly speaking not the case. \cite{lan32} puts forward the hypothesis of ``pathological cores'' made of densely packed protons and electrons, in which quantum mechanics fails, inside every star. A concept that, from a physical and astrophysical point of view, is far removed from real neutron stars. Moreover, Landau prepared and submitted this work even before the neutron was discovered (see e.g. \citealt{yhbp13} for details).} 
  The first detailed calculations of NS structures  \citep{tol39,ov39} resulted in estimates of the maximum NS mass between 0.7 and a few solar masses. With diminutive radii of $\mathcal{O}(12 \, \mathrm{km})$ and masses similar to that of the Sun, their central densities are higher than that of atomic nuclei. Owing to their compactness, the calculation of NS structures requires a fully general-relativistic hydrostatic equilibrium equation, together with detailed information on the macroscopic properties (especially pressure) of nuclear matter at these high densities, its equation of state (EoS) \citep{tol39,ov39}.
  Even today, the unknown state of matter at these densities implies that there are still significant uncertainties in its EoS, which result in $\mathcal{O}(20\%)$ uncertainties in the  maximum mass, radius and moment of inertia (MoI) of NSs, making this a very active topic of research with implications not only for many branches of astrophysics \citep{of16} but also for the understanding of the strong nuclear force \citep{lat21}.
\item The association of the Vela and Crab pulsars with well-known supernova (SN) remnants 
  (the Crab nebula is associated with a historical SN observed in the constellation Taurus in 1054 AD, with detailed records from China and Japan, \citealt{sg02})  implied that some types of core-collapse SNe mark the birth of a NS \citep{jan12,bv21}. Thus, NSs represent the end products of the evolution of some types of massive stars, with the collapse of their cores (and the associated neutrino burst) powering the SN explosion, as had been suggested since the 1930s \citep{bz34,zwi38}.
\end{enumerate}


\subsection{Characteristics of the radio emission}

Despite all the previous work on NSs, the discovery of pulsars was surprising because no one predicted their radio emission. This is still the case today: 56 years after the discovery of PSR~B1919+21, the radio emission from pulsars remains poorly understood (but see \citealt{fts20}). However, this does not hinder in the least the use of pulsars for the experiments described below.

In the radio sky, pulsars appear as faint, point-like radio sources. This faint emission is generally observed at frequencies of a fraction to a few GHz, i.e., decimetric to metric wavelengths. It is a broadband phenomenon, without any recognisable spectral features, with a maximum of emission around a couple hundred MHz and a steep decrease in power at higher frequencies, which means that the vast majority of pulsars become undetectable at frequencies above a few GHz. However, about 10\% of known radio pulsars have associated gamma-ray emission \citep{saa+23}.

One of the most prominent characteristics of the radio emission is its high degree of polarisation, with percentages of both linear and circular polarisations that are much higher than in other radio sources.
This points to another fundamental characteristic of this emission, its \emph{coherence}: the effective temperatures required from an object produce the observed radio emission via thermal blackbody emission (i.e., its \emph{emission temperature}) are of the order of $10^{25}\, \rm K$ for most pulsars, which is too high.
For additional details, see \cite{lg12,lk12}.


\subsubsection{The lighthouse model}

Despite the lack of a good model for the radio emission, a geometric model was firmly established soon after the discovery of PSR~B1919+21. The pulsar is a magnetised NS, and the radio emission is highly anisotropic, emerging mostly from regions close to open magnetic field lines in the vicinity of the magnetic poles. Because the magnetic axis is generally misaligned with the spin axis, distant observers only detect radio emission for the time (within each rotation of the NS) that some part of the emission beam is pointing to the Earth. The overall effect is similar to the apparent pulsations of a lighthouse.
This hypothesis is confirmed by the observation that the position angle (PA) of the linear polarisation is seen to change with the spin phase in a way that is consistent with the changing orientation of the magnetic field lines directly along the line of sight to the observer \citep{rc69c}. From this change of the PA with spin phase the rotation geometry can be determined, especially the angle between the magnetic field and the spin axis and the angle between the spin axis and the line of sight (for details, see \citealt{lk12}).


\subsubsection{Time domain signals}

Figure~\ref{fig:CP1919} illustrates a fundamental aspect of most pulsar observations: that they are  \emph{time-domain} observations, where a 1-dimensional time signal from a restricted  location in the sky is recorded with high time resolution, instead of a 2-dimensional spatial signal as when an image of the sky is being produced. 

This fundamental difference implies that while imaging observations of many radio sources require the use of \emph{radio interferometers}, like the Very Large Array (VLA) near Socorro, New Mexico (USA), the MeerKAT 64-dish array near Carnarvon, South Africa \citep{MeerKAT}, and the upgraded Giant Metrewave Radio telescope (uGMRT) near Narayangaon, India \citep{uGMRT}, or {\em scans} with single dish telescopes, sometimes using multiple receivers in the focal plane of the telescope (``multi-pixel" receivers), the use of such imaging techniques is not required for most pulsar observations, which can also be made by single radio dishes with single-pixel receivers---the poorer imaging resolution is in this case completely irrelevant.

Given the faintness of pulsar emission, what matters above all for pulsar observations is the sensitivity of the radio telescope. Currently, the most sensitive radio telescope in the world for pulsar observations is a single dish, the Five hundred meter Aperture Spherical Telescope (FAST), in Guizhou province, China \citep{FAST}, a design inspired on (and the first to surpass the sensitivity of) the 305-m  William E.\ Gordon radio telescope near Arecibo, Puerto Rico, USA, which is no longer in operation. Other telescopes used extensively for pulsar searches are the largest fully steerable radio telescopes in the world, the 100-m Green Bank radio telescope in Green Bank, West Virginia, the 100-m Effelsberg radio telescope near Effelsberg, Germany, the 75-m telescope at Jodrell Bank near Manchester, UK and the 64-m Murriyang radio telescope near Parkes, Australia. Although not fully steerable, the Nan\c{c}ay radio telescope in Nan\c{c}ay, France is also extensively used for pulsar observations.

When observing radio pulsars (and many other types of sources), these radio telescopes use high sensitivity, broadband single-pixel receivers. These produce two voltage streams, one for each radio polarisation; these should be proportional to the electric field (the \emph{amplitude})  of the incoming wave. In modern observing systems, these voltages are sampled and digitised at rates of at least twice the bandwidth being registered, which typically implies a few times $10^9$ samples per second (e.g., \citealt{MeerKAT}). These two voltage streams---and the local time references---provide all the information in pulsar observations. From here on we describe how these data are processed and reduced.

From these two digitised voltage streams, a digital spectrometer (also called in radio instrumentation a ``filterbank") produces spectra, either using fast Fourier transforms (FFTs) or, for better channelisation, polyphase filterbank techniques. The number of frequency channels in the spectra varies according to the needs of different observations: a few hundred to a few thousand for pulsar observations, tens of thousands for observations of radio spectral lines. 

In pulsar observations, at least one spectrum is recorded, for the total intensity of the signal ($\mathcal{I}$), obtained from the addition of the squares of the voltages from each of the two polarisations. This process is called \emph{detection}: $\mathcal{I}$ should be proportional to the \emph{power} of the incoming radio waves.

However, four spectra can be computed and recorded at the same time, one for each of the {\em Stokes parameters}: in addition to $\mathcal{I}$ these include $\mathcal{Q}$, $\mathcal{U}$ and $\mathcal{V}$ \citep{lk12}, thus preserving the polarisation characteristics of the signal. This is important for pulsars, since as mentioned above they are highly polarised sources. The Stokes parameters are the components of the 4-dimensional {\em Stokes vector}.
The rationale for describing the polarisation characteristics of the signal using Stokes vectors is that they can, like any other vectors, be added as necessary. This means that we can add all polarisation measurements occurring at a particular spin phase of the pulsar to derive a high S/N measurement of the average polarisation of the radio emission at that phase. The correct derivation of the Stokes parameters requires careful calibration of the two voltage streams.

The limiting time resolution within each spectral channel is the inverse of its bandwidth; for pulsar observations the latter is of the order of MHz, thus the time resolution is typically of the order of microseconds ($\mu$s). Consecutive samples are integrated in all channels, and the resulting spectra are recorded, typically once every few tens of $\mu$s; this is generally done in order to reduce the data rates. In any case, this timescale is much faster than for most other astronomical observations.


\subsubsection{Dispersion and dedispersion}
\label{sec:dispersion}

Since pulsars have no narrow spectral features, one might wonder about the need for obtaining spectral information. This is primarily because of the phenomenon of \emph{dispersion}, and secondarily for the purposes of the rejection of radio frequency interference, which tends to appear in a limited number of spectral channels.

Dispersion happens because the group velocity of radio waves in a cold, diluted plasma is smaller than the speed of light in vacuum, $c$. If the radio frequency $F$ is much higher than the plasma frequency anywhere along the line of sight, then the accumulated delay for the wave arriving at the Earth is given by the expression (in cgs units):
\begin{equation}
  \Delta t_\mathrm{dis}(F) = \frac{e^2}{2 \pi m_\mathrm{e} c} \frac{1}{F^2} \,
    \int_\mathrm{pulsar}^\mathrm{Earth} n_\mathrm{e}(l) \, dl \,,
\end{equation}
where $F$ is the radio frequency, $e$ and $m_\mathrm{e}$ are the charge and mass of the electron, and $n_\mathrm{e}$ is the density of free electrons of the interstellar medium (ISM) at a distance $l$ along the propagation path. If $l$ is specified in parsecs and $n_\mathrm{e}$ in $\rm cm^{-3}$, then the integral---the electron \emph{column density} between the pulsar and the Earth---is expressed in $\rm cm^{-3}\, pc$ and is known as the {\rm dispersion measure} (DM). Specifying $F$ in GHz, we obtain:
\begin{equation}
  \Delta t_\mathrm{dis}(F) = 4.1488 \, \frac{\rm DM}{F^2} \, \mathrm{ms} \,.
\label{eq:disp}
\end{equation}
If the dispersive delays are not subtracted, the pulsar signal will be smeared in time, and the pulsar signal lost \citep{lk12}.

This subtraction can be done in two ways, both requiring the use of channelisation by a spectrometer. The simpler is to move the detected signal of each spectral channel at frequency $F$ forward by $\Delta t_\mathrm{dis}(F)$, this is known as \emph{incoherent dedispersion}. The advantage is the simplicity and small amount of computing effort required, while the disadvantage is that the dispersive smearing within each channel is not removed. The method used for most modern timing observations is to remove the effect coherently, \emph{before} detection: after the two voltage streams are Fourier transformed, a rotation proportional to $\Delta t_\mathrm{dis}(F)$ is applied to the complex number in the Fourier spectrum at frequency $F$. Fourier transforming this signal back to the time domain gives two voltage streams that are apparently unaffected by the ISM. From these voltage streams, the Stokes parameters are computed as described above. This process is known as \emph{coherent dedispersion} \citep{han71}. The advantage is that it eliminates the dispersive smearing within each channel, while the disadvantage is the large computational power required and the need for precise \emph{a priori} knowledge of the DM.

After dedispersion, all spectral channels can be added in frequency, producing a data set recording the variation of the Stokes parameters with time---a \emph{time series}---which is the final product of the dedispersion procedure.


\subsubsection{Individual pulses and the average pulse}
\label{sec:pulses}

The lighthouse model implies that by keeping track of the radio pulsations, we can in principle measure precisely how the number of rotations of the NS $N$ changes with the proper time $T$ of the reference frame of the pulsar (see Sect.~\ref{sec:isolated}).\footnote{The proper time of the pulsar, as for instance measured on its surface, is proportional to $T$. The corresponding (constant) factor is irrelevant for pulsar timing observations.} 
To leading order, the relation is given by:
\begin{equation}
  N(T) = \frac{\phi(T) - \phi_0}{2 \pi} = \nu (T - T_0) + \frac{1}{2} \dot{\nu}(T - T_0)^2 \,,
\label{eq:spin}
\end{equation}
where $\phi(T)$ and $\phi_0$ are the spin phases at time $T$ and a reference time $T_0$ (these phases are not limited to the interval between 0 and $2\pi$, they increase continuously with time), $\nu$ is the spin frequency of the NS and $\dot{\nu}$ is its time derivative, both measured at $T_0$. For spin-powered pulsars, the rotation slows down with time (as mentioned above for the Crab pulsar), implying that $\dot{\nu}$ is negative.

In the intensity time series of the first pulsar (Fig.~\ref{fig:CP1919}, and Fig.~1 of \citealt{hbp+68}) it was already apparent that there is considerable variation not only in the strength, but also in the particular shape of each pulse. This means that, by observing individual pulses, it is difficult to measure $N(T)$. However, if in Fig.~\ref{fig:CP1919} we add all individual pulses on top of each other---a process known as \emph{folding}---we recover a stable average pulse profile. Thus, folding is vitally important, not only to increase the S/N of the signal, but also because it makes timing possible.

To fold a time series properly, one needs a good \emph{a priori} estimate of spin frequency of the pulsar as seen at the receiving radio telescope at the local time $\tau$\footnote{This is generally measured with a local time standard, like a Hydrogen maser---clock corrections are generally added later by the timing programme to convert this to a well-known time scale, like Universal coordinated time (UTC).} 
when the observation is occurring, $\tau_\mathrm{obs}$. This local spin frequency $\nu_\mathrm{obs}$ changes constantly because of the constantly changing conversion factor between $T$ and $\tau$ (see Sect.~\ref{sec:timing}). If this is not taken into account, the phase of the radio emission will appear to drift with time. The prediction of $\nu_\mathrm{obs}$ is made by a timing programme using the best available pre-existing timing model for a pulsar, its \emph{ephemeris}, and the best description of the motion of the radio telescope relative to the Solar System barycentre (SSB).

For most pulsars, there are significant deviations from the simple spin-down given by Eq.~(\ref{eq:spin}), which are generally categorised as either ``timing noise'' or ``glitches'' \citep{lk12}. However, for some types of pulsars---especially the \emph{recycled} pulsars, described in the next section---the rotation can be described by Eq.~(\ref{eq:spin}) to a very good approximation. This means that they have an extremely stable rotation, making them ideal tools for the types of experiments described in this review.


\subsubsection{The time of arrival}
\label{sec:ToA}

This stable profile allows new measurements of $N(\tau)$ from the data of the observation. This could be done by measuring $N$ at specific values of $\tau$, but instead the convention used in pulsar astronomy is to measure $\tau$ at specific values of $N$, in particular its integer values. This is how it is done: a high S/N version of the pulse profile is used as a \emph{template}, with its zero phase representing the reference longitude on the NS. Like longitudes on Earth, this has an arbitrary element to it; it can be set, for instance, at the peak of radio emission. However, once this convention is established, it is very important to use it consistently for the same pulsar, or at least for measurements taken with the same instrument and frequency. This template is then correlated with individual pulse profiles obtained, in each observation, after dedispersion and folding of its data to derive the so-called \emph{time of arrival} (ToA), $\tau_i$ (for details, see e.g. \citealt{tay92}).

Although the ToA uncertainties can be very small for some pulsars (e.g., $0.1 \, \mu$s for the timing of PSR~J1909$-$3744, \citealt{lgi+20}), such precision is very rare. The limitations stem from several of the aforementioned characteristics of the pulsar radio signal. The faint signal implies that the ToA precision of most recycled pulsars is limited by the S/N of the detections. Furthermore, small DM variations introduce additional frequency-dependent noise into the timing. Another problem, which dominates in the few cases when the S/N of the detections is very high, is \emph{pulse jitter}. This is caused by the randomness of the individual pulses, which can, in some cases, take a long time to average into ``the'' average profile of the pulsar. Furthermore, the lack of sharp features in the profile can further limit the timing precision. Finally, long-term deviations from Eq.~(\ref{eq:spin})---the timing noise and glitches---can also degrade the timing of some recycled pulsars.


\subsection{Pulsar evolution and binary pulsars}
\label{sec:evolution}

The vast majority of known pulsars are found in our Galaxy and its retinue of globular clusters: the most distant pulsars known to us are located in two satellite galaxies of the Milky Way, the LMC and SMC, located about 50 and 60 kiloparsec (kpc) away from Earth, respectively. Because of their intrinsic faintness, the sensitivity of the observing instruments remains as the main limitation in the discovery of pulsars: the known pulsar population (3473 pulsars at the time of writing, \citealt{mhth05}) is thought to represent only a few percent of the likely population of active pulsars in our own Galaxy, and a tiny fraction of its $\sim 10^9$ NSs.

\begin{figure}[ht]
    \centering
    \includegraphics[width=0.9\textwidth]{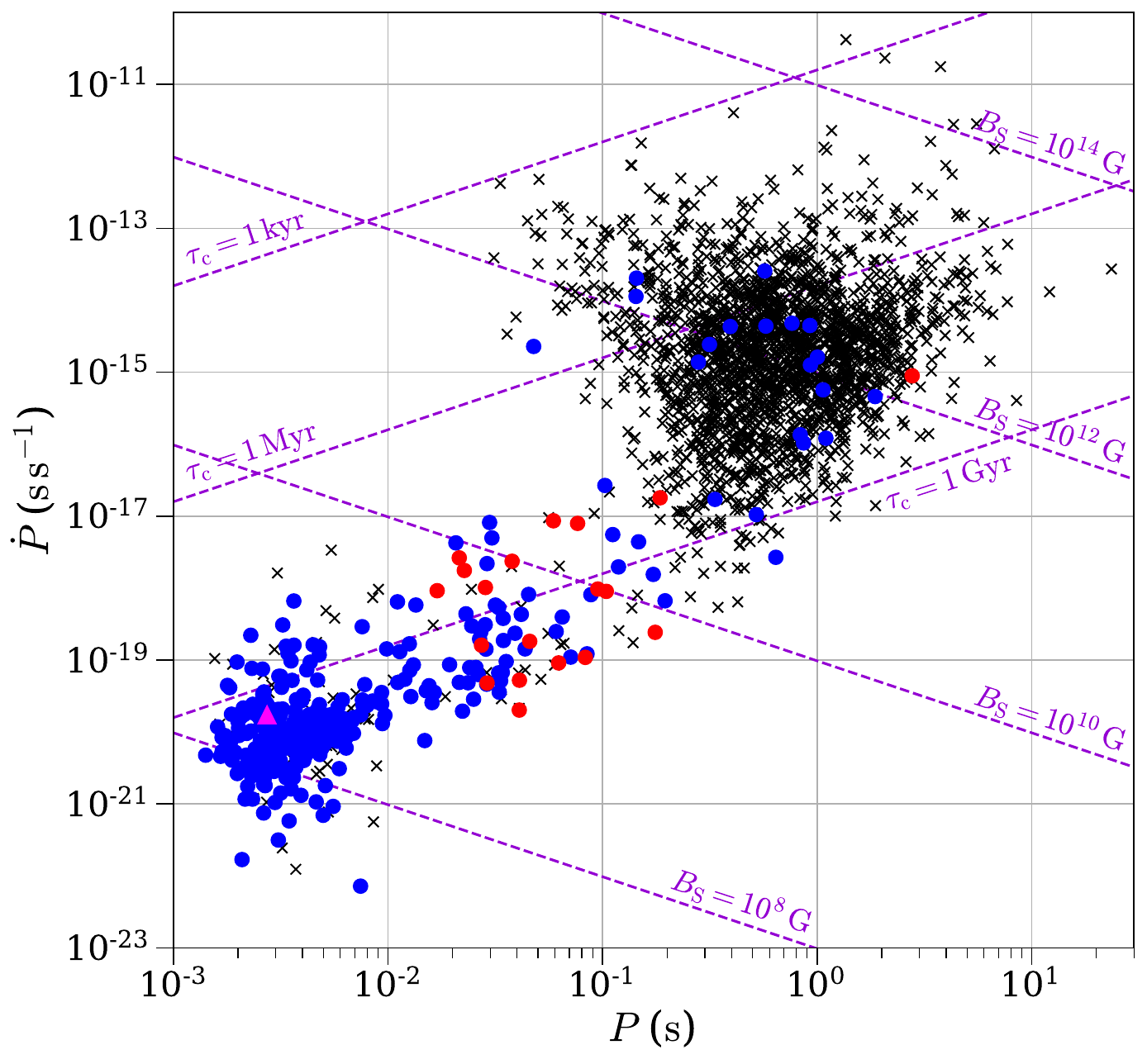}
    \caption{Period--(intrinsic) period derivative diagram for radio pulsars, taken from \cite{mhth05}. Double NS systems are highlighted by red and all other binary pulsars by blue circles. The pulsar in the stellar triple system is marked by a magenta triangle. The rest, i.e.\ the isolated pulsars, are marked with black crosses. Dashed lines indicate constant characteristic age $\tau_\mathrm{c}$ and surface magnetic field strength $B_\mathrm{S}$ (labeled accordingly).}
    \label{fig:ppdot}
\end{figure}


\subsubsection{Normal and recycled pulsars}

In Fig.~\ref{fig:ppdot}, we plot the spin period ($P = \nu^{-1}$, horizontal axis) and the spin period derivative ($\dot{P} = - \dot{\nu} \nu^{-2}$, vertical axis) for about 3000 rotation-powered pulsars (black crosses) for which these parameters have been measured \citep{mhth05}. From these two parameters, the age, magnetic field and energetics of the pulsars can be inferred (dotted lines) using a variety of assumptions, like in this case that the spindown is caused by the emission of a low-frequency electromagnetic (EM) wave, as one would expect from a rotating magnetic dipole, and that the dipole is orthogonal to the spin axis \citep{lk12}. For the rate of rotational energy loss,
\begin{equation}
  \dot{E}_\mathrm{rot} = -4 \pi^2 I \frac{\dot{P}}{P^3} \,,
\label{eq:E-dot}
\end{equation}
a MoI ($I$) of $10^{45} \rm \, g \, cm^{2}$ is generally assumed \citep{lk12}.

Clearly, there are two main groups of pulsars in this diagram: the ``normal'' pulsars form the central, more numerous group, with $0.1\, < \, P \, < 5$ s and $10^{-17}\, < \,\dot{P} \, < \, 10^{-11} \rm \, s \, s^{-1}$. In the lower left, with values of $P$ and $\dot{P}$ (and magnetic fields) three orders of magnitude smaller, are the ``recycled'' pulsars. Unlike the normal pulsars, they were spun up (and their magnetic fields degraded) by accretion of mass from a stellar companion after they formed. This is indicated by the fact that about 80\% of these pulsars are in binaries (indicated by the filled circles), while among the normal pulsars this is less than 1\%.


\subsubsection{Binary pulsars}
\label{sec:binpsr}

The two main types of binary pulsars are indicated in Fig.~\ref{fig:ppdot} by the coloured dots: red for pulsars with NS companions and blue for pulsars with WD (and other types of) companions.

How do these systems form? This is important for understanding why some systems are more useful than others for GR tests. A detailed account of their evolution is given by \cite{tv23}, and in what follows we present a very summarised outline. For simplicity, we start with a binary system consisting of two main sequence (MS) stars. This is a common occurrence, especially for massive stars, the majority of which form in binary and multiple systems.

As the more massive star (the primary) evolves, it eventually explodes as a SN, forming a normal pulsar. The system will then very likely disrupt owing to the large kick\footnote{The average kick for young isolated radio pulsars is $400-500 \, \mathrm{km\,s^{-1}}$, but ranges from barely measurable to more than $1500 \, \mathrm{km\,s^{-1}}$ \citep{hllk05}.} 
and mass loss associated with this SN. We know this not only from the aforementioned fact that $>99\%$ of normal pulsars are not in binary systems, but also from the observation that the few surviving normal pulsar - massive MS star systems (e.g., \citealt{jml+92,lsk+15}) have very high orbital eccentricities (generally $e \sim 0.9$), indicating a near disruption.

Eventually, the pulsar might cease emitting radio waves. The secondary will also evolve and become a giant star. Its large size will cause several effects, the first being tidal circularisation of the orbit. When the secondary fills its Roche lobe, the transfer of matter to the NS starts, and the system becomes an X-ray binary. The transfer of matter can become unstable, in which case the system might go through a common envelope phase. At this stage, the NS is slowly spun up, and its spin will become aligned with the orbital angular momentum. It is thought that during this process the magnetic flux density at the surface of the NS is ablated, becoming much smaller.

What happens next depends on the mass of the secondary:
\begin{itemize}
\item If the secondary is light, then it will slowly evolve into a WD star, in this case the 
  system will retain both the very low orbital eccentricity and the alignment between the NS spin and orbital angular momentum acquired during the recycling phase.
\item If it is massive enough, it will also go through a SN explosion, the system then either 
  disrupts (a likely occurrence) or forms a double NS system (DNS, for a detailed discussion, see \citealt{tkf+17}). In most cases we observe only the recycled pulsar because their NS companions are only observable as normal pulsars for a short amount of time. However, in one case (the PSR~J0737$-$3039A/B system), we see both NSs as pulsars: as predicted by the laws of stellar evolution, one of them (the first-formed NS) is a recycled pulsar (PSR~J0737$-$3039A, with $P = 22.7 \, \mathrm{ms}$, \citealt{bdp+03}) and the second-formed NS (PSR~J0737$-$3039B, with $P = 2.77 \, \mathrm{s}$) is a normal pulsar \citep{lbk+04}.
\end{itemize}

After the demise of the companion, the primary is now spinning fast and emitting radio waves again - this is why it is known as a recycled pulsar. 

Fully recycled pulsars (the MSPs) appear overwhelmingly in the lower left corner of Fig.~\ref{fig:ppdot}, with spin periods of a few ms. Most of them have low-mass companions. The pulsars in DNSs have spin periods that are one order of magnitude larger. The reason is the time they spend as X-ray binaries: this is much shorter for the NSs with high-mass companions, because the latter evolve much faster.

The high risk of disruption makes the DNSs relatively rare ($<30$ known, representing less than 10\% of the binary pulsar population and less than 1\% of the total pulsar population). These systems are necessarily eccentric because of the mass loss and the kick associated with the second SN\footnote{The range of kicks could be smaller for NSs born in closely interacting binary systems, as is the case for the second-formed NS in DNS systems, but nevertheless many systems are still expected to disrupt at this stage} 
\citep{tlm+13,tlp15,tkf+17}. Because of this kick, the orbital plane after the SN might become very different from what it was before the explosion, causing a misalignment between the spin of the recycled pulsar and post-SN orbital angular momentum.

The energy loss due to GW damping in compact DNSs observed for the first time in PSR~B1913+16 (see Sect.~\ref{sec:psrB1913}) has an inevitable consequence: every year the pulsar and companion come about 3.5\,m closer, and in 300 million years---a short time compared to the age of the Universe---this system will coalescence in a NS-NS merger, producing extreme amounts of GW emission. This discovery meant that future ground-based observatories will have a secure source of GWs. This was repeatedly confirmed by the discovery of DNSs with even smaller coalescence times in our Galaxy \citep{bdp+03,cck+18,sfc+18}. The observation by the Advanced LIGO and Virgo of precisely such a NS-NS merger event in August 2017, GW170817 \citep{LVC2017} and its EM counterpart \citep{aaa+17} in the galaxy NGC~4993 not only confirmed beyond doubt the connection between (a certain class of) gamma-ray bursts with the collision of two NSs, but represented the fulfillment of the promise brought by the discovery of PSR~B1913+16.


\section{Timing and relativistic effects in binary pulsars}
\label{sec:timing}

Most (but not all) of the applications discussed below rely on a technique where pulsars excel, timing. This technique yields highly precise measurements of many astrophysical parameters, and is more precise (and therefore powerful) for recycled, fast-spinning pulsars.

In this technique, we use the ToAs to infer fundamental characteristics of the pulsar. This is done using a \emph{timing programme}: by far, the most commonly used programmes are \textsc{tempo} \citep{tempo}, \textsc{tempo2} \citep{tempo2_1,tempo2_2} and ``PINT is not tempo'' (\textsc{pint}, \citealt{pint}). These programs use the pulsar ephemeris and Eq.~(\ref{eq:spin}) to calculate predictions for ToAs ($T_{i,p}$) that are within less than half a spin period from the observed ToAs $T_i$. The programme determines the optimal timing parameters by minimising the sum of the squares of the normalised residuals:
\begin{equation}
  \chi^2 = \sum_{i=1}^{n} \left( \frac{T_i - T_{i, p}}{\sigma_i} \right)^2 \,,
\end{equation}
where $n$ is the number of ToAs and $\sigma_i$ is the measurement uncertainty of $T_i$. If all residuals are small compared to the spin period and if the normalised residuals have a normal distribution around zero, this is evidence that the ephemeris is precise enough to determine, without ambiguity, the number of rotations between any two ToAs, i.e, it is a \emph{phase-coherent ephemeris}. Without unambiguous knowledge of these rotation numbers, no reliable ephemerides can be derived. In a good timing model, the value of $\chi^2$ is similar to the number of degrees of freedom in the fit, which is the number of ToAs minus the number of parameters being fitted.

As can be deduced from Eq.~(\ref{eq:spin}), any small error in $\nu$ will produce a linearly increasing residual, and a small error in $\dot{\nu}$ will produce a quadratically increasing residual. Thus, by adjusting $\nu$ and $\dot{\nu}$, the timing programme can make any such trends in the residuals disappear.


\subsection{Timing isolated pulsars}
\label{sec:isolated}

As discussed above, Eq.~(\ref{eq:spin}) is calculated assuming a time ($T$) in the reference frame of the pulsar. However, as discussed below, we do not know the line-of-sight velocity of the pulsar, so $T$ is not directly accessible to us. Nonetheless, a suitably Doppler-shifted version of this time can be calculated as \citep{tay94,tempo2_2}:
\begin{equation}
  T' = \underbrace{\tau_\mathrm{obs} - \Delta_\mathrm{E, \odot}}_{t_\mathrm{obs}} - \, t_0
       - \Delta_\mathrm{R, \odot} - \Delta_\mathrm{S, \odot} - \Delta t_\mathrm{dis}(F) \,,
\label{eq:delays}
\end{equation}
where $t_\mathrm{obs}$ is the coordinate time (in TCB) of the ToA, $t_0$ is a reference epoch (in TCB), $\Delta t_\mathrm{dis}(F)$ is given by Eq.~(\ref{eq:disp}) and the multiple $\Delta_\odot$ terms correspond to three different delays:
\begin{enumerate}
\item $\Delta_\mathrm{E, \odot}$, the \emph{Einstein delay} represents the time part of the 
  full four-dimensional transformation between the SSB coordinate time $t_\mathrm{obs}$ and the proper time of the observer $\tau_\mathrm{obs}$, along the observer's world-line. To leading order, it is a result of the special-relativistic time dilation due to the motion of the observer in the SSB and the gravitational redshift caused by the masses in the Solar System (\citealt{iau2000,tempo2_2}; note that there are different sign conventions for this term in the literature). This delay is independent of the direction to the source.
\item $\Delta_\mathrm{R, \odot}$, the \emph{``R{\o}mer'' delay}, corresponds (in the Newtonian limit) to the geometric light-propagation delay. 
  This corresponds to the projection, along the direction of the incoming wave from the pulsar (unit vector $\hat{\mathbf{n}}$), of the position vector of the receiving radio telescope in the frame of the SSB at time $t_\mathrm{obs}$. This is the sum of two (coordinate-position) vectors: the first is the position vector of the radio telescope relative to the Earth's centre; to calculate it, accurate coordinates for the radio telescope are necessary, together with accurate information on the orientation of the Earth. The second, the position vector of the Earth's centre relative to the SSB, is calculated using a Solar System ephemeris. 
  Finally, a direction to the pulsar (Right Ascension $\alpha$ and Declination $\delta$, unit vector $\hat{\mathbf{K}}_0 = -\hat{\mathbf{n}}$) must be assumed in order for the timing programme to calculate the projection correctly. 
\item $\Delta_\mathrm{S, \odot}$, the \emph{Shapiro delay} \citep{sha64}, is a relativistic 
  light propagation delay caused by the curvature of the Solar System’s spacetime. This has been measured extensively with the aid of planetary radar and space probes moving in the Solar System (e.g., \citealt{bit03}). Like $\Delta_\mathrm{R, \odot}$ it is also direction-dependent and can be calculated precisely once $\alpha$, $\delta$ are specified.
\end{enumerate}
Small errors in $\alpha$, $\delta$ will produce a nearly sinusoidal 1-year trend in the residuals, which indicate an incorrect subtraction of Earth's motion. Adjusting $\alpha$ and $\delta$ the timing programme can make such trends disappear. When a growing sinusoidal in the residuals is observed, then additional linear variations of these terms ($\dot{\alpha}$, $\dot{\delta}$) can be fitted, from which we can derive the \emph{proper motion} of the pulsar ($\mu_{\alpha} = \dot{\alpha} \cos \delta$, $\mu_{\delta} = \dot{\delta}$). Finally, if a pulsar is close enough to the Solar System and its timing precision is good enough, then the \emph{parallax} ($\varpi = 1/d[\mathrm{au}]$, where $d$ is the distance to the pulsar measured in astronomical units) can also be measured from the timing. These ``astrometric'' parameters of the pulsar ($\alpha,\delta,\mu_{\alpha},\mu_{\delta},\varpi$) can also be measured with very long baseline interferometry (VLBI, see examples in \cite{dbl+13,ddf+21,ksm+21,dds+23}).

\begin{figure}[ht]
\centering  
\includegraphics[width=11cm]{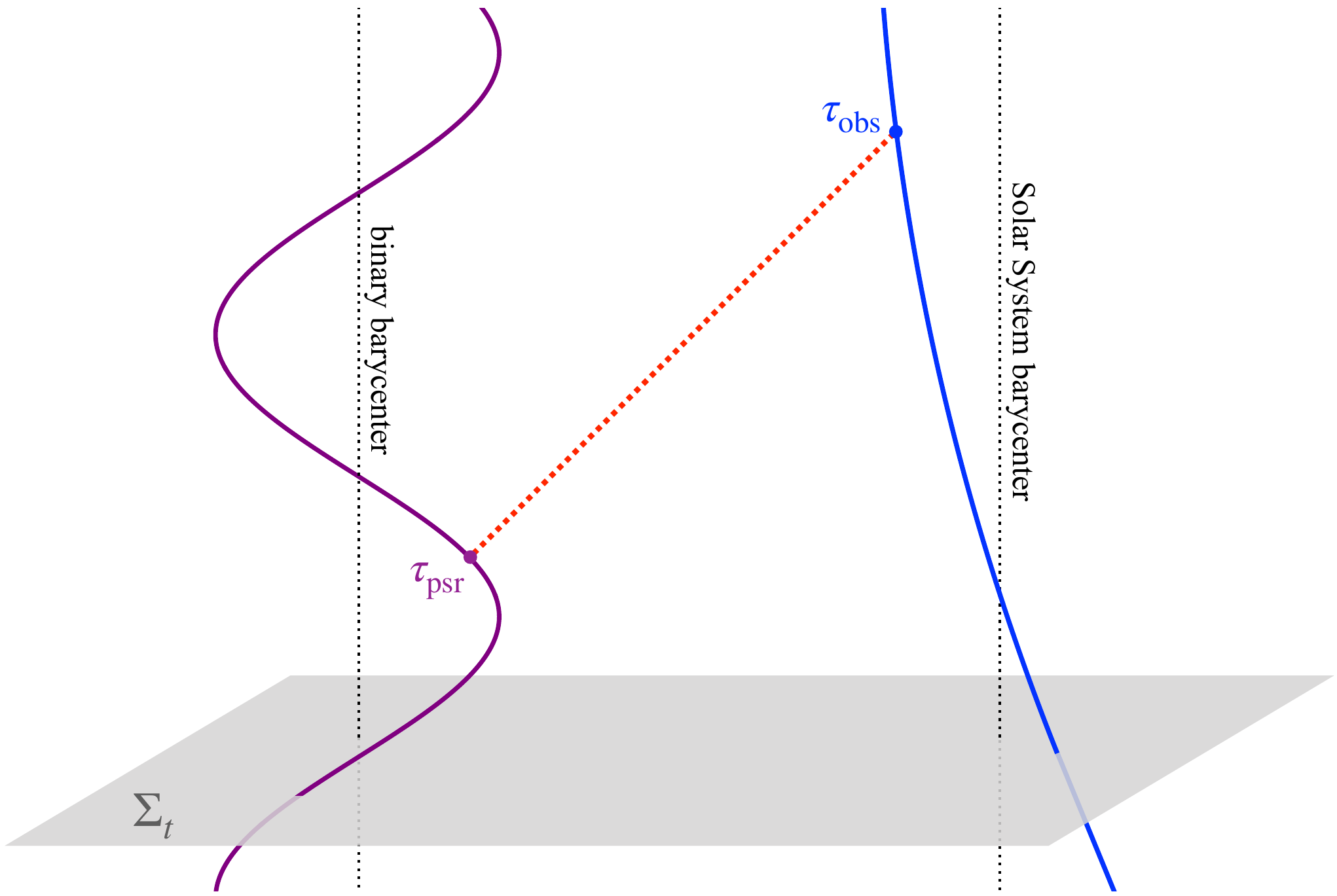}  
\caption{Spacetime perspective of timing observations of a binary pulsar. Depicted are the world lines of pulsar (purple) and the observer (blue). A photon is emitted by the pulsar at the pulsar's proper time $\tau_\mathrm{psr}$ (which is proportional to $T$ in Eq.~\ref{eq:delays2}) and arrives at the observer at the observer's proper time $\tau_\mathrm{obs}$. For simplicity, we have assumed that the barycentres of the binary and the Solar System are at rest with respect to each other. $\Sigma_t$ depicted a hyper-surface of constant coordinate time $t$ (corresponds to TCB).}
\label{fig:spacetime}
\end{figure}


\subsection{Timing binary pulsars}

In Fig.~\ref{fig:spacetime}, we depict the more complex (and more interesting) situation that occurs when the pulsar itself is in a binary system. In this case, Eq.~(\ref{eq:delays}) becomes (where we ignore the constant factor between $T$ and $T'$):
\begin{equation}
T = \tau_\mathrm{obs} - \Delta_\mathrm{E,\odot} - t_0 
    - \Delta_\mathrm{R,\odot} - \Delta_\mathrm{S,\odot} - \Delta t_\mathrm{dis}(F) 
    - \Delta_\mathrm{R} - \Delta_\mathrm{S} - \Delta_\mathrm{E} - \Delta_\mathrm{A} \,,
\label{eq:delays2}
\end{equation}
where the four last terms are, respectively, the R{\o}mer, Shapiro, Einstein and aberration delays of the binary. These delays are described in detail in the following section. We now make a few general considerations about them:

\begin{itemize}
\item Any quantities with units of mass, length or time (for instance, the delays $\Delta_\mathrm{ R}$, $\Delta_\mathrm{ S}$, $\Delta_\mathrm{ E}$, $\Delta_\mathrm{A}$, the spin period of the pulsar, $P_\mathrm{int}$) measured in the reference frame of its centre of mass (CM) are related to the same quantity measured at the SSB ($P_\mathrm{obs}$) by $P_\mathrm{int} = D P_\mathrm{obs}$, where $D$ is the special-relativistic Doppler factor \citep{dd86}. 
To leading order, $P_\mathrm{obs}$ is given by \citep{dt92}:
\begin{equation}
  P_\mathrm{obs} = D^{-1} P_\mathrm{int} 
  \simeq \left(1 + \frac{\mathbf{v}_\mathrm{CM} \cdot \hat{\mathbf{K}}_0}{c}\right) P_\mathrm{int} \,,
\label{eq:doppler}
\end{equation}
where $\mathbf{v}_\mathrm{CM}$ the velocity of the CM of the binary relative to the SSB (its ``systemic'' velocity). However, because $P_\mathrm{int}$ is unknown, we cannot calculate the systemic radial velocity $v_{r, \mathrm{CM}} = \mathbf{v}_\mathrm{CM} \cdot \hat{\mathbf{K}}_0$ from $P_\mathrm{obs}$.\footnote{The situation is different in astronomical spectroscopy because, unlike pulsars, all atoms of the same element are identical, so their lines in stellar spectra have well-known ``rest'' wavelengths measured in laboratories that can be compared to the observed wavelengths to measure $v_{r, \mathrm{CM}}$.}

For the tests described below the lack of knowledge of $v_{r, \mathrm{CM}}$ is not a problem, precisely because all physical quantities re-scale coherently with $D$ \citep{dd86,dt92}. However, a  relative acceleration between the pulsar system and the SSB leads to (apparent) secular changes in the dimensional parameters of the binary pulsar system, as discussed in detail in Sect.~\ref{sec:Pbdot_ext}.

\item Pulsar timing or Doppler measurements only provide information for the motion along the line of sight. For most binary pulsars, the transverse component of the orbital motion is not measurable by any other means like VLBI\footnote{There is a single case where the orbital motion of a pulsar can be detected with VLBI, PSR~J2222$-$0137, a nearby binary pulsar with a massive WD companion \citep{dbl+13,gfg+21}. Scintillation measurements can provide constraints on the transverse orbital motion of some pulsars (e.g., \citealt{rcn+14}).}
--- only the transverse component of the systemic velocity can be measured with VLBI because its effects increase linearly with time.

\item The measurement of the binary-related delays in Eq.~(\ref{eq:delays2}) allows a much more precise measurement of the pulsar's orbital motion than is possible using Doppler shift techniques. Although Doppler measurements are possible for binary pulsars---and indeed, quite useful for the initial measurement of its orbital parameters---their precision is limited by the length of the individual observations. Once a phase-coherent ephemeris is obtained, the precision of the measurement of the orbital motion increases by factors of thousands to millions (larger factors for wider orbits, where the Doppler shifts are smaller and the delays are larger)---while using the same radio data. This is especially relevant for the detection of small relativistic effects, and is the main reason why binary pulsars can be excellent laboratories for testing gravity theories.
\end{itemize}


\subsection{Binary models}

At the time of the discovery of the Hulse--Taylor pulsar no timing models for binary pulsars were available, for the simple reason that no binary pulsars had been found before. After this discovery, the development of such models became vital in order to materialise the precision of the measurement of the orbital motion promised by the discovery of a binary pulsar. 

\begin{center}
\begin{table}[ht!]
\caption{List of Keplerian and PK parameters of the DD86 model. The units are those specified in the input and output files of standard pulsar software.\protect\\
$^\dagger$ These parameters are only in the DD86 model. $\delta_r$ is not separately measurable \citep{dd86,dt92}.
\label{tab:params}}
\begin{tabular}{|l|l|l|}
\hline
Parameter & Physical meaning & Unit \\
\hline
\multicolumn{3}{| c |}{\bf Keplerian parameters} \\
\hline
$P_\mathrm{b}$  & Orbital period & day \\
$x$             & Projected semi-major axis of the pulsar's orbit & second (s) \\
$e$             & Orbital eccentricity & \\
$\omega$        & Longitude of periastron & degree \\
$T_0$           & Time of passage through periastron & Modified Julian Date\\
\hline
\multicolumn{3}{| c |}{\bf PK parameters} \\
\hline
$\dot\omega$    & Rate of advance of periastron & degree per year \\
$\gamma$        & Amplitude of Einstein delay & s \\
$\dot{P}_\mathrm{b}$  & Rate of change of the orbital period & $10^{-12}$ s s$^{-1}$ \\
$r$             & ``Range'' parameter of the Shapiro delay   & s \\
$s$             & ``Shape'' parameter of the Shapiro delay   &  \\
$\delta_r$      & Relativistic radial deformation of orbit$^\dagger$  & $10^{-6}$ \\
$\delta_\theta$ & Relativistic angular deformation of orbit$^\dagger$ & $10^{-6}$ \\
$\dot{x}$       & Time derivative of $x$    & $10^{-12}$ s s$^{-1}$ \\
$\dot{e}$       & Time derivative of $e$    & $10^{-12}$ s$^{-1}$ \\
\hline
\end{tabular}
\end{table}
\end{center}

One of the first models to appear was developed by \cite{bt76}, henceforth BT76. This model describes the observed orbital motion as a Keplerian motion with small relativistic perturbations. This ``quasi-Newtonian'' model was first used in the first phase-coherent ephemeris for PSR~B1913+16 \citep{thf+76}. In their Table~1, we see that some orbital parameters derived from the timing analysis are already two orders of magnitude more precise than the same parameters estimated from the Doppler analysis; for the spin period the improvement is 6 orders of magnitude.

A more complete timing formula was published in \cite{dd86}, henceforth DD86. This is based on their elegant analytical, quasi-Keplerian solution to the first PN equations of motion of a binary system in harmonic coordinates \citep{dd85}.

In both models, the orbital motion along the radial direction is described by five Keplerian parameters. In addition, there are a few PK parameters that quantify relatively small relativistic deviations from a Keplerian orbit. All are listed in Table~\ref{tab:params}. Some notes on these parameters:

\begin{itemize}
    \item There are two additional Keplerian parameters, that we generally cannot determine only with measurements of the radial motion. The first is the orbital inclination $i$ (but see Sect.~\ref{sec:pkparms}). Since
\begin{equation}
  x = \frac{a_\mathrm{p}}{c}\, \sin i \,,
\label{eq:x}
\end{equation}
the lack of knowledge of $i$ and $\sin i$ implies that the semi-major axis of the pulsar's orbit, $a_\mathrm{p}$, is also unknown. However, changes in the orbital plane w.r.t. the plane of the sky
(which is perpendicular to the line of sight and passes through the CM of the system) can cause a change in $i$, which is detectable as a change in $x$ ($\dot{x}$, see Sect.~\ref{sec:kopeikin} and  \ref{sec:SOcoupling}).
\item The second unknown Keplerian parameter is the PA of the ascending node, $\Omega$. The ascending node is the point in the orbit where the pulsar, while moving away from us, crosses the plane of the sky. A change in $\Omega$ represents a rotation of the orbital plane of the system around the line of sight, which for most cases leaves the timing practically unchanged.
\item Since $\omega$ is the angle periastron--CM--ascending node, it is (like $i$) generally affected by a change of the orbital plane w.r.t.\ the plane of the sky. Moreover, the change of the position of the nodes caused by the proper motion causes a change of the ascending node (and on $\omega$) that depends on $\Omega$ (Sect.~\ref{sec:kopeikin}).
\item Unlike $P_\mathrm{b}$, $x$ and $e$, the parameters $\omega$, $i$, $\Omega$ and $T_0$ have no astrophysical meaning, merely specifying the orientation of the system and a particular occurrence of the time of periastron as seen by the observer.
\end{itemize}

A major advantage of these models is that they describe the orbital motion for, at least, any fully conservative, boost-invariant gravity theories that have a Newtonian limit. This means that the PK parameters can be understood as phenomenological parameters which can be measured in a theory-independent manner, providing a \emph{parameterised PK (PPK) framework} (see \citealt{dd86,dt92} for more details).

We now discuss the DD86 timing formula. For each time $T$, there are two auxiliary parameters that must be calculated. The first one is the eccentric anomaly, $u$. To relate $u$ to $T$, we use an analogue of Kepler's equation, which must be solved numerically:
\begin{equation}
  u - e \sin u = 2\pi \left[ \left( {\frac{T-T_0}{P_\mathrm{b}}} \right) -
    {\frac{\dot P_\mathrm{b}}{2}} \left( {\frac{T-T_0}{P_\mathrm{b}}} \right)^2 \right]. 
    \label{eqn:eccen_anom}
\end{equation}
We now calculate $\omega$ at time $T$. To do this, one needs a second auxiliary parameter, the true anomaly, $A_{e}(u)$; this is the angle pulsar--CM--periastron:
\begin{equation} 
  A_e(u) = 2 \arctan \left[ \left( {\frac{1+e}{1-e}}\right)^{1/2} \tan {\frac{u}{2}} \right], \label{eqn:true_anom} 
\end{equation}
from which $\omega$ can be calculated directly via a proportionality constant $k$:
\begin{equation} 
  \omega = \omega_0 + k A_e(u),
\label{eq:k}
\end{equation}
where $\omega_0$ is the reference value of $\omega$ at time $T_0$. The time-averaged periastron advance $\langle\dot{\omega}\rangle$ is thus $2 \pi k$ radians for each orbital period $P_\mathrm{b}$: 
\begin{equation}
  \langle\dot{\omega}\rangle = \frac{2 \pi}{P_\mathrm{b}} \, k \,.
\label{eq:omdot_k}
\end{equation}
Equation~(\ref{eq:k}) is the main difference between the DD86 and the BT76 orbital models, in the latter $\omega$ changes linearly with time, like $x$ and $e$:
\begin{equation}
  \omega = \omega_0 + \dot{\omega}(T - T_0).
\label{eq:omega}
\end{equation}
Equation~(\ref{eq:k}) describes the observed evolution of $\omega$ in binary pulsars significantly better than Eq.~(\ref{eq:omega}) \citep{dt92}; for this reason the use of the DD86 model is to be preferred. The meaning of $\dot{\omega}$ in the latter model, and in the discussions below, is really as a scaled version of $k$ according to Eq.~(\ref{eq:omdot_k}).

With $u$ and $A_e(u)$, plus the Keplerian and PK parameters, we can calculate the time delays associated with the binary orbit in Eq.~(\ref{eq:delays2}).


\subsubsection{R{\o}mer delay}

In the DD86 model, the R{\o}mer delay is given by the following equation:
\begin{equation} 
  \Delta_\mathrm{R} = 
    x \sin\omega \big[\cos u -e(1+\delta_r)\big] + 
    x \cos\omega \big[1-e^2(1+\delta_{\theta})^2\big]^{1/2} \sin u \,.
\label{eq:deltaR}
\end{equation}
In the BT76 model, neither $\delta_r$ nor $\delta_\theta$ are taken into account. This is the second main difference between the BT76 and DD86 models.


\subsubsection{Shapiro delay}
\label{sec:shapiro_delay}

In a binary pulsar, the Shapiro delay is caused solely by the gravitational field of the companion. To leading order it is given by:
\begin{equation} 
  \Delta_\mathrm{S} = -2 \, r \ln \left\{ 1-e\cos u - s \left[
    \sin\omega (\cos u - e) + (1-e^2)^{1/2} \cos\omega \sin u\right] \right\} \,. 
\label{eq:deltaS} 
\end{equation}
which is calculated assuming a static configuration \citep{bt76,dd86}. The next-to-leading order term caused by the motion of the companion during the propagation of the signal \citep{ks99} has to be taken into account in the case of the Double Pulsar \citep{ksm+21}.

\cite{fw10} introduced an ``orthometric'' parameterisation for the Shapiro delay based on its Fourier expansion. The orthometric amplitude $h_3$ and the orthometric ratio $\varsigma$ replace $r$ and $s$ via the relations
\begin{equation}
    \varsigma = \frac{s}{1 + \sqrt{1 - s^2}} \,, \quad h_3 = r\,\varsigma^3 \,.
\label{eq:stig_h3}
\end{equation}
This parameterisation is particularly advantageous for systems where $i$ is far from edge-on. In such cases the Shapiro delay is much smaller and $r$ and $s$ are strongly correlated, while $h_3$ and $\varsigma$ are not.


\subsubsection{Einstein delay}
\label{sec:einstein_delay}

Unlike in the Solar System, the Einstein delay represents only the variable part of the relativistic time dilation experienced by the pulsar---to leading order this can be understood as the result of the second order Doppler effect due to the motion of the pulsar in the centre-of-mass frame and the gravitational redshift caused by the companion---as seen by distant observers. It is given by:
\begin{equation} 
  \Delta_\mathrm{E} = \gamma \sin u \,, 
\label{eq:deltaE} 
\end{equation}
where we see that, unlike $\Delta_\mathrm{R}$ and $\Delta_\mathrm{S}$, this term is independent of the orientation of the binary relative to the observers ($\omega$, $i$, $\Omega$). Note that this effect can only be measured in a timing baseline where $\omega$ changes significantly. If that is not the case, then the effect of $\gamma$ is re-absorbed in a small change from $x$ and $\omega$ to $x'$ and $\omega'$ (e.g., \citealt{wj98}):
\begin{eqnarray}
  x'      &=& x      + \frac{\gamma  }{\sqrt{1 - e^2}} \, \cos\omega \,, \\
  \omega' &=& \omega - \frac{\gamma/x}{\sqrt{1 - e^2}} \, \sin\omega \,,
\end{eqnarray}
where we ignored $\delta_{\theta}$, which is generally of order $10^{-6}$. $x'$ and $\omega'$ are the parameters in DD86 ephemerides of eccentric binary pulsars for which no $\gamma$ is being either measured nor assumed.


\subsubsection{Aberration delay}
\label{sec:aberration_delay}

As the pulsar moves in its orbit, one needs to account for aberration, i.e.\ the velocity-dependent transformation between the co-moving frame of the pulsar and the centre-of-mass frame of the binary for the propagation of the radio signal \citep{sb76,dd86}. This causes, at any given time, small changes to the direction of the emission beam. For the emission to be directed to Earth, more (or less) time---called the aberration delay---must be allowed for the rotation of the pulsar to compensate for these small changes. This delay would be zero if the pulsar were an actual pulsating, non-rotating source. In most cases, this effect is very small and is absorbed by re-definitions of $x$, $e$, $\delta_{\theta}$ and $\delta_r$ \citep{dd86,dt92}.
However, if the spin-orientation of the pulsar changes due to geodetic precession (Sect.~\ref{sec:SOcoupling}), then aberration becomes time-dependent and needs to be properly accounted for \citep{dt92}.

In the case of the Double Pulsar, additional changes to the direction of emission of the beam of PSR~J0737$-$3039A originate from light deflection near PSR~J0737$-$3039B when the latter is passing near the line of sight; its effect on the timing cannot be absorbed by a re-definition of the Keplerian or PK parameters. The observed magnitude of the effect \citep{ksm+21} matches the detailed GR calculations presented there and in the references therein if PSR~J0737$-$3039A is rotating in the same sense as the orbital motion (see details in Sect.~\ref{sec:psrJ0737}). 


\subsubsection{Gravitational radiation and the orbital decay}
\label{sec:PBdot_experimental}

For the more compact double-degenerate binaries, the variation of the orbital period ($\dot{P}_\mathrm{b}$) used in Eq.~(\ref{eqn:eccen_anom}) is dominated by the orbital decay caused by GW damping, but has other contributions (Sect.~\ref{sec:Pbdot_ext}). While the orbit shrinks due to the loss of energy through GWs (an effect that has not yet been measured directly), the orbital period shortens according to Kepler's third law (see Eq.~\ref{eq:Kepler_3} below). As a result, one gets a quadratic effect in the evolution of the orbital phase (see Eq.~\ref{eqn:eccen_anom}) which leads to a prominent effect in the ToAs that builds up quickly with time.


\subsection{Keplerian and post-Keplerian parameters in GR}
\label{sec:pkparms}


\subsubsection{The mass ratio and the mass function}
\label{sec:keplerian_parms}

If we denote the semi-major axis of the relative orbit (the ``separation'' of the binary, $a$) in time units ($\bar{a} \equiv a/c$), we can write Kepler's third law (to Newtonian order) as:
\begin{equation}
  \left( \frac{P_\mathrm{b}}{2 \pi} \right)^2 = \frac{\bar{a}^3}{(\Tsun M)} \,,
\label{eq:Kepler_3}
\end{equation}
where $\Tsun\equiv (\mathcal{GM})_{\odot}^{\rm N}/c^3 = 4.925490947641\dots\,\mu$s is an exact quantity, the \emph{nominal Solar mass parameter}\footnote{This quantity was defined by the IAU 2015 Resolution B3 \citep{mamajek2015iau} to be $(\mathcal{GM})_{\odot}^{\rm N} \equiv 1.3271244 \times 10^{26}\,{\rm cm^3\,s^{-2}}$. It is very close to $G M_{\odot}$, where $G$ is Newton's gravitational constant and $M_{\odot}$ is the mass of the Sun. Despite its high precision, $G M_{\odot}$ is tied to the mass of the Sun, which is physically variable, but also subject to measurement uncertainty, while $(\mathcal{GM})_{\odot}^{\rm N}$ is defined exactly in the international system of units (SI), which is more relevant for timing measurements.} 
in time units and $M$ is the (dimensionless) total mass of the system expressed in units of the \emph{nominal Solar mass}, which we denote by $\Msun$ in this review (see \citealt{Prsa_2016} for details, who in their Table~2 suggest the symbol $\mathcal{M}_\odot^\mathrm{N}$). The reason for this is that the precision of the total mass in mass units (like grams),
\begin{equation}
  m = (\Tsun M) \, \frac{c^3}{G} \,,
\end{equation}
is limited by the comparably low precision of Newton's gravitational constant $G$, while $M$ is tied to the precision of the measurements of $P_\mathrm{b}$ and $\bar{a}$ in Eq.~(\ref{eq:Kepler_3}), which can be much higher\footnote{This is why $G M_\odot$ is known much more precisely than either $G$ or $M_\odot$.}. From Eq.~(\ref{eq:Kepler_3}), we obtain
\begin{equation}
  \bar{a} =  \left( \frac{P_{\mathrm {b}}}{2 \pi} \right)^{2/3} (\Tsun M)^{1/3} \,,
\end{equation}
where we see that if $M$ is known, $\bar{a}$ can be estimated independently of the orbital inclination. Furthermore,
\begin{equation}
    \beta_\mathrm{O} \equiv \left( \frac{P_{\mathrm {b}}}{2 \pi} \right)^{-1/3} (\Tsun M)^{1/3}
\end{equation}
is the relative orbital velocity, divided by $c$, for a circular binary of separation $a$. This quantifies how relativistic the orbit is and will appear implicitly in many equations below.

Since $a_\mathrm{p}$ is the semi-major axis of the pulsar's orbit, we obtain for the companion $a_\mathrm{c} = a - a_\mathrm{p}$. In this case, the ratio of the mass of the pulsar and the mass of the companion is given by
\begin{equation}
  R \equiv \frac{M_\mathrm{p}}{M_\mathrm{c}} 
    = \frac{a_\mathrm{c}}{a_\mathrm{p}} 
    = \frac{x_\mathrm{c}}{x} \,,
\label{eq:R}
\end{equation}
where $x_\mathrm{c} = (a_\mathrm{c}/c) \sin i$. Since $M = M_\mathrm{p} + M_\mathrm{c}$, we obtain $a = a_\mathrm{p} (M / M_\mathrm{c})$. From this expression and Eqs.~(\ref{eq:x}) and (\ref{eq:Kepler_3}), one derives the (Newtonian) \emph{mass function equation}:
\begin{equation}
  f(M_\mathrm{p}, M_\mathrm{c}, i) 
    = \frac{(M_\mathrm{c} \sin i)^3}{(M_\mathrm{p} + M_\mathrm{c})^2} 
    = \frac{x^3}{\Tsun} \left( \frac{P_\mathrm{b}}{2 \pi} \right)^{-2} ,
\label{eq:fm}
\end{equation}
which has on the right all the relevant information available to an observer that cannot measure the additional mass constraints described below. Although in this case we have a single equation for three unknowns, if one assumes a value for $M_\mathrm{p}$ and $i = 90^\circ$ we can immediately derive a minimum value for $M_\mathrm{c}$ (for that $M_\mathrm{p}$) by solving Eq.~(\ref{eq:fm}) numerically. This is the only available mass constraint for $\sim 80 \%$ of binary pulsars.

To measure the component masses, two additional mass constraints are necessary. Besides measuring PK parameters (see next section), there are two mass constraints of relevance for the systems discussed below (Sects.~\ref{sec:psrJ0737}, \ref{sec:psrJ1738}, \ref{sec:psrJ1909}):
\begin{enumerate}
\item Measurement of $R$: If the companion is sufficiently bright in the optical band, then 
  phase-resolved spectroscopy can yield many measurements of $v_r$. These are the sum of $v_{r, \mathrm{CM}}$ and the orbital component of $v_r$, which is generally parameterised by optical spectroscopists using $P_\mathrm{b}$, $T_0$, $e$, $\omega_c = \omega + \pi$ and the orbital velocity amplitude, $K_\mathrm{c}$. The latter relates to $x_\mathrm{c}$ via:
  \begin{equation}
    x_\mathrm{c} = \frac{K_\mathrm{c}}{c} \frac{P_\mathrm{b}}{2\pi} \sqrt{1 - e^2} \,.
  \end{equation}
  In the case of the Double Pulsar, $x_c$ is available directly from the radio timing of PSR~J0737$-$3039B. 
\item Measurement of $M_\mathrm{c}$: For systems with optically bright WD companions, the width of the Balmer lines allows the determination of the surface gravity. This can then be combined with the mass-radius relation for WDs to determine the WD mass and the radius (see e.g., \citealt{avk+12,afw+13,lgi+20}).
\end{enumerate}
These constraints are very rarely available, and in the case of $M_\mathrm{c}$ they depend on several assumptions and detailed modeling. Including the first PN corrections of the Damour \& Deruelle solution, the simple relation $a = a_\mathrm{p} + a_\mathrm{c}$ and in particular Eq.~(\ref{eq:R}) still hold, not only in GR but for any Lorentz-invariant gravity theory (\citealt{dd85}; see also the detailed discussion in \citealt{dam09}). The spectroscopic determination of $M_\mathrm{c}$, in case of a WD companion, is performed in the well-tested Newtonian limit. For this reason, these constraints are of particular importance for tests of alternatives to GR (see Sect.~\ref{sec:altGrav}).

Finally, it should be noted that the Newtonian expression (\ref{eq:fm}) has so far been sufficient for all binary pulsars with the exception of the Double Pulsar, for which first PN corrections to Eq.~(\ref{eq:fm}) are necessary (more in Sect.~\ref{sec:psrJ0737}).


\subsubsection{Post-Keplerian parameters}

In GR, to leading order, the PK parameters depend simply on the well-measured, astrophysically meaningful Keplerian parameters ($P_\mathrm{b}$, $e$ and in one case $x$) and the masses $M_\mathrm{p}$ and $M_\mathrm{c}$, they are given by \citep{dd86,tw89,dt92}:
\begin{eqnarray}
\dot{\omega} & = & 3 \left(\frac{P_\mathrm{b}}{2\pi}\right)^{-5/3}
    (\Tsun M)^{2/3} \, (1-e^2)^{-1}\,, \label{eq:omdot} \\
  \gamma &=& \left(\frac{P_\mathrm{b}}{2\pi}\right)^{1/3}
    (\Tsun M)^{2/3}\,X_\mathrm{c} (1 + X_\mathrm{c}) \, e \,, \\
  \dot P_\mathrm{b} &=& -\frac{192\pi}{5}
    \left(\frac{P_\mathrm{b}}{2\pi}\right)^{-5/3}
     (\Tsun M)^{5/3}\, X_\mathrm{p} X_\mathrm{c}\,
    \frac{\left(1 + \frac{73}{24} e^2 + \frac{37}{96} e^4 \right)}{(1-e^2)^{7/2}}\,, \label{eq:pbdot} \\
  r &=& (\Tsun M) X_\mathrm{c} = \Tsun M_\mathrm{c}\,, \\
  s &=& x \left(\frac{P_\mathrm{b}}{2\pi}\right)^{-2/3}
    (\Tsun M)^{-1/3}\,X_\mathrm{c}^{-1}\,, \label{eq:s} \\
  \delta_\theta &=& \left(\frac{P_\mathrm{b}}{2\pi}\right)^{-2/3}
    (\Tsun M)^{2/3}\,\left(\frac{7}{2}\,X_\mathrm{p}^2 + 6\,X_\mathrm{p}\,X_\mathrm{c} + 2\,X_\mathrm{c}^2\right) \,, \label{eq:d_theta} \\
  \delta_r &=& \left(\frac{P_\mathrm{b}}{2\pi}\right)^{-2/3}
    (\Tsun M)^{2/3}\,\left(3\,X_\mathrm{p}^2 + 6\,X_\mathrm{p}\,X_\mathrm{c} + 2\,X_\mathrm{c}^2\right) \,, \label{eq:d_r} 
\end{eqnarray}
where $s\equiv\sin i$, $X_\mathrm{p} = M_\mathrm{p}/M$ and $X_\mathrm{c} = M_\mathrm{c}/M$. Equations~(\ref{eq:stig_h3}) can then be used to derive the correspondent values of $h_3$ and $\varsigma$.

This ``PK simplicity'' is a consequence of a fundamental property of GR, the \emph{effacement} of the internal structure of a self-gravitating mass \citep{dam87,wil18_book}. Consequently, the motion of bodies in a $N$-body system is independent of their internal structure.\footnote{This all has to be understood in an approximate sense. As soon as tidal effects (at fifth PN order) start to play a role, also in GR the internal structure of a body becomes relevant.} 
The PK simplicity has the following implications:
\begin{itemize}
\item The measurement of two PK parameters leads to a determination of both masses. If none of 
  these PK parameters is $s$ or $\varsigma$ (which yields $\sin i$ directly), we can then use Eq.~(\ref{eq:s}) (a re-write of Eq.~\ref{eq:fm}) to determine $\sin i$ from the masses.
\item Measuring additional PK parameters we can test the validity of GR from a test of the 
  self-consistency of these equations. This is the main method used in Sect.~\ref{sec:GR_tests}.
\item This also means that additional NS properties, like their radius, tidal deformability or 
  $I$ cannot be measured with this technique, at least with only the leading order terms in the PK equations above. This would imply that the only way pulsar timing can constrain the EoS is via the measurement of large NS masses (see e.g.\ \citealt{fcp+21}). Measurements of other NS bulk parameters must therefore come from other types of measurements: radii come from X-ray observations, in some cases aided by radio measurements \citep{of16,mld+21,vsw+24} and tidal deformabilities come from the observation of GWs from a DNS merger \citep{LIGO2018}.
\end{itemize}
However, for the Double Pulsar (Sect.~\ref{sec:psrJ0737}) higher order contributions to Eq.~(\ref{eq:omdot}) become relevant for the description of the timing data, specifically the second PN correction $\dot\omega_\mathrm{2pN}$ as well as the Lense-Thirring (LT) contribution $\dot\omega_\mathrm{LT}$, which results from the coupling between the orbital motion and the rotation of pulsar A (relativistic spin-orbit coupling). More generally, with relativistic spin-orbit coupling, the spins of the two binary components enter the equations of motion, which leads to an additional spin (magnitude and orientation) and thus $I_\mathrm{p}$ and $I_\mathrm{c}$ dependence of some PK parameters (cf.\ discussion in Sect.~\ref{sec:SOcoupling}).

The PK simplicity is captured by the DDGR timing model, a version of the DD86 model that assumes the validity of GR (and thus of Eqs.~\ref{eq:omdot}--\ref{eq:d_r}): This model describes the timing using only the measurable Keplerian parameters, $M$ and $M_\mathrm{c}$ \citep{dd86,tw89}. This \emph{theory-dependent} model is very useful for many applications, where then GR is assumed to be the correct theory of gravity.

Other theories of gravity, such as those with one or more scalar fields in addition to a tensor field, have different mass dependencies for the PK parameters. In addition, there is (unlike in the case of GR) also a dependence on the internal structure of the bodies, meaning the PK parameters depend on the EoS of the pulsar and the EoS of the companion, which, for instance, is very different for a WD companion. Some specific examples of such theories will be discussed in Sect.~\ref{sec:altGrav} below.


\subsection{``Contaminations'' of the post-Keplerian parameters}

In this sub-section, we do not aim at an exhaustive listing of terms that can affect the measurement of the PK parameters; only a short discussion of the effects relevant for the systems described in Sect.~\ref{sec:GR_tests}.


\subsubsection{Variation of the orbital period}
\label{sec:Pbdot_ext}

Re-writing Eq.~(\ref{eq:doppler}) for $P_\mathrm{b}$, and then differentiating it, we obtain:
\begin{equation}
  \left( \frac{\dot{P}_\mathrm{b}}{P_\mathrm{b}} \right)_\mathrm{int} =
  \left( \frac{\dot{P}_\mathrm{b}}{P_\mathrm{b}} \right)_\mathrm{obs} -
  \frac{\mathbf{a}_\mathrm{CM} \cdot \hat{\mathbf{K}}_0} {c} - \frac{\mu^2 d}{c} \,,
\label{eq:pbdot_ext}
\end{equation}
where $\mathbf{a}_\mathrm{CM}$ is the (small) acceleration of the CM of the pulsar binary relative to the SSB (caused by the gravitational field of the Milky Way) and $\mu = \sqrt{\mu_{\alpha}^2 + \mu_\delta^2}$ is the magnitude of the proper motion. The same relations apply to any other derivative of a time-like quantity of the binary system, like $\dot{P}$ and $\dot{x}$.

The term on the left is what we generally want to measure in most binary pulsars: the intrinsic variation of the orbital period, caused by orbital decay due to GW damping. Additional intrinsic terms are possible if the system is, for instance, losing mass, but for most DNSs the mass loss term given by Eq.~(\ref{eq:E-dot}) (with $\dot{m}_\mathrm{p} = \dot{E}_\mathrm{rot}/c^2$) \citep{dt91} can be safely ignored, except in the case of the extreme precision achieved for the Double Pulsar \citep{ksm+21}.

The first term on the right generally improves fast with additional observations (Sect.~\ref{sec:PBdot_experimental}).

The second term is generally dominated by the difference of Galactic accelerations of the SSB and the pulsar \citep{dt91}, this can be estimated from the position of the pulsar in the Galaxy (which requires an accurate distance) and a model of the Galactic potential (e.g., \citealt{mcm17}). For many pulsars, additional contributions come from the gravitational fields of globular clusters and in a few cases of nearby stars. 

The third term is known as the Shklovskii effect \citep{shk70}; this again requires knowledge of $d$ and $\mu$ (the latter is generally easier to measure from the timing, especially for pulsars with good timing precision and stability).

Thus, lack of precise knowledge of the second and third terms on the right (the ``extrinsic'' terms) limits the precision of the measurement of $\dot{P}_\mathrm{ b, int}$. The situation
can be improved with precise measurements of $d$ and more accurate Galactic potential models.


\subsubsection{Other effects of the proper motion}
\label{sec:kopeikin}

Differentiating Eq.~(\ref{eq:x}) in time and then dividing the result by that equation, we obtain, in the absence of a change in $a_\mathrm{p}$ (for instance by GW damping):
\begin{equation}
  \frac{\dot{x}}{x} = \cot i \, \frac{\mathrm{d}i}{\mathrm{d}t} \,,
\label{eq:xdot}
\end{equation}
where the variation of the inclination $\mathrm{d}i/\mathrm{d}t$ can be due to a change in our viewing angle of the binary (this section) or an intrinsic change of the orbital plane (Sect.~\ref{sec:SOcoupling}).
For nearly edge-on systems ($i \approx 90^{\circ}$), $\cot{i}\ll 1$ and $\mathrm{d}i/\mathrm{d}t$ is much more difficult to measure.

The proper motion of the system causes a change in our viewing angle of the binary, which can change $i$ as seen from Earth. This was first calculated by \cite{ajrt96} and, in greater detail by \cite{kop96}. The latter's expressions can be re-written as:
\begin{equation}
  \left(\frac{\mathrm{d}i}{ \mathrm{d} t}\right)_\mu = \mu \, \sin(\Theta_{\mu} - \Omega) \,,
\end{equation}
where $\Theta_{\mu}$ is the PA of the proper motion. Like any change of the viewing angle, the proper motion also produces a change in the line of nodes, and therefore a change in $\omega$:
\begin{equation}
  \dot{\omega}_{\mu} = \frac{\mu}{ \sin i} \, \cos(\Theta_{\mu} - \Omega) \,.
\end{equation}

These equations are valid in any right-handed coordinate system, like that described in \cite{dt92} (where PAs are measured clockwise starting from East through North, where a positive spin along the line of sight points away from us) or the ``observer's convention'' (PAs measured anti-clockwise from North through East, where a positive spin along the line of sight points towards us) that is implemented in most timing programs \citep{tempo2_2}.

Normally, a measurement of the Shapiro delay and $\mathrm{d}i/\mathrm{d}t$ allows four ``degenerate'' solutions for $i$ and $\Omega$. However, the orbital motion of the Earth also changes the viewing angle of the binary pulsar, leading to yearly variations of $x$ and $\omega$, known as the ``annual orbital parallax'' \citep{kop95}. These are measurable in the timing of some nearby binary pulsars, especially wider systems with excellent timing precision (for early examples, see \citealt{sbb+01,sns+05,zdw+19,sfa+19}, see also \citealt{gfg+21}); these measurements allow the determination of $i$ and $\Omega$.
The $\dot{\omega}_\mu$ term can introduce significant corrections to the total mass derived from $\dot{\omega}$ \citep{sfa+19}.

One way of taking all these effects into account in a self-consistent way is to introduce the orbital orientation ($i$, $\Omega$) and all its effects into the timing model, which are then fit like all other parameters, as done in the ``T2'' orbital model implemented in \textsc{tempo2} \citep{tempo2_2}.


\subsubsection{Relativistic spin-orbit coupling}
\label{sec:SOcoupling}

A fundamental property of a curved spacetime is that the spin (proper angular momentum) of a freely-falling rotating body changes its direction with respect to a distant observer. This is certainly the case for the spin of pulsars moving in an orbit with a companion. The \emph{geodetic precession} is the leading order contribution to such a spin precession which results from the relativistic spin-orbit coupling \citep{bo75b}. As a consequence, the spin of the pulsar precesses around the conserved total angular momentum of the system at a rate $\Omega^\mathrm{geod}_\mathrm{p}$ that depends, like the PK parameters in Eqs.~(\ref{eq:omdot})--(\ref{eq:d_r}), simply on the masses and the Keplerian parameters \citep{bo75b,ber75,dam78}:\footnote{In binary pulsars we have the situation that the orbital angular momentum is generally several orders of magnitude larger than the pulsar spin. For that reason, the orbital angular momentum is nearly aligned with the conserved total angular momentum of the system.}
\begin{eqnarray}
\Omega^\mathrm{geod}_\mathrm{p} 
   &=& \frac{1}{2} \left( \frac{P_\mathrm{b}}{2 \pi} \right)^{-5/3}
       (\Tsun M)^{2/3} (1-e^2)^{-1} \, \left(3 + X_\mathrm{p}\right) X_\mathrm{c}         \nonumber\\
   &=& \frac{1}{6}  (3 +  X_\mathrm{p}) X_\mathrm{c} \, \dot{\omega} \,.
\end{eqnarray}
This ``simplicity'' means that a measurement of $\Omega^\mathrm{geod}_\mathrm{p}$ can be used as a test of GR.
If $M_\mathrm{c} = M_\mathrm{p}$, then $X_\mathrm{c} = X_\mathrm{p} = 1/2$ and $\Omega^\mathrm{geod}_\mathrm{p} = \frac{7}{24}\,\dot{\omega} \simeq 0.29\,\dot{\omega}$.

It is detectable mostly by the fact that, as the spin axis of the pulsar precesses, different parts of the pulsar's emission beam become visible from Earth, thus causing a long-term change in the pulse profile. In some cases, the pulsar might even become completely undetectable from Earth, as in the case of pulsar B in the Double Pulsar system \citep{pmk+10}. It is generally difficult to convert the observed changes in the pulse profiles into a qualitative measurement of $\Omega^\mathrm{geod}_\mathrm{p}$, but several techniques can help, particularly if polarimetric measurements are available, since these can tell us about the changing angle between the spin axis and the line of sight. We will give various examples in Sect.~\ref{sec:GR_tests} where geodetic precession has been not only observed but also quantified in binary pulsars, providing additional GR tests.

Given that the total angular momentum must be conserved (up to 2PN approximation), a change in the direction of the spin of a pulsar (and that of a rotating companion) must correspond to a change in the direction of the orbital angular momentum---the LT precession of the orbital plane. This causes an intrinsic change in $i$ that is, when averaged over one orbit, given in GR by \citep{ds88}:
\begin{equation}
  \left\langle\frac{\mathrm{d}i}{\mathrm{d}t}\right\rangle^\mathrm{LT} = 
    \frac{1}{2} \left( \frac{P_\mathrm{b}}{2 \pi} \right)^{-2} 
    (\Tsun M) \, (1-e^2)^{-3/2} \, 
    \sum_{j = \mathrm{p,c}} (3 + X_j)X_j \, (\boldsymbol{\chi}_j \cdot \hat{\mathbf{I}}) \,,
\label{eq:didtLT}
\end{equation}
where $\hat{\mathbf{I}}$ is a unit vector pointing from the CM of the system to the ascending node of the pulsar orbit and the \emph{dimensionless spin} of body $j$ is given by:
\begin{equation}
  \boldsymbol{\chi}_j \equiv \frac{c}{G m_j^2} \, \mathbf{S}_j
    = \frac{G}{c^5}(\Tsun M_j)^{-2} \, \mathbf{S}_j \,,
\end{equation}
where $\mathbf{S}_j$ is its spin angular momentum (``spin''). If its spin frequency ($\nu_j$) is known, we can relate it to its spin magnitude $S_j \equiv \lvert\mathbf{S}_j\rvert$ via:
\begin{equation}
  S_j =  2 \pi \nu_j \, I_j,
\end{equation}
where $I_\mathrm{j}$ is, by definition, its MoI. For NSs and BHs, $\chi \equiv \lvert\boldsymbol{\chi} \rvert$ varies between 0 and the maximum value for an astrophysical BH, 0.998 \citep{Lo_2011,Thorne_1974}. For the fastest-spinning NSs, we obtain, for reasonable values of $I$, $\chi \lesssim 0.5$.

In DNSs, where we can find misaligned spins and measure $\left\langle \mathrm{d}i / \mathrm{d}t \right\rangle^\mathrm{LT}$, the recycled pulsar is found to be in a range $\chi \sim 0.01 \dots 0.03$ \citep{sfc+18}. This is generally much larger than the non-recycled pulsar; this means that only one spin in the system matters in Eq.~(\ref{eq:didtLT}). On the long term, $i$ will then change periodically with a period of $2 \pi / \Omega^\mathrm{geod}_\mathrm{p}$. This is quite long for all DNSs discovered to date: the shortest, PSR~J1946+2052 \citep{sfc+18}, has a geodetic precession cycle of about 48 years (assuming equal masses).

The spin-orbit coupling also has an effect on $\dot{\omega}$---the LT contribution to the observed longitude of periastron. The detailed expressions are given in \cite{ds88}, which simplify considerably in a DNS where only the spin of the recycled pulsar matters. In the case of the Double Pulsar, discussed in Sect.~\ref{sec:0737_periastron}, the spin of the recycled pulsar (A) is nearly aligned with the orbital angular momentum \citep{fsk+13}, further simplifying the effect on $\omega$, which becomes a ``secular'' increase given by \citep{ksm+21,hkc+22}:
\begin{equation}
  \langle\dot{\omega}\rangle^\mathrm{LT} = 
    -\left( \frac{P_\mathrm{b}}{2 \pi} \right)^{-2} (\Tsun M) \,
    (1-e^2)^{-3/2} \, (3 + X_\mathrm{A}) X_\mathrm{A} \, \chi_\mathrm{A} \,,
\label{eq:omdot_LT}
\end{equation}
where the ``A'' subscripts indicate parameters of pulsar A.

Neither $\langle\mathrm{d}i/\mathrm{d}t\rangle^\mathrm{LT}$ nor $\langle\dot{\omega}\rangle^\mathrm{LT}$ due to a NS spin have been measured precisely for any binary pulsar\footnote{\label{footnote:1141}However, in the case of the PSR~J1141$-$6545 binary pulsar, an intrinsic $\langle\mathrm{d}i/\mathrm{d}t\rangle$ of 14 arcseconds per year has been measured using the DDGR timing model. This is caused by spin-orbit coupling, with some unknown percentage being due to the LT effect \citep{vbv+20}. This measurement implies that the companion WD spins very fast and its spin axis is not aligned with the orbital angular momentum, which beautifully confirms our understanding of the evolution of this system \citep{ts00}.}, owing to the fact that the effects are next-to-leading order and therefore very small. Nevertheless, the current upper limit on $\langle\dot{\omega}\rangle^\mathrm{LT}$ in the Double pulsar system already introduces an independent and useful upper limit on the MoI for J0737$-$3039A ($I_\mathrm{A} < 3.0 \times 10^{45}\, \rm g \, cm^2$, 90\% C. L., \citealt{ksm+21}); improving this measurement remains an important goal because a precise estimate of $I$ and $M_\mathrm{p}$ for the same pulsar would represent a powerful constraint of the EoS \citep{ls05,hkw+20,hf24}.


\subsubsection{Measurability of relativistic effects}

The reason why only $\sim$20\% of binary pulsars have one or more measured PK
parameters\footnote{\label{footnote:list}An up-to-date list of binary pulsars with measured PK parameters is maintained at \url{https://www3.mpifr-bonn.mpg.de/staff/pfreire/NS_masses.html}}
is that this requires several conditions which are often not fulfilled.
A fundamental issue for all measurements, discussed in Sect.~\ref{sec:ToA}, is the intrinsic faintness of most pulsars and the resulting limits on the timing precision, which highlights the importance of sensitive radio telescopes for these experiments.

The measurement of $\dot{\omega}$ and $\gamma$ requires a reasonably eccentric orbit; the measurement of $\gamma$ generally requires, in addition, a large change in $\omega$ during the timing baseline, otherwise, as discussed in Sect.~\ref{sec:einstein_delay}, it is not separable from the Keplerian parameters (but see discussion in~\citealt{rfgr19}).
The fractional uncertainty on both parameters decreases as $\Delta T^{-3/2}$, where $\Delta T$ is the timing baseline (see Tab.~II of \citealt{dt92}), but in the case of $\gamma$, the uncertainty decreases more slowly after a significant fraction of a precession cycle is completed.

The measurement of the Shapiro delay parameters ($r$, $s$ or $h_3$, $\varsigma$) is strongly favoured by nearly edge-on orbits and large companion masses. Because the Shapiro delay is a periodic effect, the fractional uncertainty of these parameters decreases slowly with time (as $\Delta T^{-1/2}$), so their measurement depends, more than other parameters, on timing precision. Measurements of this effect are also helped by an eccentric, precessing orbit, which help separate it effectively from the R{\o}mer delay; a large rate of precession might result in a faster improvement of the measurements, especially as the superior conjunction moves through periastron (see Sect.~\ref{sec:psrB1913}).

The measurement of $\dot{P}_\mathrm{b}$ requires relatively short orbital periods, because the contribution to $\dot{P}_\mathrm{b}$ from GW damping is proportional to $P_\mathrm{b}^{-5/3}$, while the kinematic contribution described in Sect.~\ref{sec:Pbdot_ext} scales with $P_\mathrm{b}$. The detection of $\delta_\theta$ requires, as in the case of $\gamma$, a large value of $e$, a large sweep in $\omega$ and, more than even $\dot{P}_\mathrm{b}$, a short orbital period (see Tab.~II of \citealt{dt92}); for $\dot{P}_\mathrm{b}$ and $\delta_\theta$, the fractional uncertainty decreases as $\Delta T^{-5/2}$, but as in the case of $\gamma$, the uncertainty of $\delta_\theta$ decreases more slowly once a significant fraction of a precession cycle is completed. In the case of the Double Pulsar, one of the two systems where there is a hint of a detection of $\delta_\theta$ (Sects.~\ref{sec:psrB1913}, \ref{sec:psrJ0737}), this parameter is strongly correlated with $\gamma$ \citep{ksm+21}.

The measurement of the geodetic precession requires a pulsar spin that is sufficiently misaligned with the orbital angular momentum. Generally, this is only seen in DNS systems (see Sect.~\ref{sec:binpsr})\footnote{But see the pulsar-WD system PSR~J1141$-$6545 \citep{mks+10,vbv+19}.}.
The measurement of $\left\langle \mathrm{d}i / \mathrm{d}t \right\rangle^\mathrm{LT}$ requires, in addition, that a misaligned NS has a sufficiently large spin. The measurement of $\langle\dot{\omega}\rangle^\mathrm{LT}$ does not require such a misalignment, but instead extremely precise mass measurements from other PK parameters (to calculate and subtract the non-LT contributions to $\dot\omega$ with sufficient precision, see Sect.~\ref{sec:0737_periastron}).

Because of their high spin frequency, MSPs have the best timing precision; however, their nearly circular and generally wider orbits severely restrict the number of cases where $\dot{\omega}$ and especially $\gamma$ can be measured. Furthermore their (generally) low companion masses restrict the number of Shapiro delay measurements. The alignment of the spin with the orbital angular momentum (Sect.~\ref{sec:binpsr}) means that geodetic precession cannot be measured in these systems. It is for these reasons that, despite their much smaller numbers, DNS systems have been so important for tests of gravity theories.


\section{Strong-field GR tests with pulsars}
\label{sec:GR_tests}

To test GR in the presence of strong gravitational fields, i.e. in spacetimes that deviate  (at least in some regions) significantly from Minkowski spacetime, binary pulsars provide some of the best and most precise experiments. Here, pulsars with compact companions, such as another NS or a WD, are of particular interest. In such binaries, the separation between the two bodies is still large compared to their size. Meaning, we can study the gravitational dynamics of two essentially point-like masses, without complicating contributions like tidal interaction. Pulsar-WD systems are of particular interest for testing alternative theories of gravity, due to their high asymmetry in gravitational self-energy. 

In this section we will introduce pulsar tests of GR, before we then discuss tests of specific deviations from GR and alternative gravity theories in Sect.~\ref{sec:beyond_GR}. A particularly clear way of illustrating these different GR tests is the \emph{mass-mass diagram}: because of the PK simplicity (see Sect.~\ref{sec:pkparms}), each measurement of a PK parameter leads to a constraint in the $m_\mathrm{p}-m_\mathrm{c}$ parameter space. Hence, if all PK parameters agree on a common region in the $m_\mathrm{p}-m_\mathrm{c}$ plane GR has passed the test.


\subsection{\texorpdfstring{
The Original System: PSR B1913$+$16 }{
The Original System: PSR B1913+16   }}
\label{sec:psrB1913}

It is certainly fair to say that the whole field of experimental gravity with radio pulsars started with the discovery of pulsar B1913$+$16 in July 1974 by Russell Hulse and Joseph Taylor \citep{ht75a,hul94,tay94} using the Arecibo telescope. PSR~B1913$+$16, now called the `Hulse--Taylor pulsar', is a 59\,ms pulsar in an eccentric ($e=0.62$)  7.75-hour orbit with an unseen compact companion. Based on the deduced masses (assuming GR), with $M_\mathrm{p} = 1.44$ and $M_\mathrm{c} = 1.39$, and considering evolutionary scenarios (as discussed in Sect.~\ref{sec:evolution}), it became soon evident that the companion is almost certainly also a NS. Soon after the discovery a significant advance of periastron ($\dot\omega$) of about $4.2^\circ$ per year was detected \citep{thf+76}, and by 1978 the time dilation $\gamma$ and a decay in the orbital period $\dot{P}_\mathrm{b}$ were measured \citep{tm80,tfm76}.

\begin{figure}[ht]
\begin{center}  
\includegraphics[width=10cm]{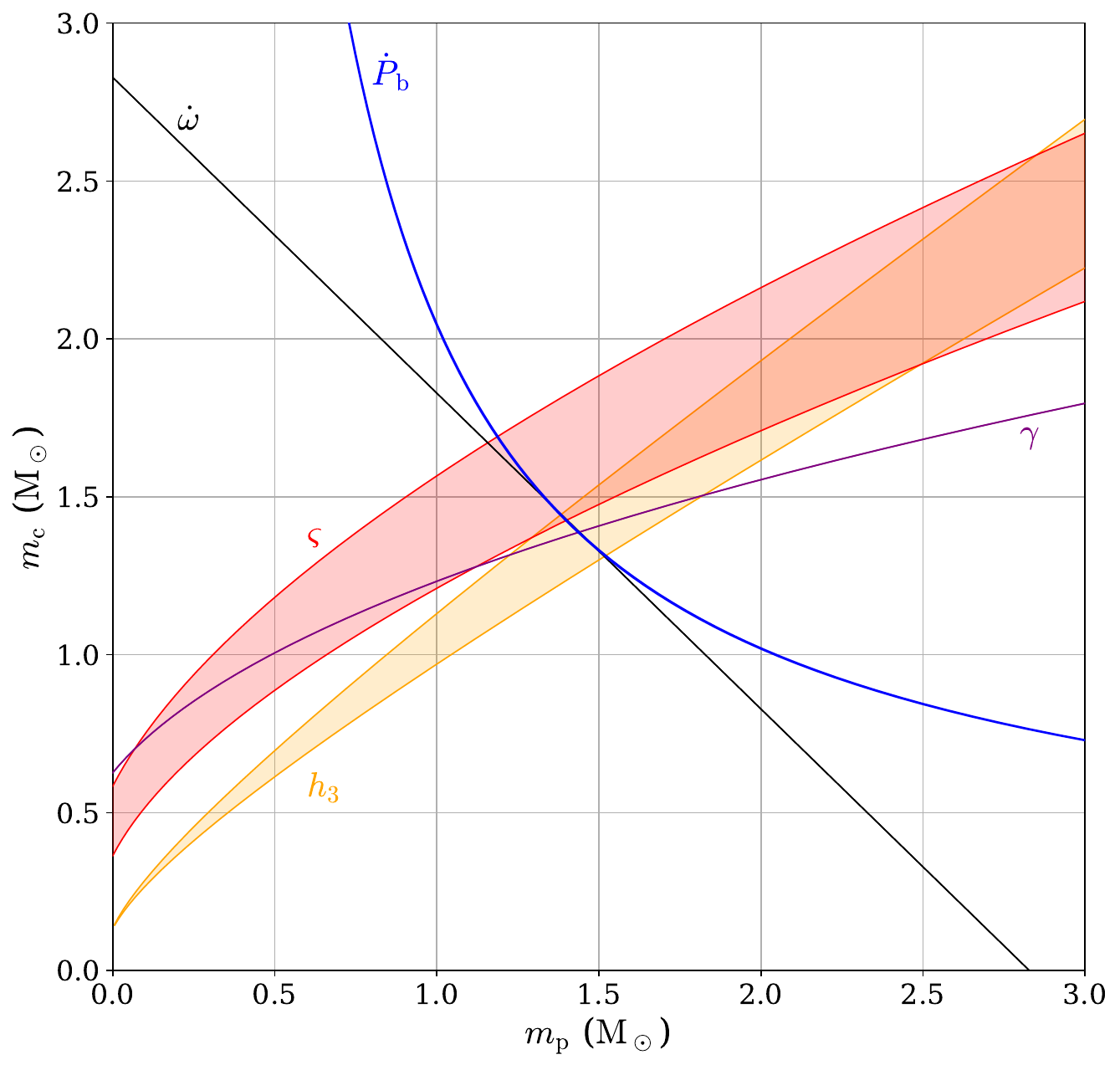}  
\caption{Mass-mass diagram for the PSR~B1913+16 system based on GR, using
  the PK parameter $\dot\omega$, $\gamma$, intrinsic $\dot{P}_\mathrm{b}$, and the orthometric Shapiro parameters $\varsigma$ and $h_3$. All curves agree on a small common region in the mass-mass plane. Parameter values have been taken from \cite{wh16}. In their Fig.\ 4 they display $r$ and $s$ for the Shapiro delay. As is obvious from the comparison of the two figures, the $h_3$-test is much more stringent than the $r$-test, which is one of the key advantages of the orthometric parameterisation of \cite{fw10}. In this and all other mass-mass diagrams, the width of each curve represents the one-sigma (68.3\% C.L.) uncertainty in the corresponding PK parameter. In addition, \cite{wh16} reports a $\sim$2-$\sigma$ measurement of the PK parameter $\delta_\theta$, which is not shown here.}
  \label{fig:m1m2_B1913}
\end{center}
\end{figure}

The agreement of those three PK parameters with GR confirmed, for the first time, the existence of GWs as predicted by GR. More generally, it confirmed the validity of GR for the gravitational interaction between two strongly self-gravitating masses. Regular observations of this system over the ensuing decades have lead to a steady improvement of that $\dot\omega$-$\gamma$-$\dot{P}_\mathrm{b}$ test (mixture of quasi-stationary and radiative aspects) \citep{tw82,tw89,dt91,wnt10}, where in the meantime there is even a detection of the Shapiro delay in that system \citep{wh16} (see also Fig.~\ref{fig:m1m2_B1913}). Despite the increasing timing base-line and precision, the testing of the GW emission by the Hulse--Taylor system has stagnated at around 0.3\% (95\% C.L.) for some time. The reason for this is the imprecise knowledge of the system's distance, which limits our ability to correct for the extrinsic contributions to the observed $\dot{P}_\mathrm{b}$ (cf.\ Sect.~\ref{sec:Pbdot_ext}) \citep{dt91,wnt10,wh16,dwnc18}.

Soon after the discovery of the Hulse--Taylor pulsar, its was pointed out that a pulsar in binary is subject to geodetic precession (see Sect.~\ref{sec:SOcoupling}) and that the Hulse--Taylor pulsar is generally expected to show changes in its emission and polarisation geometry with a precession rate of about $1.2^\circ\,\mathrm{yr}^{-1}$ \citep{dr74,bo75b,dr75,ber76,dt92}. Such changes have indeed been observed, as the pulsar's rotational axis is misaligned ($\sim 21^\circ$, \citealt{Graikou_PhD}) with respect to the orbital angular momentum (a result of the SN kick; see Sect.~\ref{sec:evolution}) \citep{wrt89,kra98,wt02}. However, until today these observations could not be converted into a quantitative test of geodetic precession. Other systems turned out to be more useful for such a test, as we discuss below.

Finally, the discovery of the Hulse--Taylor pulsar was of great importance for the development of ground-based GW detectors, as explained in detail in Sect.~\ref{sec:binpsr}.

\subsection{\texorpdfstring{
The Double Pulsar system: PSR J0737$-$3039A/B  }{
The Double Pulsar system: PSR J0737-3039A/B    }}
\label{sec:psrJ0737}

In 2003, a truly remarkable binary pulsar system was discovered in a Parkes survey of the Galactic anti-centre, the \emph{Double Pulsar} PSR~J0737$-$3039A/B, which up to date is the only known binary system consisting of two active radio pulsars that are detectable from Earth \citep{bdp+03,lbk+04}. Pulsar A is a mildly recycled 23\,ms pulsar which is in a mildly eccentric ($e = 0.089$) 2.5-hour orbit with a non-recycled 2.8\,s pulsar B. The system is significantly more relativistic than the Hulse--Taylor pulsar (Sect.~\ref{sec:psrB1913}) showing, e.g., an advance of periastron ($\dot\omega$) of 16.9 degrees per year. In addition, the system is seen nearly edge on, which leads to a very prominent Shapiro delay of about 130\,$\mu$s in the timing of pulsar A near its superior conjunction \citep{lbk+04,ksm+06,ksm+21}. The masses are somewhat smaller than in the Hulse--Taylor pulsar, amounting to $M_\mathrm{A} = 1.338$ and $M_\mathrm{B} = 1.249$, when assuming GR \citep{ksm+06,ksm+21}.

From timing pulsars A and B one obtains the projected semi-major axes of both of the pulsar orbits, $x_\mathrm{A}$ and $x_\mathrm{B}$; from this one directly obtains the mass ratio $R = M_\mathrm{A}/M_\mathrm{B} = 1.0714 \pm 0.0011$ \citep{lbk+04,ksm+06}. The uncertainty in $R$ is dominated by the uncertainty in $x_\mathrm{B}$. Therefore, any improvement is currently limited by the fact that due to geodetic precession ($5.07^\circ\,{\rm yr}^{-1}$) pulsar B's emission moved out of our line of sight in 2008 \citep{pmk+10}. More model based approaches, in terms of the emission geometry of pulsar B (which is strongly affected by the wind of pulsar A), can lead to a further reduction in the uncertainty of $R$ (see e.g. \citealt{ndk+20}).

\begin{figure}[ht]
\begin{center}
\includegraphics[width=10cm]{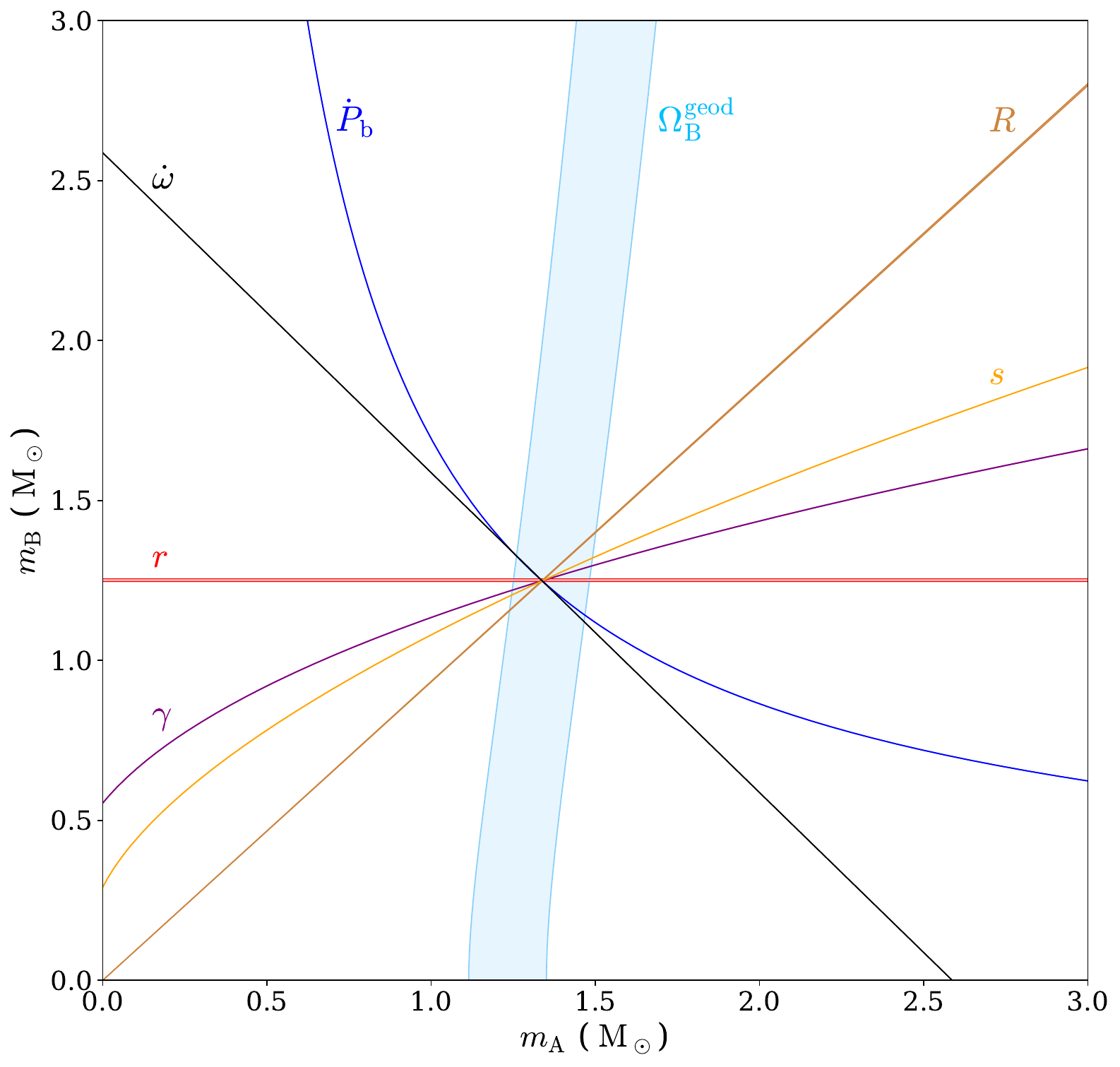}
\caption{Mass-mass diagram for the Double Pulsar based on GR. The PK parameters 
  $\dot\omega$, $\dot{P}_\mathrm{b}$, $\gamma$, $r$, and $s$ are from timing observations of pulsar A. The mass ratio $R$ comes from the observed projected semi-major axes of A and B. The rate of geodetic precession of B ($\Omega_\mathrm{B}^\mathrm{geod}$) is the result of modelling the eclipses of A near superior conjunction. Timing of pulsar A also allows for a one-sigma detection of the PK parameter $\delta_\theta$, which is not shown here. All curves agree on a small region of masses at the centre of the plot, meaning that GR provides a consistent description of the Double Pulsar system. The figure is based on the parameters and the analysis in \cite{ksm+21}. The values for $\Omega_\mathrm{B}^\mathrm{geod}$ are taken from \cite{lks+23}.}
\label{fig:0737mm}
\end{center}
\end{figure}

Fortunately, the (integrated) pulse profile of the fast rotating pulsar A remained stable over the past years, due to the fact that A's spin is closely aligned with the orbital angular momentum \citep{fsk+13}. As a consequence, all its timing parameters significantly improved since its discovery in 2003, allowing for the measurement of a total of six(!) PK parameters, some of them known to many significant digits, and all of them in perfect agreement with GR (see \citealt{ksm+21} and Fig.~\ref{fig:0737mm}). Furthermore, next-to-leading order (NLO) contributions of the advance of periastron, the Shapiro and the aberration delay had to be taken into account, in order to obtain a correct modelling and interpretation of these effects. This will be explained in more detail below. In the following we will have a closer look at four different predictions by GR tested in the Double Pulsar: GW damping, periastron advance, signal propagation, and geodetic precession.

\subsubsection{Gravitational wave damping}
\label{sec:0737_GW_damping}

\begin{figure}[ht]
\begin{center}
\includegraphics[width=10cm]{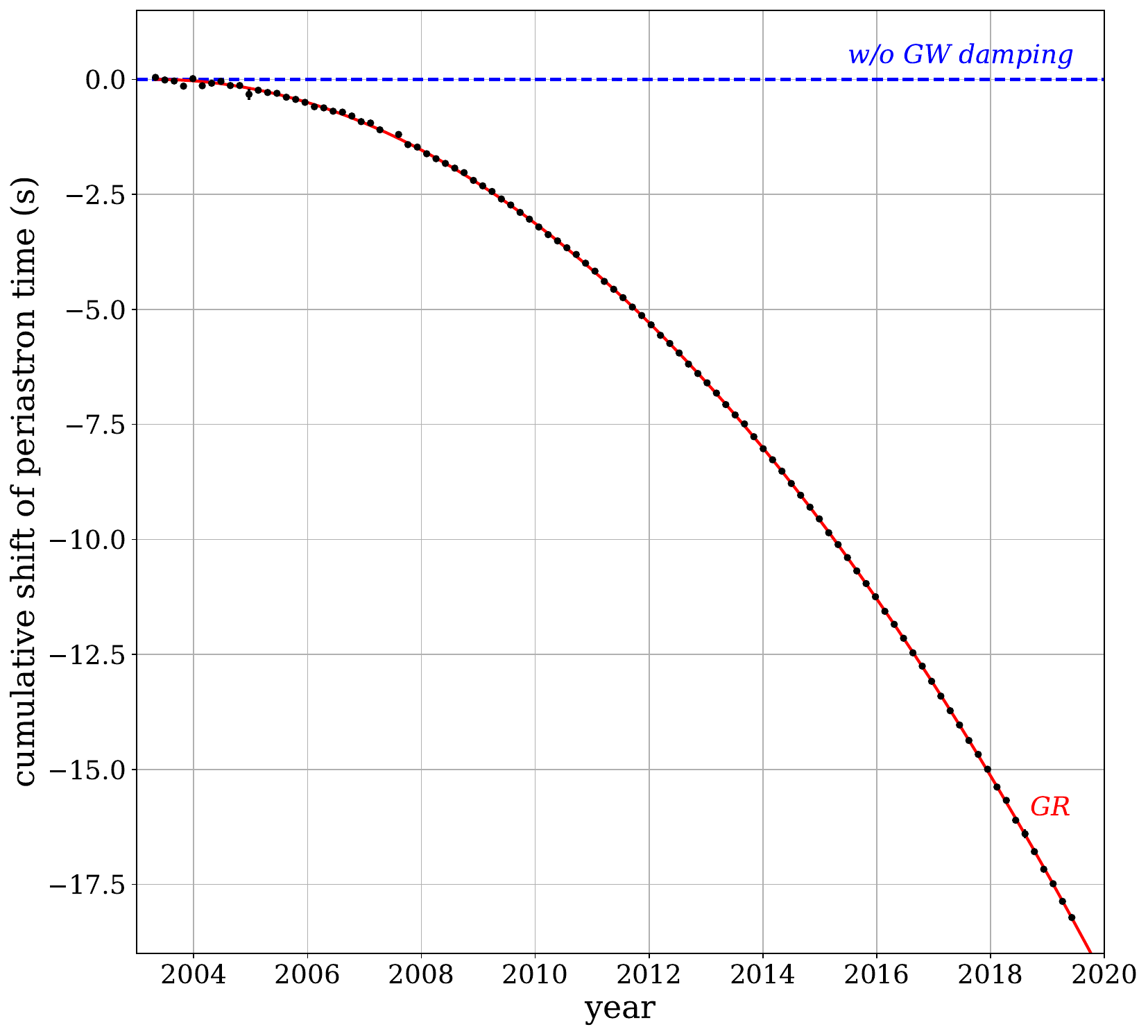}
\caption{Cumulative shift in the time of periastron passage (``GW damping parabola'') for the Double Pulsar. The red curve is the prediction by GR when using the pulsar masses calculated from the PK parameters $\dot\omega$ and $s$. The figure is based on the data and the analysis in \cite{ksm+21}.}
\label{fig:0737parabola}
\end{center}
\end{figure}

Like the Hulse--Taylor pulsar (Sect.~\ref{sec:psrB1913}), the Double Pulsar shows a significant decrease in the orbital period due to the emission of GWs (PK parameter $\dot{P}_\mathrm{b}$). As for the Hulse--Taylor pulsar, also here the observed $\dot{P}_\mathrm{b}$ is ``contaminated'' by external effects (cf.~Sect.~\ref{sec:Pbdot_ext}). Fortunately, the Double Pulsar is about ten times closer to us which allows for a direct parallax measurement of its distance: $735 \pm 60$ pc \citep{ksm+21}. As a result, the extrinsic terms in Eq.~(\ref{eq:pbdot_ext}) can be determined with sufficient precision so that they do not limit the GW test, like in the case of the Hulse--Taylor pulsar \citep{dbt09,ksm+21}. The intrinsic change of the orbital period obtained in this way ($\dot{P}_\mathrm{b}^{\rm int} = -1.247752(79) \times 10^{-12}$)  is in perfect agreement with the prediction by GR \citep{ksm+21}:
\begin{equation}
  \dot{P}_\mathrm{b}^{\rm int} / \dot{P}_\mathrm{b}^{\rm GR} = 0.999963(63) \,,
\label{eq:0737_PbdotRatio}  
\end{equation}
where the GR prediction $\dot{P}_\mathrm{b}^{\rm GR}$ is based on the masses $M_\mathrm{A}$ and $M_\mathrm{B}$, calculated from the PK parameters $\dot\omega$ and $s$.\footnote{Note, to obtain the result (\ref{eq:0737_PbdotRatio}), for completeness \cite{ksm+21} have accounted for a change in the orbital period due to the spindown mass loss by pulsar A (see Eq.~4.2 in \citealt{dt91}) and the NLO contribution in $\dot{P}_\mathrm{b}^{\rm GR}$ \citep{bs89}. Both contributions are still below the measurement uncertainty in the observed $\dot{P}_\mathrm{b}$.}
Figure~\ref{fig:0737parabola} shows a comparison of the observed and the calculated shift in the time of periastron passage, as a result of the accelerated evolution in orbital phase due to the emission of GWs.\footnote{Other effects of GW damping, like the shrinkage of the orbit, i.e. $\dot{x}_\mathrm{A} < 0$, or a decrease in the orbital eccentricity, i.e. $\dot{e} < 0$, are so far not significant.}
The result (\ref{eq:0737_PbdotRatio}) is by far the most precise test of GR's quadrupole formula for the leading-order emission of GWs by accelerated masses. Moreover it confirms the validity of the quadrupole formula in the presence of two strongly self-gravitating NSs, for which GR predicts the effacement of their internal structure \citep{dam87}. The corresponding merger time of the system is just 86\,Myr.

\subsubsection{Periastron advance}
\label{sec:0737_periastron}

The advance of periastron (PK parameter $\dot\omega$) is the most precisely measured PK parameter for the Double Pulsar and therefore, in principle, should give by far the narrowest constraint curve in the mass-mass plane of Fig.~\ref{fig:0737mm} (see also the inset in Fig.~13 of \citealt{ksm+21}). At a fractional precision of about $8 \times 10^{-7}$ for the observed $\dot\omega$, terms of the second PN (2PN) order \citep{ds88} must be considered to obtain the correct mass constraints, since the 2PN correction $\dot\omega_\mathrm{2pN} = +4.39 \times 10^{-4}\ {\rm deg\,yr^{-1}}$ is about 35 times larger than the measurement uncertainty of this PK parameter \citep{ksm+21}. Furthermore, the LT contribution to $\dot\omega$ from the coupling between the orbital motion and the spin of pulsar A has a contribution comparable in magnitude to the 2PN correction (with a negative sign, since the spin of pulsar A is aligned with the orbital angular momentum) \citep{ds88,kg06,ior09,hkw+20}.\footnote{The LT contribution from the slowly rotating pulsar B is two orders of magnitude smaller and therefore totally negligible.}
While the 2PN contribution is completely determined for given masses (and Kepler parameters $P_\mathrm{b}$ and $e$), the determination of the LT contribution also requires knowledge of the MoI $I_\mathrm{A}$ (see Eq.~\ref{eq:omdot_LT}). However, the calculation of $I_\mathrm{A}$ comes with an uncertainty related to our imperfect knowledge of the EoS of NS matter at supranuclear densities (see e.g. \citealt{lp01}). Based on the latest multimessenger constraints for the radius of a NS, \cite{ksm+21} give a LT contribution of $\dot\omega_\mathrm{LT} = -4.83^{+0.29}_{-0.35} \times 10^{-4}\ {\rm deg\,yr^{-1}}$. Although the uncertainty in this contribution is about a factor of three larger than the measurement uncertainty for $\dot\omega$, it is still not the limiting factor in the GW test outlined in Sect.~\ref{sec:0737_GW_damping} above, where the masses are estimated from $\dot\omega$ and $s$.
In turn, in the near future, in particular an improved precision for the intrinsic $\dot{P}_\mathrm{b}$ (in combination with the Shapiro shape measurement) should allow to put interesting constraints on the EoS for NSs \citep{ls05,kw09,kwkl18,hkw+20}. The current uncertainty for the intrinsic $\dot{P}_\mathrm{b}$ results in constraints on $I$ which are not competitive to current constraints from other observations (see \citealt{ksm+21} and references therein). 


\subsubsection{Signal propagation}
\label{sec:0737_signal_prop}

As it turns out, currently the Double Pulsar is the most edge-on binary pulsar observed. The orbital plane is tilted by only $0.64^\circ \pm 0.03^\circ$ with respect to the line of sight \citep{hkc+22}. At superior conjunction, the pulsar signal, on its way to the observer on Earth, comes within $\sim 10\,000$~km of the companion NS, resulting in a strong Shapiro delay of $\sim 130\,\mu$s (see Eq.~\ref{eq:deltaS}). In particular the timing of pulsar A resulted in a precise determination of the Shapiro parameters $r$ and $s$, requiring PN corrections to the mass function Eq.~(\ref{eq:s}) for a consistent conversion into constraints on the masses \citep{ksm+21,hkc+22}. In addition, it became necessary to account for the motion of the ``lens'' (pulsar B) while the radio signal is propagating across the binary system (retardation effect, \citealt{ks99,rl06}) in order to properly model the Shapiro delay near conjunction \citep{ksm+21,hkc+22} (see also Sect.~\ref{sec:shapiro_delay}).

In addition to the Shapiro propagation delay, the radio signal of pulsar A is also deflected by the curved spacetime of pulsar B. The deflection near superior conjunction reaches up to $\sim 0.03$ degrees. This deflection adds a higher order correction to the aberration of flat spacetime \citep{dk95,rl06_ApJ,rl06} which is well measured in the Double Pulsar \citep{ksm+21} (see also Sect.~\ref{sec:aberration_delay}). Based on precise timing with the MeerKAT telescope, \cite{hkc+22} give the so far best test of this {\it ‘longitudinal/rotational deflection delay} with a precision of 15\%. While this uncertainty is certainly large compared to deflection experiments in the gravitational field of the Sun \citep{wil18_book}, it is the first such experiment in the gravitational field of a NS, i.e.\ a strongly self-gravitating material body. In this context, it should be noted that the Double Pulsar tests described in this subsection probe the propagation of photons in a spacetime curvature that is orders of magnitude larger than in any other photon propagation experiment, even when compared to the shadow of the supermassive black hole (BH) Sgr A$^\ast$ \citep{EHT_2022} imaged with the Event Horizon Telescope (see Fig.~7 in \citealt{wk20}).  


\subsubsection{Geodetic precession}
\label{sec:0737_geod_prec}

The close edge-on orientation even leads to short intermittent eclipses of pulsar A by the plasma-filled magnetosphere of pulsar B during A's superior conjunction \citep{lbk+04}. While this reduces the number of ToAs and consequently the timing precision near conjunction, modelling the eclipse pattern changing over time could be used to measure the geodetic spin precession (cf. Sect.~\ref{sec:SOcoupling}) of pulsar B with moderate precision:  $4.77_{-0.65}^{+0.66}{}^\circ \, \mathrm{yr^{-1}}$ \citep{Breton_2008}, $5.16_{-0.34}^{+0.32}{}^\circ \, \mathrm{yr^{-1}}$ \citep{lks+23}. These values agree well with the prediction by GR: $5.074^\circ \, \mathrm{yr^{-1}}$. Compared to the Solar System, where geodetic precession of a gyroscope has been tested with high precision \citep{GPB}, such a test with a radio pulsar is different in two ways: it tests geodetic precession in a binary system with nearly equal masses, and more importantly, it tests geodetic precession with a strongly self-gravitating ``gyroscope''.


\subsection{Other relativistic double neutron-star systems}
\label{sec:psr-ns}

Apart from the Hulse--Taylor pulsar and the Double Pulsar, there are several additional DNS binary pulsars which have at least three measured PK parameters\footnote{See footnote~\ref{footnote:list}.}. In the following we discuss those systems which so far have played an important role in testing GR and/or alternative gravity theories.


\subsubsection{\texorpdfstring{
PSR B1534$+$12}{
PSR B1534+12   }}
\label{sec:psrB1534}

PSR~B1534$+$12, discovered in 1990 also with the Arecibo telescope \citep{wol91}, was the second DNS system suitable for testing GR and alternative gravity theories \citep{twdw92}. It is a 37.9-ms radio pulsar in an eccentric ($e=0.27$) 10.1-hour orbit. Like for the Hulse--Taylor pulsar, the three PK parameters $\dot\omega$, $\gamma$ and $\dot{P}_\mathrm{b}$ are measured in this system. Moreover, compared to the Hulse--Taylor pulsar system, where $i = 47.2^\circ$ ($s = 0.734$), this system is seen much more edge-on with an orbital inclination $i = 77.2^\circ$ ($s = 0.975$). As a result, the timing observations show a very prominent Shapiro delay allowing for the measurement of two more PK parameters, the Shapiro shape $s$ and the Shapiro range $r$ \citep{sac+98,sttw02,fst14}. PSR~B1534$+$12 thus provided the first precision test of the Shapiro delay in the gravitational field of a strongly self-gravitating mass. 

For a long time, the $\dot{P}_\mathrm{b}$ test was limited by the large (systematic) uncertainty in the estimation of the distance to PSR~B1534$+$12, which was mainly based on models for the Galactic distribution of free electrons. As a result, this led to large uncertainties in the correction for the Shklovskii contribution to the observed $\dot{P}_\mathrm{b}$ (cf.~Sect.~\ref{sec:Pbdot_ext}). Fairly recently, a reliable distance measurement was obtained with the help of VLBI ($d = 0.94^{+0.07}_{-0.06}\,\mathrm{kpc}$) leading to a robust 4\% (95\% C.L.) radiative test with this system \citep{ddf+21}.

As with the Hulse--Taylor pulsar, the kick of the second SN explosion in this system also led to a misalignment between the pulsar's rotational axis and the orbital angular momentum ($\sim 30^\circ$). Although the precession rate predicted by GR ($\Omega_\mathrm{p}^\mathrm{geod} = 0.514^\circ\,\mathrm{yr}^{-1}$) is significantly smaller than the one for the Hulse--Taylor pulsar, the high S/N ratio obtained for the (integrated) pulse profile and polarisation of PSR~B1534$+$12 not only allowed for a detection of geodetic precession at an early stage but also for the first time provided a quantitative test of the precession rate of a pulsar \citep{sta04}, with the latest value given in \cite{fst14}: $0.59_{-0.08}^{+0.12}{}^\circ\,\mathrm{yr}^{-1}$. 

\begin{figure}[ht]
\begin{center}
\includegraphics[width=10cm]{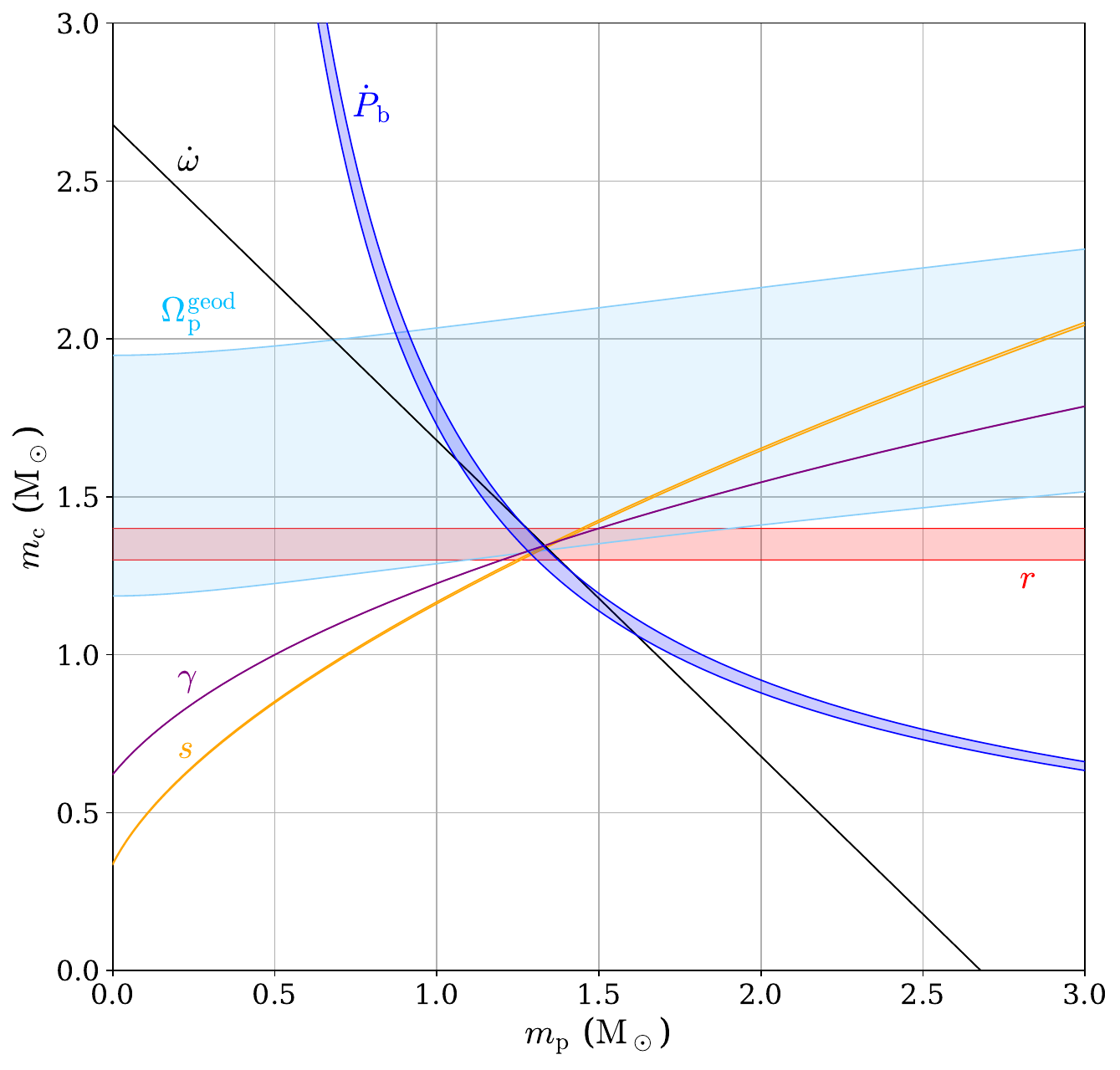}  
\caption{Mass-mass diagram for the PSR~B1534$+$12 system based on GR. PSR~B1534$+$12 provides constraints from six different PK parameters, including the geodetic precession of the pulsar ($\Omega_\mathrm{p}^\mathrm{geod}$). Data are taken from \cite{fst14,ddf+21}.}
  \label{fig:m1m2_B1534}
\end{center}
\end{figure}

Figure~\ref{fig:m1m2_B1534} shows the result of the test of GR with the six PK parameters (5 quasi-stationary, 1 radiative) of PSR~B1534$+$12. 

\subsubsection{\texorpdfstring{
PSR J1906$+$0746 }{
PSR J1906+0746   }}
\label{sec:psrJ1906}

\begin{figure}[ht]
\begin{center}
\includegraphics[width=10cm]{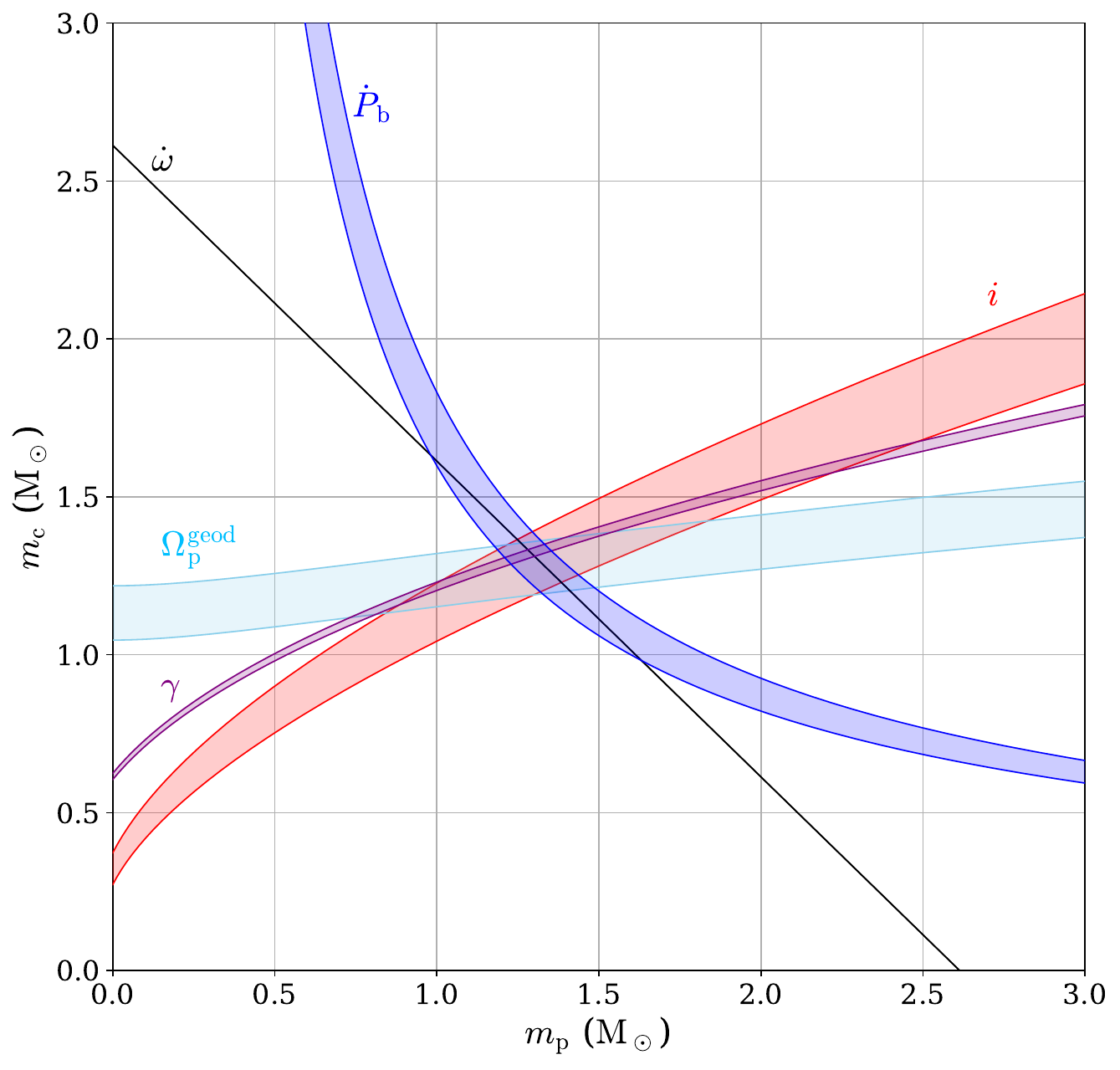}  
\caption{Mass-mass diagram for the PSR~J1906$+$0746 system based on GR. The three PK parameters $\dot\omega$, $\gamma$ and $\dot{P}_\mathrm{b}$ were measured in timing observations \citep{vks+15}. The geodetic precession rate $\Omega_\mathrm{p}^\mathrm{geod}$ and orbital inclination $i$ were obtained from modelling changes in the pulsar polarisation due to the spin-precession of the pulsar \citep{dkl+19}.}
\label{fig:m1m2_J1906}
\end{center}
\end{figure}

Binary pulsar J1906+0746, discovered with the Arecibo telescope, has a relatively slow spin period (144 ms) and a comparably large (inferred) surface magnetic field \citep{lsf+06}. Therefore it is most likely a non-recycled young pulsar in a mildly eccentric ($e = 0.085$) 4.0-hour orbit with a NS companion.\footnote{It has been argued that the companion could in principle also be a high-mass ONeMg WD \citep{tkf+17}. However, we consider it as more likely that this is a DNS system like the Double Pulsar (similar masses and eccentricity) where we only see the pulsar that was formed during the second SN explosion.} 
Timing observations allowed the determination of three PK parameters ($\dot\omega$, $\gamma$, $\dot{P}_\mathrm{b}$) with good precision, allowing for a 5\% radiative test \citep{vks+15}. This alone does not make this system particularly interesting for gravity tests. However, the pulsar's spin axis has a large misalignment of $104(9)^\circ$ with respect to the orbital angular momentum and shows a very characteristic polarisation pattern which changed over time due to geodetic precession. Crucially, these changes were observed not only for the main pulse but also for the interpulse; the observation of both magnetic poles results in unusually robust estimates of the geometry of the magnetic field and the direction of the spin axis. By modelling these polarisation changes, the precession rate of the pulsar spin could be determined to be $2.17(11)^\circ\,\mathrm{yr^{-1}}$, which is in perfect agreement with the GR value: $2.234(14)^\circ\,\mathrm{yr^{-1}}$. This is so far the best test of the geodetic precession of a spinning NS (cf.\ Sects.~\ref{sec:0737_geod_prec}, \ref{sec:psrB1534}). Figure~\ref{fig:m1m2_J1906} shows the GR mass-mass diagram for PSR~J1906$+$0746.

\subsubsection{\texorpdfstring{
PSR J1757$-$1854}{
PSR J1757-1854   }}
\label{sec:psrJ1757}

PSR~J1757$-$1854 is currently (in certain aspects) the most relativistic binary pulsar with which GR has been tested. It is a 21.5\,ms pulsar in a highly eccentric ($e = 0.606$) DNS system ($M_\mathrm{p} = 1.34$, $M_\mathrm{c} = 1.39$) \citep{cck+18}. In a sense, it is a more relativistic version of the Hulse--Taylor pulsar. Although its orbital period of 4.4 hours is clearly larger than that of the Double Pulsar, it is the high eccentricity that leads to a significantly stronger decrease in the orbital period due to GW damping ($\dot{P}_\mathrm{b} = -5.3 \times 10^{-12}$) and a correspondingly shorter merger time of 76\,Myr. At periastron the two NSs have a separation of just $0.75\,{\rm R}_\odot$, leading to a relative velocity of 1060\,km/s, the largest for known binary pulsars.

\begin{figure}[ht]
\begin{center} 
\includegraphics[width=10cm]{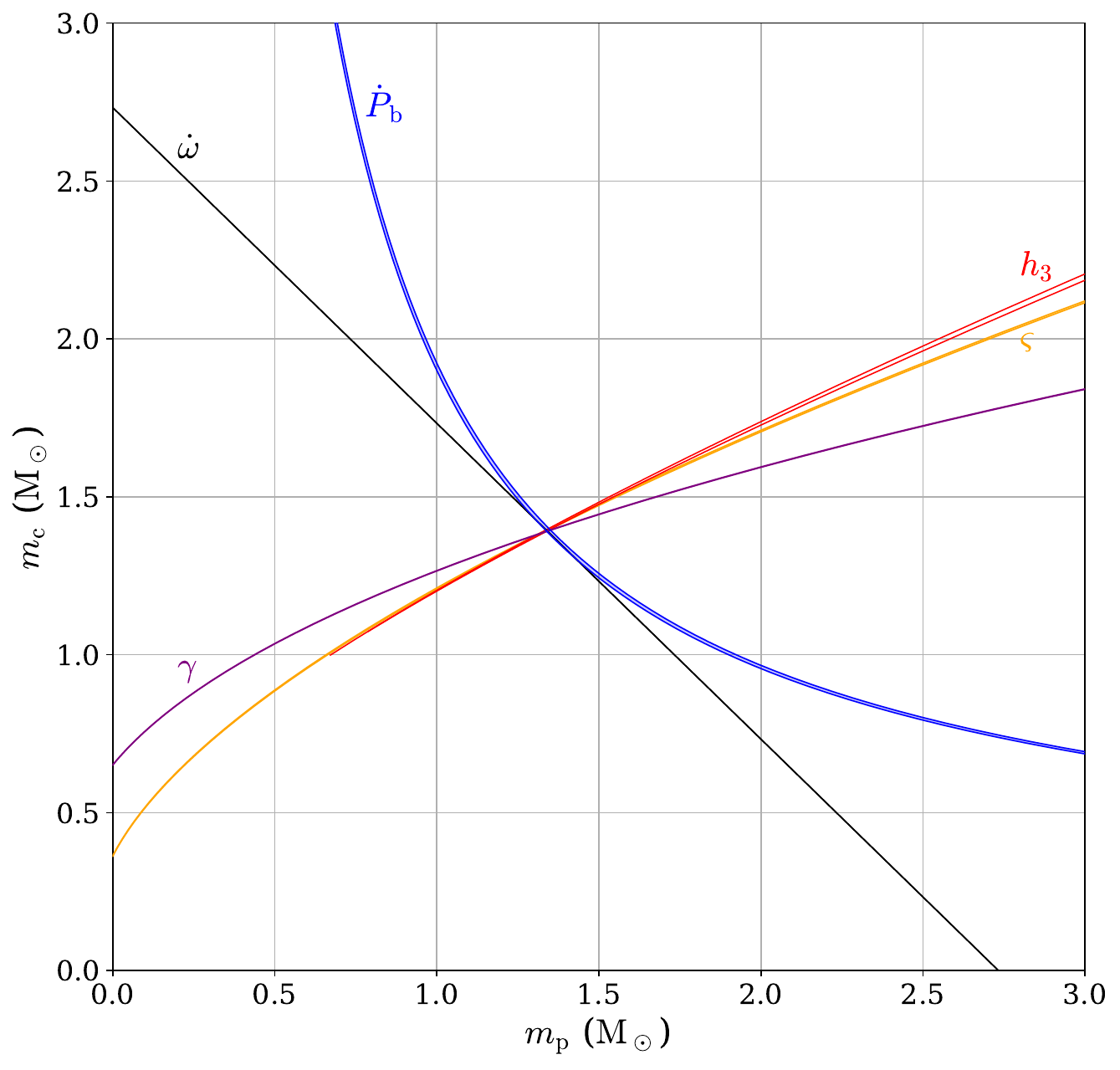}  
\caption{Mass-mass diagram for the PSR~J1757$-$1854 system based on GR. Parameter values are taken from \cite{cbc+23}}
  \label{fig:m1m2_J1757}
\end{center}
\end{figure}

So far, there are five PK parameters measured for PSR~J1757$-$1854, including the two orthometric parameters of the Shapiro delay (see Fig.~\ref{fig:m1m2_J1757}). All parameters agree on a common region in the GR mass-mass plane, meaning GR has also passed this test. However, there is a large systematic uncertainty in the intrinsic $\dot{P}_\mathrm{b}$. The reason is the unknown distance to the system, and therefore a large uncertainty in the extrinsic contributions to the observed $\dot{P}_\mathrm{b}$ (Sect.~\ref{sec:Pbdot_ext}, see also Fig.~13 in \cite{cbc+23}).\footnote{The only $\dot{P}_\mathrm{b}$-independent distance estimation, currently available, is based on the column density of free electrons obtained from the dispersion measure. Unfortunately, different models for the distribution of free electrons in our Galaxy give vastly different results (see \citealt{cbc+23}).}
If one assumes GR and a model for the Galactic gravitational potential, then the $\dot{P}_\mathrm{b}$ can be used to estimate the distance of the system to about 8 -- 13\,kpc.

While it is still unclear how to obtain better constraints on the distance without assuming GR, a promising additional GR test for the near future---due to the large eccentricity---is the relativistic deformation of the orbit $\delta_\theta$ \citep{cbc+23}, which is only barely significant in the Hulse--Taylor pulsar (Sect.~\ref{sec:psrB1913}) and the Double Pulsar (Sect.~\ref{sec:psrJ0737}). In principle, PSR~J1757$-$1854 could also be an excellent pulsar binary system for testing the LT precession of the orbital plane (PK parameter $\dot{x}$; see Eq.~\ref{eq:didtLT}). However, the analysis of the pulse structure and the polarisation of PSR~J1757$-$1854 leads to the conclusion that  the spin is oriented such that $\dot{x}_\mathrm{LT}$ is too small to be measurable in the near future.


\subsubsection{\texorpdfstring{
PSR J1913$+$1102}{
PSR J1913+1102   }}
\label{sec:psrJ1913}

\begin{figure}[ht]
\centering
\includegraphics[width=10cm]{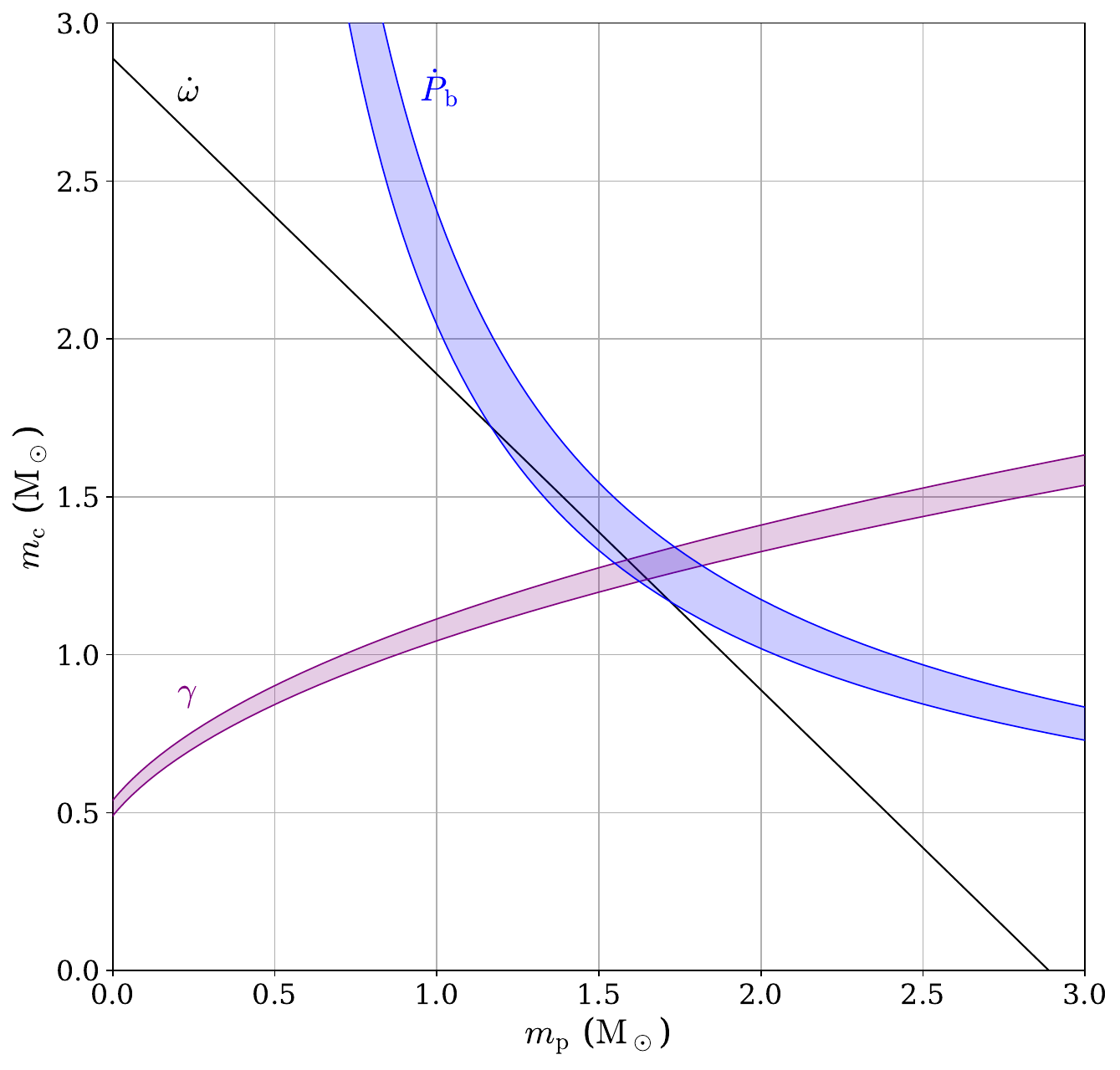}  
\caption{Mass-mass diagram for the PSR~J1913$+$1102 system based on GR. For that pulsar three 
  PK parameters have been measured: advance of periastron ($\dot\omega$, black), time dilation ($\gamma$, purple), and change of the orbital period due to GW damping ($\dot{P}_\mathrm{b}$, blue). They agree a common mass-mass region, meaning that GR has passed this test with an asymmetric DNS system (cf.\ Fig.~1 in \citealt{ffp+20}).}
  \label{fig:m1m2_J1913}
\end{figure}

PSR~J1913$+$1102 is a 27-ms pulsar which is in a mildly eccentric ($e = 0.09$) 5.0-hour orbit with a NS companion \citep{lfa+16}. Regular timing observations since its discovery in 2012, in particular with the 305-m Arecibo telescope, allowed the measurement of three PK parameters: $\dot\omega = 5.6501(7)\,{\rm deg\,yr^{-1}}$, $\gamma = 0.471(15)\,{\rm ms}$, and $\dot{P}_\mathrm{b} = -0.480(30) \times 10^{-12}$. Assuming GR, one obtains a pulsar mass of $m_\mathrm{p} = 1.62(3)\,\Msun$ and a companion mass of $m_\mathrm{c} = 1.27(3)\,\Msun$ \citep{ffp+20}. With a mass ratio of $R \equiv m_\mathrm{p}/m_\mathrm{c} = 1.28(4)$, this is the most asymmetric DNS system reported so far that shows significant GW damping, i.e.\ a significant (intrinsic) $\dot{P}_\mathrm{b}$. The observed GW damping agrees with GR, providing a 6\% test (see Fig.~\ref{fig:m1m2_J1913}). This by itself cannot compete with most of the other GW tests presented so far. However, the comparably large asymmetry in the NS masses, and the correspondingly large asymmetry in the (fractional) gravitational binding energy makes this DNS system interesting for tests of dipolar GW emission, predicted by many alternatives to GR (see Sect.~\ref{sec:altGrav}). 

\subsection{Relativistic pulsar-white dwarf systems}
\label{sec:psr-wd}

Currently, about 400 binary pulsars are known, more than half of which have a WD as a companion \citep{mhth05}. Several of these pulsar-WD systems have orbital periods of less than one day. Many of the most precise ``pulsar clocks'' are found in such systems (see  discussion in Sect.~\ref{sec:evolution}). As a result of the mass transfer, a circularisation of the orbit took place leading to very low eccentricities. As a consequence of this, often neither the advance of periastron (PK parameter $\dot\omega$) nor the Einstein delay (PK parameter $\gamma$) has been measured in these systems. In some cases the system is seen sufficiently edge on, so that a significant Shapiro delay is present in the timing data. For quite a few pulsar-WD systems, the (intrinsic) change in the orbital period (PK parameter $\dot{P}_\mathrm{b}$) is the only measurable relativistic effect. If then mass estimates can be obtained through alternative channels, e.g.\ high-resolution spectroscopy observations of the WD companion, then such systems can still provide valuable tests of GR. Pulsar-WD binaries are of particular interest for the study of alternatives to GR, since such systems exhibit a large asymmetry in the (fractional) gravitational binding energy, which in many alternative gravity theories leads to the prediction of strong dipolar GWs that do not occur in the GR (see Sects.~\ref{sec:dipolar} and \ref{sec:altGrav}). In the following, we will give a list of pulsar-WD systems that so far have been of particular importance in gravity tests, most notably for constraining alternatives to GR. For all of these systems, the masses can be determined (either from timing alone, or with the help of optical observations) and at least one (additional) PK parameter has been measured, so that the system is over-constrained.


\subsubsection{\texorpdfstring{
PSR J1738$+$0333 }{
PSR J1738+0333   }}
\label{sec:psrJ1738}

\begin{figure}[ht]
\begin{center}
\includegraphics[width=10cm]{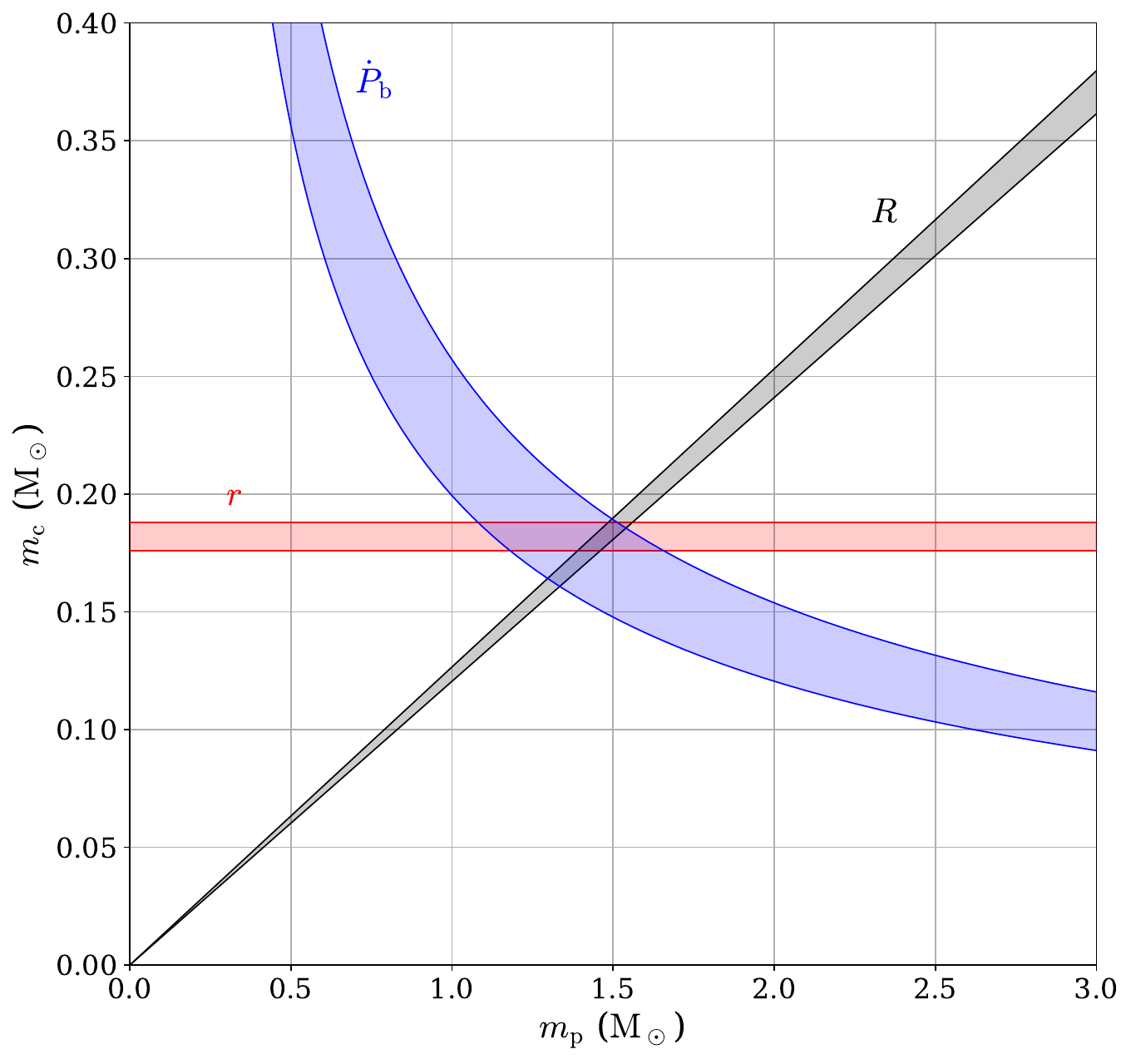}
\caption{Mass-mass diagram for PSR~J1738$+$0333, assuming GR. Bands denote one-sigma ranges in the parameters. The mass ratio $R = M_\mathrm{p}/M_\mathrm{c}$ (black) and the WD mass $M_\mathrm{c}$ (red) are obtained from optical observations and WD models. The blue band is a result of Eq.~(\ref{eq:pbdot}) with $\dot{P}_\mathrm{b} = \dot{P}_\mathrm{b}^{\rm int}$. All three bands agree on a common region in the mass-mass plane, meaning GR has passed this pulsar-WD test.}
\label{fig:m1m2_J1738}
\end{center}
\end{figure}

PSR~J1738$+$0333 is a fully recycled pulsar with a rotational period of 5.9 ms and an optically bright low-mass Helium-core WD as a companion \citep{jac05,jbo+07}. The two stars orbit each other in about 8.5 hours in a nearly circular orbit ($e < 4 \times 10^{-7}$). Regular timing observations since 2003, in particular with the 305-m William E.\ Gordon Arecibo radio telescope, allowed eventually the determination of a significant change in the (intrinsic) orbital period \citep{fwe+12}. With the latest distance measurement from VLBI, needed to correct for the extrinsic $\dot{P}_\mathrm{b}$ contributions, one finds $\dot{P}_\mathrm{b}^{\rm int} = -26.1 \pm 3.1~{\rm fs\,s^{-1}}$ \citep{dds+23}. This change in the orbital period due to GW damping is the only PK parameter known so far in this system. Fortunately, masses can be obtained independently from  optical observations of the WD companion. High-resolution spectroscopy combined with WD models lead to a mass ratio of $R = 8.1 \pm 0.2$ and a WD mass of $m_\mathrm{c} = 0.181_{-0.005}^{+0.007}\,\Msun$, which converts into $m_\mathrm{p} = 1.47_{-0.06}^{+0.07}\,\Msun$ \citep{avk+12}. With pulsar and companion masses at hand, Eq.~(\ref{eq:pbdot}) can be used to determine the GR value for the change in orbital period ($-27.7_{-1.9}^{+1.5}~{\rm fs\,s^{-1}}$) \citep{fwe+12}, which agrees within the uncertainties well with the observed value (see also Fig.~\ref{fig:m1m2_J1738}). The precision of this test is orders of magnitude weaker than the GW test with the Double Pulsar (see Sect.~\ref{sec:0737_GW_damping}). Nevertheless, due to the high asymmetry in the compactness between the pulsar and the companion WD, this limit converts into tight constraints on dipolar radiation as predicted by alternatives to GR (see Sects.~\ref{sec:dipolar} and \ref{sec:altGrav} for details).  

The GW test is not the only gravity test with PSR~J1738$+$0333. It has also been used, for instance, in (generic) tests of preferred frame effects in the gravitational interaction, as discussed in Sect.~\ref{sec:alpha1}.


\subsubsection{\texorpdfstring{
PSR J1909$-$3744 }{
PSR J1909-3744   }}
\label{sec:psrJ1909}

PSR~J1909$-$3744 is a fully recycled pulsar with a rotational period of 2.9 ms \citep{jbv+03}. Like PSR~J1738$+$0333, it has an optically bright low-mass Helium-core WD as a companion that allows for high resolution spectroscopy \citep{jbv+03}. With its orbital period of 1.53 days it is considerably less relativistic than PSR~J1738$+$0333. For that reason, and despite of its exquisite timing precision (one of the most precise timers known, \citealt{IPTA2019}), there is to date no significant value for the intrinsic change in the orbital period. The limitation comes mainly from the uncertainties in the correction for the Shklovskii contribution to $\dot{P}_\mathrm{b}$ \citep{lgi+20}. Still, this system could be used in a GW test, providing important constraints on the strong-field scalarisation of NSs, as we discuss in Sect.~\ref{sec:altGrav}.

\begin{figure}[ht]
\begin{center}
\includegraphics[width=10cm]{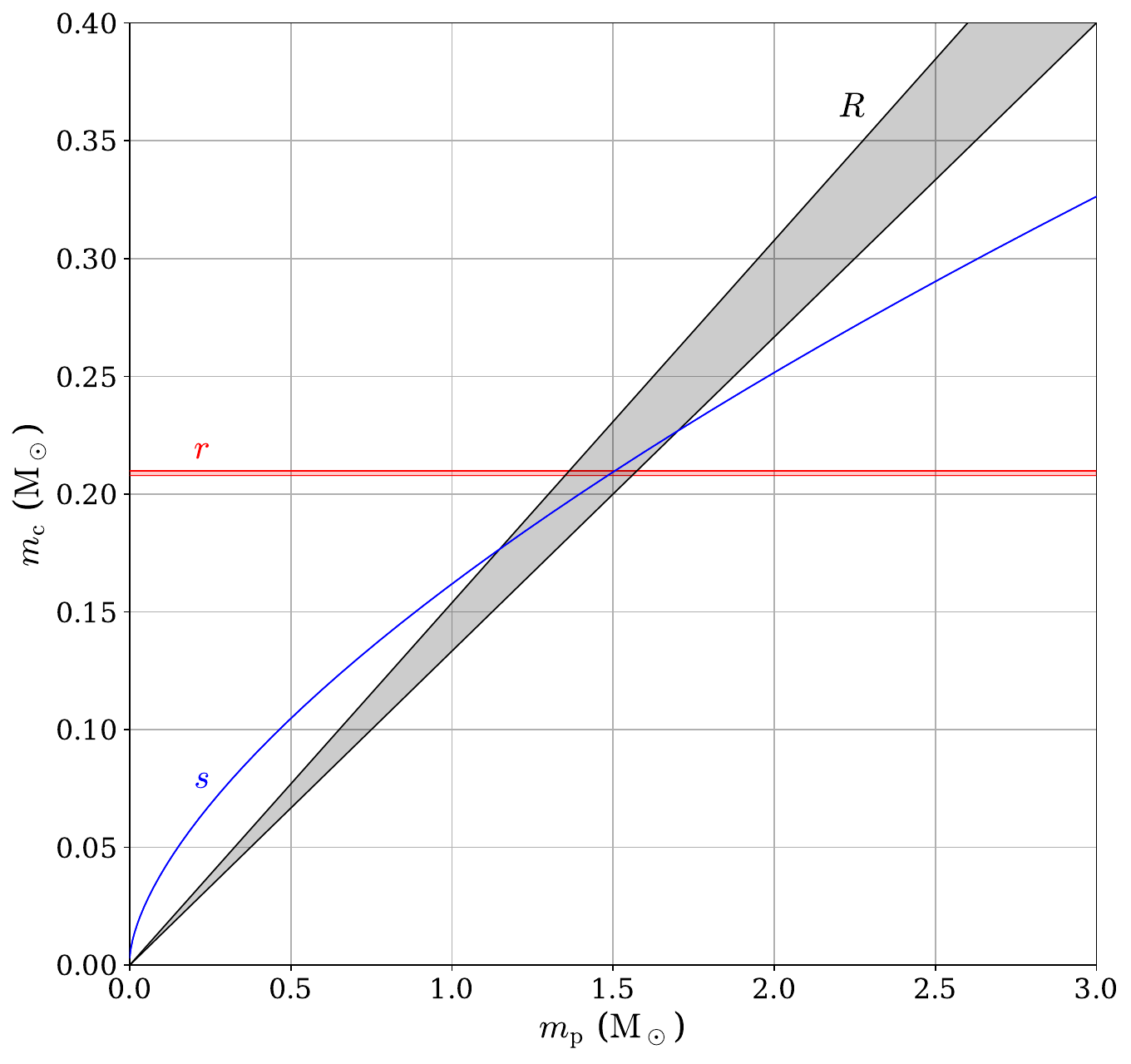}
\caption{Mass-mass diagram for PSR~J1909$-$3744, assuming GR. Band denote one-sigma ranges in the parameters. The mass ratio $R = M_\mathrm{p}/M_\mathrm{c}$ (black) is obtained by combining pulsar timing with radial velocity measurements of the WD by optical spectroscopy. The other two curves are from the detection of the Shapiro delay in the timing data: Shapiro range in red and Shapiro shape in blue. All three bands agree on a common region in the mass-mass plane, meaning GR has passed this pulsar-WD test. As is obvious, assuming GR, the Shapiro delay gives precise masses for pulsar ($m_\mathrm{p} = 1.492(14)\,\Msun$) and companion ($m_\mathrm{c} = 0.209(1)\,\Msun$). See \cite{lgi+20} for details.}
\label{fig:m1m2_J1909}
\end{center}
\end{figure}

Many years of high-precision timing of PSR~J1909$-$3744 allowed a precise measurement of two PK parameters. Due to its near edge-on orientation ($i = 86.4^\circ$ or $93.6^\circ$) with respect to the line of sight, precise measurements of the shape ($s$) and the range ($r$) of the Shapiro delay were possible (see \citealt{lgi+20} and references therein). In combination with the mass ratio (from a combination of optical and radio observation) one can do a PK parameter test (see Fig.~\ref{fig:m1m2_J1909}). That test is not of particular interest for GR and its alternatives. More importantly, the precise determination of the WD mass via the Shapiro delay makes it possible to test different WD models \citep{ant13}, which in turn is important for gravity tests in other systems (see e.g. Sect.~\ref{sec:psrJ1738}).

PSR~J1909$-$3744 has  been used in (generic) tests of gravitational preferred-frame effects and is currently providing the most stringent limits on $\hat\alpha_1$ (see Sect.~\ref{sec:alpha1}). 

Last but not least, due to its high timing precision, PSR~J1909$-$3744 is of prime importance for all PTAs and their efforts to detect nano-Hz GWs \citep{IPTA2019}.


\subsubsection{\texorpdfstring{
PSR J2222$-$0137 }{
PSR J2222-0137   }}
\label{sec:psrJ2222}

\begin{figure}[ht]
\begin{center}
\includegraphics[width=10cm]{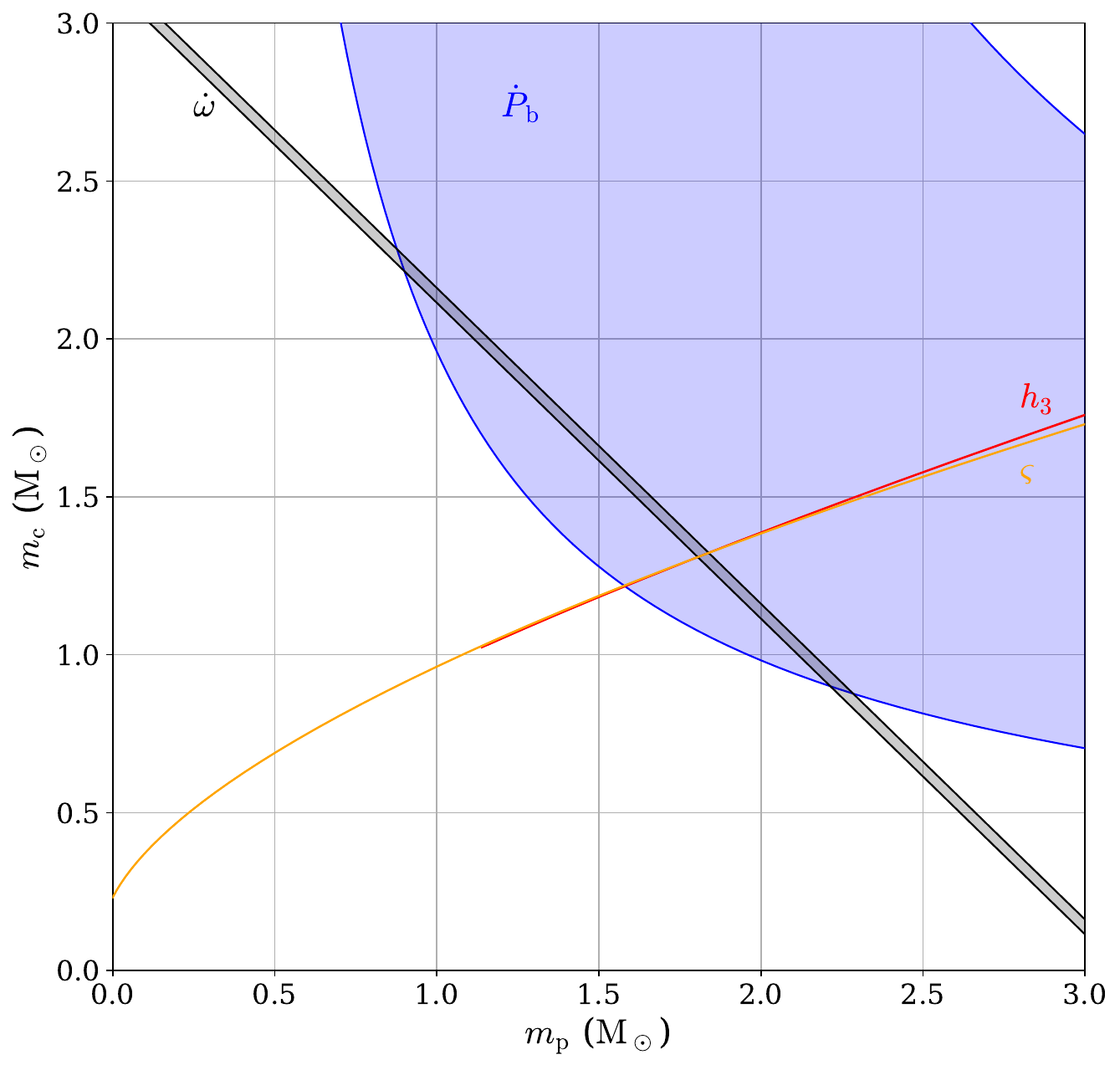}
\caption{Mass-mass diagram for PSR~J2222$-$0137, assuming GR. Bands denote one-sigma ranges in the parameters. Two parameters are associated with secular changes in the orbit: advance of periastron ($\dot\omega$, black) and change of orbital period due to GW damping ($\dot{P}_\mathrm{b}$, blue). The parameters $h_3$ (red) and $\varsigma$ (orange) are part of the orthometric parametrisation of the Shapiro delay \citep{fw10}, replacing the PK parameters $r$ and $s$ according to Eqs.~(\ref{eq:stig_h3}). See \cite{gfg+21} for details.}
\label{fig:m1m2_J2222}
\end{center}
\end{figure}

PSR~J2222$-$0137 is a mildly recycled binary pulsar with a spin period of 32.8 ms and an orbital period of 2.45 days \citep{blr+13}. The low orbital eccentricity ($e = 3.8 \times 10^{-4}$) indicates that the companion is a massive WD. Despite this low eccentricity, the relativistic advance of periastron could be measured with high precision: $\dot\omega = 0.09605(48)\,{\rm deg\,yr^{-1}}$ \citep{gfg+21}. Besides $\dot\omega$, the system gives access to two more PK parameters from the measurement of the Shapiro delay. In combination, this 3-PK-parameter test leads to a $\sim 1$\% confirmation of GR (see Fig.~\ref{fig:m1m2_J2222}). The corresponding GR masses for pulsar and companion are $1.831(10)\,\Msun$ and $1.319(4)\,\Msun$ respectively \citep{gfg+21}. 

The large mass of the pulsar makes this system particularly interesting for certain non-linear deviations from GR in the strong-gravity regime of NSs. In particular, this system plays a key role in closing the mass gap of spontaneous scalarisation in DEF gravity (see Sect.~\ref{sec:altGrav} for details). Two additional aspects are of particular importance for this test. Firstly, the timing precision and the large ratio $x/P_\mathrm{b}$ allow a particularly precise measurement of $\dot{P}_\mathrm{b}$. Secondly, VLBI observations by \cite{dbl+13} obtained the most precise VLBI distance for any pulsar ($268 \pm 1$ pc) as well as precise values for the proper motion of the system. As a result, precise corrections for the extrinsic contributions to $\dot{P}_\mathrm{b}$ were possible, leading to a two-sigma significant measurement of the GW damping ($\dot{P}_\mathrm{b} = -0.0143(76) \times 10^{-12}\,{\rm s\,s^{-1}}$), although the orbital period is much larger than for any other binary pulsar that currently allows the verification of GW damping.

\subsubsection{Other pulsar-white dwarf systems}
\label{sec:other-PSR-WD}

There are a number of short orbital period pulsar-WD systems that have been used for gravity tests in the past but became less important in recent years (which in some cases may change again in the future). There are various reasons for this. In many cases, the precision of gravity tests with these systems has now simply been surpassed by other systems. The most important examples are \emph{PSR~J0348+0432} \citep{afw+13}, \emph{PSR~J1012$+$5307} \citep{lwj+09,ddf+20}, \emph{PSR~J1141$-$6545} \citep{bbv08}. In particular for the last one, an eccentric pulsar-WD system in a 4.7-hour orbit, we expect updated gravity tests in the near future that will provide the best constraints within specific parameter ranges of scalar-tensor theories (some preliminary results can already be found in \citealt{ven19}).\footnote{See details on spin-orbit coupling detected in this system in footnote~\ref{footnote:1141}. However, it is important to note that this did not provide a gravity test as the analysis in \cite{vbv+20} is entirely based on the assumption that GR is the correct theory of gravity.}

A completely different group of pulsar-WD systems relevant for gravity tests are those with very wide orbits. In general, these systems do not allow the test of an effect predicted by GR, like the pulsar-WD systems above (i.e.\ one cannot draw three curves in a mass-mass plane), but due to their orbital properties they provide important constraints on specific deviations from GR, such as a time dependent gravitational constant $G$ or a violation of the universality of free fall (see Sect.~\ref{sec:altGrav} below for more details). A particularly noteworthy pulsar in this context is the 4.6-ms pulsar \emph{J1713$+$0747}. It is in a 68-day low-eccentricity ($e = 7.5 \times 10^{-5}$) orbit with a $0.3\,\mathrm{M}_\odot$ WD. Mass measurements in this system were possible due to the detection of a Shapiro delay, the only relativistic effect measured in this system. The latest timing results for this pulsar can be found in \cite{zdw+19}. 

A truly unique pulsar-WD system is that of the 2.7-ms pulsar \emph{J0337$+$1715} \citep{rsa+14}. This pulsar is a member of a hierarchical triple, where the $1.44\,\mathrm{M}_\odot$ pulsar and a $0.20\,\mathrm{M}_\odot$ WD orbit each other in 1.63 days. This inner binary is in a 327-day orbit with an outer WD of $0.41\,\mathrm{M}_\odot$. Both orbits have a low eccentricity of $7\times 10^{-4}$ and 0.035, respectively. The only relativistic effect observed in this system so far is the (varying) special relativistic time dilation caused by the (epicyclic) motion of the pulsar in the inertial frame of the triple system. Nevertheless, as we will see in Sect.~\ref{sec:Nordtvedt} and Sect.~\ref{sec:def_grav}, this system provides some of the tightest limits on deviations from GR (generic and theory specific). For details see also Refs.~\cite{agh+18,vcf+20}.


\section{Pulsar experiments and strong field deviations from GR}
\label{sec:beyond_GR}

The fact that GR has passed all these tests presented in the previous chapter has important consequences in two ways. On the one hand, GR can bee used as a tool in binary pulsars to determine precise masses of NSs \citep{of16,fcp+21}, to get pulsar distances (see e.g.\ PSR~B1534+12 in Sect.~\ref{sec:psr-ns}), or to confirm the fast rotation of a companion WD where, besides classical spin-orbit coupling, the LT effect contributes to the precession of the orbital plane \citep{vbv+20}, just to name a few. 

On the other hand, the excellent agreement of pulsar experiments with GR also means tight constraints on deviations from GR in the presence of strongly self-gravitating bodies, i.e.\ deviations in the orbital motion, GW emission, photon propagation, etc. In the following we will highlight some of such tests conducted with radio pulsars. Our list is, however, far from complete. We distinguish between generic tests of phenomena expected on the basis of general theoretical assumptions on how deviations from GR might affect pulsar systems, and tests of specific theories of gravity. 

For generic tests in the weak-field regime, the most important framework is the parametrised PN (PPN) framework, where 10 parameters quantify deviations from GR at the first PN level \citep{wil93,wil14_LRR}. These PPN parameters assume different values in different alternatives to GR, depending also on whether they are conservative theories of gravity or not (see Tab.~\ref{tab:PPN_params}). It is important for pulsar experiments that these PPN parameters become body-dependent in a system with strongly self-gravitating bodies, which we will explain in more detail below.


\begin{table}
\caption{List of generic parameters subject to gravity tests (including the ten PPN parameters; first group); their physical meaning and predictions for them in GR and fully conservative and semi-conservative gravity theories (under ``c/sc''; the parameters $\alpha_1$ and $\alpha_2$ are 0 in fully conservative theories). Taken from the compilations by \cite{wil18_book}.
\label{tab:PPN_params}}
\begin{tabular} { l l l l }
\hline
Parameter & GR & c/sc & Physical meaning \\
 & & & \\
\hline\hline
$\gamma_\mathrm{PPN}$ & 1 & $\gamma$ & Space curvature produced by unit rest mass \\
$\beta_\mathrm{PPN}$ & 1 & $\beta$ & Non-linearity in superposition law for gravity \\ 
$\xi$ & 0 & $\xi$ & Preferred-location effects \\
$\alpha_1$ & 0 & $\alpha_1$ & Preferred-frame effects \\
$\alpha_2$ & 0 & $\alpha_2$ & Preferred-frame effects \\
$\alpha_3$ & 0 & 0 & Preferred-frame effects and non-conservation of momentum \\
$\zeta_1$ & 0 & 0 & Non-conservation of momentum \\
$\zeta_2$ & 0 & 0 & Non-conservation of momentum \\
$\zeta_3$ & 0 & 0 & Non-conservation of momentum \\
$\zeta_4$ & 0 & 0 & Non-conservation of momentum \\ \hline
$\eta$    & 0 & $\eta$ & Nordtvedt effect; combination of seven PPN parameters \citep{wil93} \\
$\kappa_\mathrm{D}$ & 0 & $\kappa_\mathrm{D}$ & Dipolar radiation coupling \\
$\dot G/G$ [${\rm yr}^{-1}$] & 0 & $\dot G/G$ & Variation of Newton's gravitational constant \\
\hline
\end{tabular}
\end{table}


If the symmetric spacetime metric $g_{\mu\nu}$ is the only gravitational field in a four-dimensional spacetime manifold then, under some plausible assumptions, GR (including a cosmological constant) emerges as the unique theory of gravity. This is the result of Lovelock's uniqueness theorem \citep{Lovelock_1972}. To get around Lovelock's theorem, one has to relax one (or more) of the assumptions that go into it (see e.g.\ Fig.~1 in \citealt{Berti_2015}). Arguably the most popular  assumption is the existence of additional (generally dynamical) gravitational fields, for instance one or more scalar fields. If non-gravitational fields do not couple directly to these additional gravitational fields (only to $g_{\mu\nu}$; universal coupling), the theory is a \emph{metric theory of gravity} and by construction fulfills the EEP (see e.g.\ \citealt{wil18_book}).  Gravitational experiments, on the other hand, are generally expected to deviate from GR due to the presence of additional gravitational fields. From a heuristic perspective (and with some degree of approximation), certain aspects of such deviations have been summarised in the \emph{``strong equivalence principle (SEP)''}, an extension of the EEP to local gravitational experiments:
\begin{itemize}
    \item Extension of the universality of free fall (UFF) to self-gravitating bodies in an external gravitational field.
    \item Absence of preferred-location effects, i.e. any local (including gravitational) experiment is independent of where and when it is performed.
    \item Absence of preferred-frame effects, i.e. any local (including gravitational) experiment is independent of motion of the (freely falling) local reference frame.
\end{itemize}
See \cite{wil14_LRR} for a detailed discussion. It is plausible that GR is the only viable metric theory of gravity that embodies SEP completely \citep{wil14_LRR,wil18_book}.\footnote{Nordstr{\"o}m's conformally-flat scalar theory also fulfills the SEP. However, this theory is already excluded by Solar System experiments, for instance by light deflection tests since it has $\gamma_\mathrm{PPN} = -1$. See \cite{Deruelle:2011} for details.}

In contrast to GR, in alternative theories featuring auxiliary gravitational fields ($\psi_a$), the structure of a self-gravitating body generally depends on its external gravitational environment. Therefore, the masses of the bodies that enter the action of a $N$-body system depend on the boundary values $\psi_a^{(0)}$. As a consequence of this, unlike in GR, the dynamics of an $N$-body system depends on the actual structure of the individual bodies, which leads to a violation of the SEP.\footnote{This statement is to be understood in an approximate sense. If there is rotation or tidal interaction then at some level the internal structure of a bodies also becomes relevant in GR.}
An important set of (body-dependent) parameters are the {\it sensitivities}, which describe the dependence of certain properties of a body on the boundary values $\psi_a^{(0)}$. For instance, the {\it sensitivities of the mass} are defined by
\begin{equation}
    s_A^{(a)} \equiv \left(\frac{\partial\ln m_A}{\partial\ln\psi_a^{(0)}}\right)_{N_b} \,,
\label{eq:sensitivity}
\end{equation}
whereby the number of baryons $N_b$ is kept constant. Sensitivities enter directly into the dynamics of a system of gravitationally interacting bodies and in this sense quantify the violation of the SEP. At the Newtonian level of the equations of motion, sensitivities enter the effective gravitational constant $G_{AB}$ for the interaction between two bodies, and therefore lead to a violation of the UFF. They further modify the (weak-field) PPN parameters of Tab.~\ref{tab:PPN_params} in a body-dependent way; for instance $\gamma_\mathrm{PPN}$ gets replaced by $\gamma_{AB}$. Furthermore, sensitivities also enter aspects of GW damping, which occurs beyond the first PN order. Sensitivities are theory dependent and depend on the structure of the body, hence its EoS. For weakly self-gravitating bodies the sensitivities are small. For instance, in mono-scalar-tensor theories $s_A$ is of the order of the fractional binding energy of body $A$ (see e.g.\ \citealt{wil18_book}), which is $\lesssim 10^{-6}$ for bodies of the Solar System. Depending on the details of the theory, the sensitivities for a strongly self-gravitating body such as a NS are generally of the order of 0.1, but can also be much larger than that. Therefore, precision pulsar experiments are ideal for searching for deviations from GR caused by the strong internal fields of NSs. Many more details on the above can be found, e.g., in \cite{wil93,wil18_book}.

In the following, we will present various theory agnostic and theory specific gravity tests based on radio-pulsar observations. Our theory agnostic discussions will mainly focus on the PPN framework and its extensions to strongly self-gravitating bodies via (body-dependent) effective PPN parameters. Table~\ref{tab:PPN_params_experimental} gives a comparison of experimental constraints on the parameters listed in Table~\ref{tab:PPN_params} from Solar System and pulsar experiments. However, there are other generic frameworks to discuss and compare gravity experiments, for instance the Standard-Model Extension (SME; \citealt{kos04}), where also pulsar experiments have provided interesting constraints on some of the parameters of this effective field theory \citep{Shao_2014,sb19,dws23}. Another popular framework, in particular in the context of GW observations, is the parametrised post-Einsteinian (ppE) framework. Pulsar experiments in the context of the ppE framework have been discussed, e.g., in \cite{yh10,ny20}.


\begin{sidewaystable}
\caption{Comparison of Solar System and binary pulsar tests for various parameters. Binary pulsars test strong-field ``effective'' PPN parameters. MESSENGER results were taken from \cite{gmg+18}, and some of the LLR results from \cite{bmt21}. See also the review by \cite{fm23}. For the sake of simplicity, we only give leading order values for all limits in this table.  \protect\\
 $^\dagger$ Limit on deviation from 1.
\label{tab:PPN_params_experimental}}
\begin{tabular} {| l | l l | l l |}
\hline
Parameter & Solar-system test & Limit & Pulsar test (eff.~PPN) & Limit \\
 & & & & \\
\hline\hline
$\gamma_\mathrm{PPN}$ & Cassini, Shapiro delay & $2\times 10^{-5}$ $^\dagger$ & Double Pulsar, Shapiro delay (§\ref{sec:0737_signal_prop}) & $7 \times 10^{-3}$ $^\dagger$ \\
$\beta_\mathrm{PPN}$ & Perihelion shift of Mercury; MESSENGER & $2\times 10^{-5}$  $^\dagger$ &---&---\\ 
$\xi$ &  Solar alignment with ecliptic  & $4\times10^{-7}$ & Two solitary pulsars (§\ref{sec:xi}) & $4 \times 10^{-9}$ \\
$\alpha_1$ & LLR & $2 \times 10^{-4}$ & J1909$-$3744 (§\ref{sec:alpha1}) & $2 \times 10^{-5}$ \\
$\alpha_2$ & Solar alignment with ecliptic  & $4\times10^{-7}$ & Two solitary pulsars (§\ref{sec:alpha2}) & $2 \times 10^{-9}$ \\
$\alpha_3$ & Perihelion shift of Earth and Mercury & $2\times10^{-7}$ & J1713$+$0747 (§\ref{sec:alpha3}) & $4\times 10^{-20}$ \\
$\zeta_1$ & Combined PPN limits & $2\times10^{-2}$ &---&---\\
$\zeta_2$ &---&---& Multiple binary pulsars (§\ref{sec:zeta2}) & $10^{-5}$ \\
$\zeta_3$ & Lunar acceleration & $10^{-8}$ &---&---\\
$\zeta_4$ & Not independent &---&---&---\\
$\eta$    & LLR  & $7 \times 10^{-5}$ & J0337+1715 (§\ref{sec:Nordtvedt}) & $2 \times 10^{-5}$ \\
$\kappa_\mathrm{D}$ &---&---& J1738$+$0333 (§\ref{sec:dipolar}) & $2 \times 10^{-4}$ \\
$\dot G/G$ [${\rm yr}^{-1}$] & LLR  & $ 10^{-14}$ & J1713$+$0747 (§\ref{sec:Gdot}) & $5 \times 10^{-13}$ \\
\hline
\end{tabular}
\end{sidewaystable}


\subsection{Strong-field Nordtvedt effect}
\label{sec:Nordtvedt}

GR is based on two basic postulates, namely a) the postulate of a \emph{universal coupling} between matter and gravity (where the Minkowski metric $\eta_{\mu\nu}$ in the laws of special relativity gets replaced by a curved spacetime metric $g_{\mu\nu}(x^\alpha)$), and b) \emph{Einstein's field equations} that define the dynamics of $g_{\mu\nu}(x^\alpha)$ (see e.g.\ \citealt{wil93,dam09,dam12}). It follows from a) that a small (sufficiently idealised) test body with negligible gravitational binding energy follows a geodesics in the curved spacetime $g_{\mu\nu}$, independent of its mass and composition (weak equivalence principle (WEP); see e.g. \citealt{eg04,dl08,sp10,dls14} and references therein). From a Newtonian point of view, this can be understood as an equivalence between inertial and (passive) gravitational mass (see e.g. \citealt{wil93,dls15}). Over the course of history, this UFF for test bodies has been verified with ever greater precision and has currently reached a precision of $\sim 10^{-15}$ \citep{MICROSCOPE}. As discussed above in the introduction to this section, for GR this UFF extends to self-gravitating bodies, including strongly self-gravitating bodies like NSs and BHs.\footnote{To prove the latter, as stated in \cite{wil18_nat}, one needs \emph{``full general relativity in all its complex glory''}.}

The postulate of universal coupling is not a unique feature of GR. As we have discussed above, it is a basic postulate for all metric theories of gravity. These theories of gravity differ from each other by the second postulate, i.e.~the field equations. It is these field equations that are responsible for a violation of the SEP by alternatives to GR in general and specifically for a violation of the UFF by the presence of gravitational binding energy (breakdown of the gravitational WEP (GWEP)). As mentioned above, to leading order in the equations of motion, instead of the Newtonian gravitational constant $G_\mathrm{N}$, we have an effective gravitational constant $G_{AB}$ that depends on the structure of the bodies. To give an example, for the class of Bergmann--Wagoner scalar-tensor theories of gravity this effective gravitational constant is given by 
\begin{equation}
    G_{AB} = G_\mathrm{N} \left.[1 - 2\zeta(s_A + s_B - 2s_A s_B) \right]  \quad (A \ne B) \,,
\label{eq:GAB}
\end{equation}
where $\zeta$ is a theory dependent constant (constrained to $\lesssim 10^{-5}$ in the Solar System \citep{bit03,wil14_LRR}) and $s_A$ is the sensitivity defined in Eq.~(\ref{eq:sensitivity}), which is body dependent and can assume very large values for NSs (see e.g. \citealt{wil18_book}).\footnote{See \cite{de92}, in particular their Eq.~(8.11), for a general expression for $G_{AB}$ in the context of the even wider class of multi-scalar-tensor theories.}
Equation (\ref{eq:GAB}) shows that interpreting a violation of the UFF in terms of body-specific `gravitational masses' is no longer meaningful in a strong-field context (see also the discussion in \citealt{dls15}). 

In case of a violation of the UFF, a binary system falling freely in the external gravitational field of a third body will experience a characteristic polarisation of the orbit (a gravitational analogue to the Stark effect in atoms exposed to an external electric field). The reason for this is the different (external) acceleration ($g$) of the two components, A and B, of the binary system in the direction of the third body C:
\begin{equation}
    \Delta_\mathrm{AB} 
    \equiv \frac{g_\mathrm{A} - g_\mathrm{B}}{\frac{1}{2}(g_\mathrm{A} + g_\mathrm{B})} 
    \simeq \frac{G_\mathrm{AC} - G_\mathrm{BC}}{G} \,.
\end{equation}
It was Kenneth Nordtvedt who first discovered such an effect in an alternative to GR, i.e.\ Jordan--Fierz--Brans--Dicke (JFBD) gravity, and suggested to test this with the help of Lunar Laser Ranging (LLR) \citep{nor68}. In the weak field of the Solar System one can approximate for the Earth(E)-Moon(M) system: $\Delta_\mathrm{EM} \simeq \eta_\mathrm{N}\left(\frac{E_\mathrm{E}^\mathrm{grav}}{m_\mathrm{E}c^2} - \frac{E_\mathrm{M}^\mathrm{grav}}{m_\mathrm{M}c^2}\right)$, where $\eta_\mathrm{N}$ is the \emph{Nordtvedt parameter}, a combination of various PPN parameters (see e.g.\ \citealt{wil18_book} for details). 

\begin{figure}[ht]
\begin{center}
  \includegraphics[width=6cm]{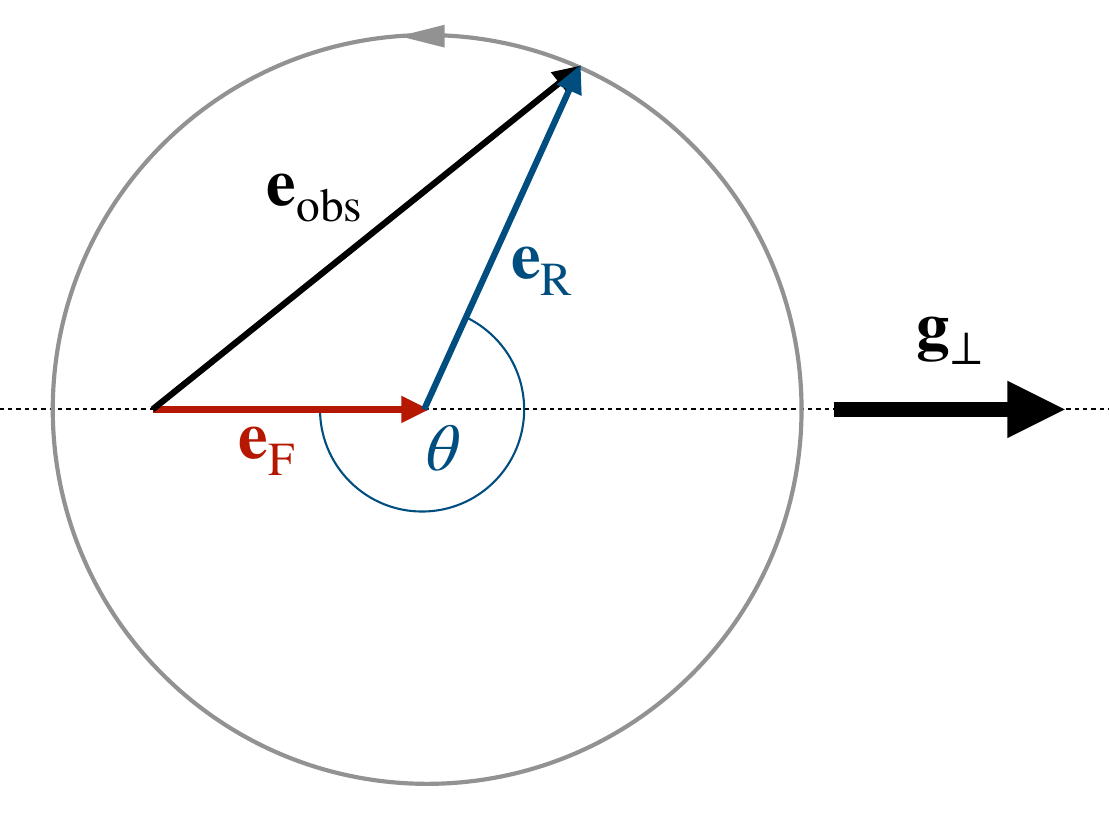}  
  \caption{``Polarisation'' of a nearly circular binary orbit under the influence of a forcing vector $\mathbf{g}$, showing the relation between the forced eccentricity, $\mathbf{e}_\mathrm{F}$, the eccentricity evolving under the general-relativistic advance of periastron, $\mathbf{e}_\mathrm{R}(t)$, and the angle $\theta = \theta_0 + \dot\omega \, t$. The actually observed eccentricity is given by $\mathbf{e}_\mathrm{obs}(t) = {\bf e}_\mathrm{F} + {\bf e}_\mathrm{R}(t)$. After \cite{wex97}.}
  \label{fig:eccrotate}
\end{center}
\end{figure}

In the meantime LLR has put tight constraints on the Nordtvedt parameter: $\lvert\eta_\mathrm{N}\rvert \lesssim 7 \times 10^{-5}$ \citep{bmt21}. However, \cite{ds91} pointed out that such a test in the weak-field regime of the Solar System is unable to constrain any higher order/strong-field contributions to the \emph{Nordtvedt effect} that might become significant in the presence of strongly self-gravitating masses. Furthermore, they proposed a test of such a strong-field Nordtvedt effect with the help of pulsar-WD systems falling in the gravitational field of the Milky Way (\emph{Damour--Sch\"afer test}). If the pulsar were to experience a different acceleration in the external gravitational field than the WD, the eccentricity of the binary pulsar system would change in a characteristic way over time (see Fig.~\ref{fig:eccrotate}). By combining several suitable pulsar-WD systems, this kind of test has lead to a limit of $\lvert\Delta_\mathrm{p0}\rvert \lesssim 5 \times 10^{-3}$ (95\% C.L., \citealt{sfl+05,gsf+11}), where the index 0 indicates that this is with respect to the (well tested) acceleration of a weakly self-gravitating mass, i.e. the WD. The best such limit from a pulsar-WD system comes from PSR~J1713$+$0747 (cf.\ Sect.~\ref{sec:other-PSR-WD}): $\lvert\Delta_\mathrm{p0}\rvert \lesssim 2 \times 10^{-3}$ (95\% C.L., \citealt{zdw+19}).


\begin{figure}[ht]
\begin{center}
  \includegraphics[width=10cm]{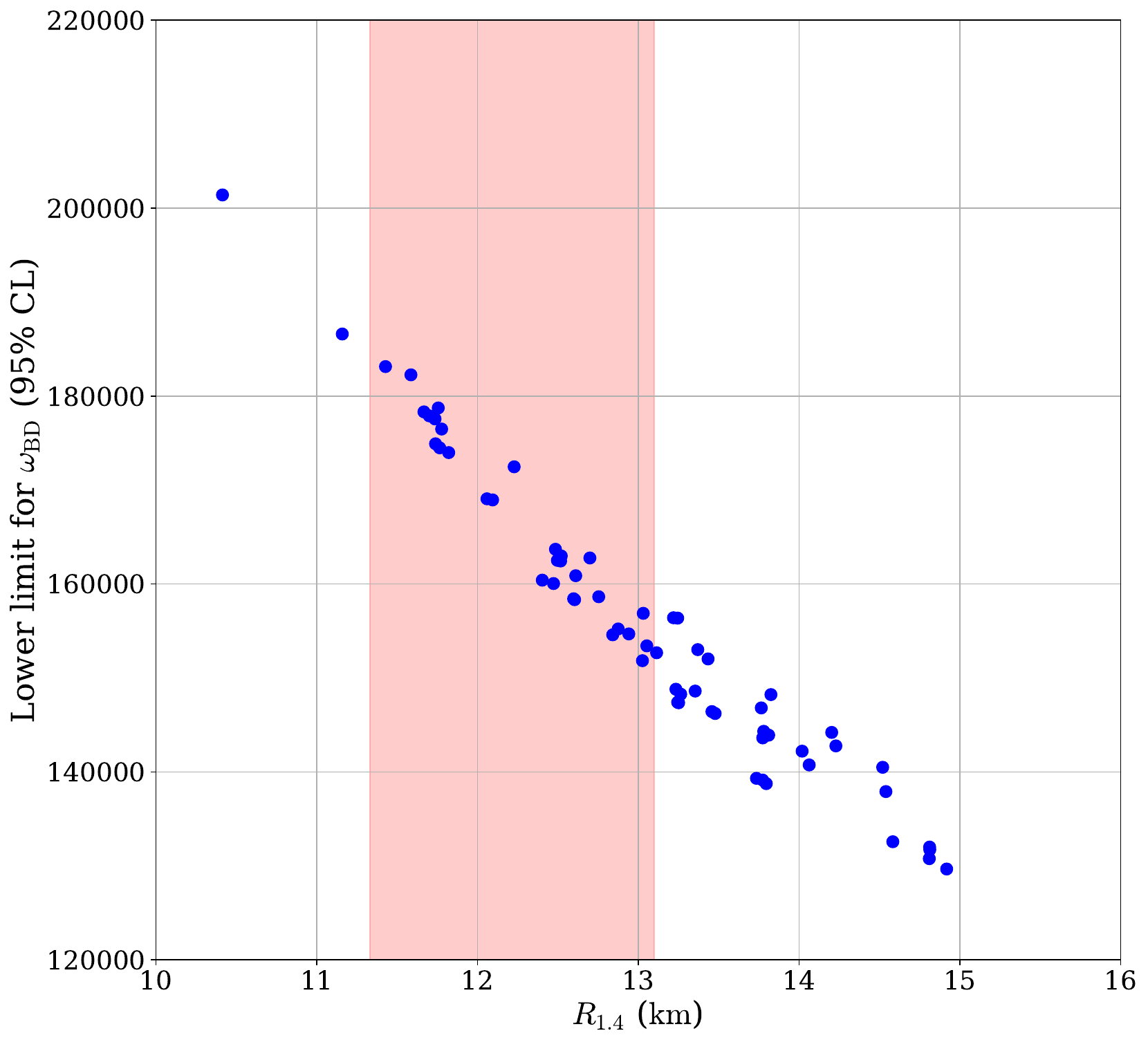}  
  \caption{95\% confidence lower limit on the Brans--Dicke parameter $\omega_\mathrm{BD}$ from the pulsar in a stellar triple system, for 64 different NS EoSs. The $x$-axis shows the radius of a $1.4\,\Msun$ NS predicted by the corresponding EoS. Obviously, stiffer EoSs give weaker limits (GR is obtained for $\omega_\mathrm{BD} \rightarrow \infty$). The red band gives the 95\% credible limit for $R_{1.4}$ given as the more conservative range in \cite{krp+24}, hence we have the conservative lower limit of $\omega_\mathrm{BD} \gtrsim 150,000$ (see \citealt{vcf+20} for more details). The 64 EoSs were taken from \cite{rlof09,kl19}, assuming the criterion of a maximum mass of more than $1.92\,\Msun$ (99\% confidence lower limit of PSR~J0740$+$6620 \citep{fcp+21}).}
  \label{fig:JFBDlimit}
\end{center}
\end{figure}

Compared to the Earth-Moon system in the Sun's gravitational field, pulsar-WD systems as described above have a decisive disadvantage in testing the Nordtvedt effect: their external acceleration in the Galactic gravitational field is about seven orders of magnitude smaller. In this sense, the discovery of PSR~J0337$+$1715 as a member of a hierarchical triple system with two WD companions (cf.\ Sect.~\ref{sec:other-PSR-WD}) was a decisive turning point. The inner pulsar-WD binary accelerates in the gravitational field of the external WD with about $0.17\,\mathrm{cm\,s^{-2}}$ which is comparable to the external acceleration of the Earth-Moon system ($\sim 0.6\,\mathrm{cm\,s^{-2}}$). Consequently, after about six years of timing observations of PSR~J0337$+$1715 the above limits from pulsar-WD binaries could be improved by three orders of magnitude: $\lvert\Delta_\mathrm{p0}\rvert < 2.6 \times 10^{-6}$ (95\% C.L.) \citep{agh+18}. In an independent approach using a different set of timing data, \cite{vcf+20} were able to slightly improve that limit to $\Delta_\mathrm{p0} = (0.5 \pm 1.8) \times 10^{-6}$ (95\% C.L.). Due to the large fractional binding energy of a NS, this limit implies the tightest limit for some alternative theories of gravity, including JFBD gravity, which originally motivated Kenneth Nordtvedt to suggest this type of test with LLR (see Sect.~\ref{sec:altGrav} for more details). Just to mention it already here, for JFBD gravity, PSR~J0337$+$1715 gives a conservative 95\% confidence lower limit for the Brans--Dicke parameter of $\omega_\mathrm{BD} \gtrsim 150,000$ (GR is obtained for $\omega_\mathrm{BD} \rightarrow \infty$), which is considerably stronger than the Solar System limit of 40,000  obtained from Cassini \citep{bit03,wil18_book} (see Fig.~\ref{fig:JFBDlimit}).


Finally, it should be noted that the triple system test has not made the tests with pulsar WD systems that fall in the gravitational field of the Milky Way completely obsolete. There are still aspects of the UFF that cannot be tested with PSR~J0337$+$1715. For instance, the limit from PSR~J1713$+$0747 can be interpreted as a test of the UFF towards dark matter in our Galaxy, which is of particular interest due to the pulsar's neutron-rich composition and the significant amount of gravitational binding energy of the pulsar (see e.g. \citealt{swk18}).


\subsection{Dipolar gravitational radiation}
\label{sec:dipolar}

GR predicts that the lowest multipole for the generation of GWs is the mass quadrupole. The absence of any time-varying mass monopole and mass dipole is closely connected to the fulfillment of the SEP and the corresponding effacement of the internal structure of bodies \citep{dam87,wil18_book}. Alternative theories of gravity, which violate the SEP are generally expected to predict gravitational radiation at lower multipole moments, linked to additional gravitational fields (e.g.\ a scalar field). The corresponding (specific) gravitational ``charge'' of a body depends on the body's internal structure. In a binary pulsar system with $M_\mathrm{p} \ne M_\mathrm{c}$ this leads, most notably, to a time-varying gravitational dipole moment \citep{wil93,gw02}.\footnote{In principle also a difference in rotation can lead to a gravitational dipole moment (see e.g.\ \citealt{pb14}), which however is practically irrelevant for binary pulsar experiments.} 
As a consequence, there is an additional damping of the orbital motion by dipolar GWs, leading to a contribution in the change of the orbital period, which in terms of the sensitivities can be written as \citep{wil93}
\begin{equation}
  \dot{P}_\mathrm{b}^{\rm Dipole} = -2\pi\left(\frac{P_\mathrm{b}}{2\pi}\right)^{-1} 
    (\Tsun M) X_\mathrm{p} X_\mathrm{c} \,
    \frac{1 + e^2/2}{(1 - e^2)^{5/2}} \, 
    \kappa_\mathrm{D}(s_\mathrm{p}-s_\mathrm{c})^2 + \mathcal{O}(c^{-5},s^3)\,,
\label{eq:pbdotdipole}
\end{equation}
where $\kappa_\mathrm{D}$ is a theory-dependent constant of the theory. The constant $G$ is Newton's gravitational constant as obtained in a Cavendish-type experiment. While in GR the damping of the binary orbit enters the equations of motion at the 2.5 PN order ($\mathcal{O}(c^{-5})$), the dipolar radiation---as can be seen from Eq.~(\ref{eq:pbdotdipole}) (see e.g. also \citealt{mw13})---already enters at the 1.5 PN order, i.e. $\mathcal{O}(c^{-3})$. For $\kappa_\mathrm{D}(s_\mathrm{p} - s_\mathrm{c})^2 \sim 1$ that means a change in the orbital period which is about six orders of magnitude larger than the GR prediction for binary pulsars. Consequently, binary pulsar experiments are generally very sensitive to any presence of dipolar radiation, in particular if there is a significant asymmetry in the sensitivities between pulsar and companion.

If there is a gravitational dipole moment, one would generally expect it to be particularly large for pulsars with a WD companion, where, in comparison to the pulsar, the WD can be considered as a body with weak self-gravity having $s_A \lesssim 10^{-3}$. For that reason, pulsar-WD systems are particularly interesting for constraining alternatives to GR (see Sect.~\ref{sec:altGrav} for a detailed discussion). From those pulsar-WD systems that allow a (mostly) theory independent determination of the masses, one can even derive quite generic limits on dipolar radiation. One such pulsar is PSR~J1738$+$0333 which has a bright WD as companion that shows very prominent Balmer lines (see Sect.~\ref{sec:psrJ1738} for more details on that system). High resolution spectroscopy gives access to the mass ratio $R$ and the mass of the WD ($M_\mathrm{c}$) \citep{avk+12}. The agreement of the (intrinsic) $\dot{P}_\mathrm{b}$ of the PSR~J1738$+$0333 system with GR can then directly be converted into the limit $\lvert\kappa_\mathrm{D}^{1/2}s_\mathrm{p}\rvert \lesssim 2 \times 10^{-3}$ (95\% confidence), for a NS of about $1.47\,\Msun$ \citep{fwe+12}.

Finally, theories that violate the SEP and predict the existence of dipolar radiation generally also are expected to predict a temporal variation of the gravitational constant $G$. A change in $G$, however, also leads to a change in the orbital period. To disentangle this effect from dipolar radiation, one needs to combine (at least two) different binary pulsars with different orbital periods (see Sect.~\ref{sec:Gdot} for more details). 


\subsection{Preferred-location effects}
\label{sec:PLE}

The SEP states that the outcome of any local experiment, including gravitational experiments with self-gravitating bodies, is independent of where and when in the universe it is performed. A violation of this local position invariance leads to a location and/or time dependence of gravitational phenomena, for instance a gravitational constant that evolves over time, with the expansion of the universe, or depends on the spatial location in a gravitating system. Scalar-tensor theories of gravity generally show such kind of preferred-location effects, for instance a change in Newton's gravitational constant (as measured in a Cavendish experiment) due to the cosmological evolution of the (background) scalar field.


\subsubsection{Time variation of Newton's gravitational constant}
\label{sec:Gdot}

That Newton's gravitational constant is not in fact a constant of nature, but decreases in value over time, was already proposed by Paul Dirac as part of his \emph{large numbers hypothesis} \citep{dir37}. Efforts to put Dirac's heuristic reasoning on a field-theoretical footing, together with the motivation to implement (certain aspects of) Mach's principle, eventually led to the development of scalar-tensor theories, with the JFBD theory of gravity (often just called Brans--Dicke gravity) as their best-known representative \citep{jor55,fie56,bd61,bra05}. 

Quite generally, theories that violate the SEP by allowing for preferred location effects are expected to permit Newton's constant, $G$, to vary over time while the universe expands. Purely heuristically, temporal variations in $G$ are expected to occur on the timescale of the age of the Universe, such that $\dot G/G \sim H_0 \sim 0.7\times 10^{-10}$\,yr$^{-1}$, where $H_0$ is the Hubble constant. Three different pulsar-derived tests can be applied to these predictions, as a SEP-violating time-variable $G$ would be expected to alter the properties of NSs and WDs, and to affect binary orbits.

The effects on the orbital period of a binary system of a varying $G$ were first considered in \cite{dgt88}, who expected:
\begin{equation}
  \left(\frac{\dot P_\mathrm{b}}{P_\mathrm{b}}\right)_{\dot G} = -2\,\frac{\dot G}{G} \,.
\label{eq:PbdotfromGdot1}
\end{equation}
Applying this equation to the limit on the deviation from GR of the $\dot P_\mathrm{b}$ for PSR~1913+16, they found a value of $\dot G/G = (1.0 \pm 2.3)\times 10^{-11}$\,yr$^{-1}$. With the latest results from the Double Pulsar (see in particular Sect.~\ref{sec:0737_GW_damping}) one obtains $\dot G/G = (-0.8 \pm 1.4)\times 10^{-13}$\,yr$^{-1}$. Applying Eq.~(\ref{eq:PbdotfromGdot1}) to the binary pulsar J1713$+$0747, which has a WD companion, gives $\dot G/G = (-0.8 \pm 2.4)\times 10^{-13}$\,yr$^{-1}$. While such an approach to obtain limits for $\dot G$ from binary pulsars provides a first order estimation, the actual numbers have to be taken with a grain of salt for several reasons. 

\citet{nor90} pointed out that for a strongly self-gravitating body a change in $G$ leads to a significant change in the mass of the body, which in turn adds significant changes to the orbital period of a binary pulsar. Taking this effect into account leads to a corrected expression for Eq.~(\ref{eq:PbdotfromGdot1})
\begin{equation}
  \left(\frac{\dot P_\mathrm{b}}{P_\mathrm{b}}\right)_{\dot G} = 
  -\left[ 2 - (X_\mathrm{p}c_\mathrm{p} + X_\mathrm{c}c_\mathrm{c}) 
           - \frac{3}{2} (X_\mathrm{p}c_\mathrm{c} + X_\mathrm{c}c_\mathrm{p}) \right]
  \frac{\dot G}{G} \, .
\label{eq:PbdotfromGdot2}
\end{equation}
The compactness $c_A \equiv -2\partial\ln m_A/\partial\ln G \approx -2 E_A^\mathrm{grav}/(m_Ac^2)$ is a measure for the change of a body's mass due to a change of the (local) gravitational constant.\footnote{Note that there is a difference of a factor of $-2$ in Nordtvedt's definition of $c_A$.} 
It is closely related to the body's sensitivity $s_A$ and one often finds $c_A \approx 2s_A$ (see e.g.\ \citealt{de92} for multi-scalar-tensor theories). The corrections by the compactnesses of pulsar and companion in Eq.~(\ref{eq:PbdotfromGdot2}) generally weaken the limits which are obtained directly from  Eq.~(\ref{eq:PbdotfromGdot1}). In fact, as already pointed out in \cite{nor90}, depending on the masses, the EoS of NS matter, and the underlying theory gravity, for a pulsar with a NS companion the expression in box brackets in Eq.~(\ref{eq:PbdotfromGdot2}) may easily be small compared to the factor 2 in Eq.~(\ref{eq:PbdotfromGdot1}). For that reason, pulsars with WD companions provide a somewhat more reliable generic test of $\dot{G}$, since there $c_\mathrm{c} \simeq 0$ and therefore
\begin{equation}
  \left(\frac{\dot P_\mathrm{b}}{P_\mathrm{b}}\right)_{\dot G} = 
  -\left[ 2 -\left(1 + \frac{1}{2} X_\mathrm{c}\right) c_\mathrm{p} \right]
  \frac{\dot G}{G} \, .
\label{eq:PbdotfromGdot2wd}
\end{equation}
For $c_\mathrm{p}$ typically of order 0.2, limits are still comparable to what one obtains from Eq.~(\ref{eq:PbdotfromGdot1}).

The approach discussed so far comes with the assumption that a limit on the deviation of the (system intrinsic) $\dot{P}_\mathrm{b}$ from its GR expectation can directly be converted into a limit for $\dot{P}_\mathrm{b}^{\dot{G}}$. This assumption is unjustified insofar as a theory that predicts a $\dot{G}$ is quite generally also predicting a $\dot{P}_\mathrm{b}$ contribution from dipolar radiation (see Sect.~\ref{sec:dipolar}) that cannot be separated from $\dot{P}_\mathrm{b}^{\dot{G}}$ in a single system.\footnote{Interestingly, in the highly dynamical strong-field regime of merging NSs this degeneracy is broken \citep{wza+22}.}
\cite{lwj+09} have therefore suggested to do a combined test with two suitable pulsar-WD systems that have a sufficiently large difference in their orbital periods, in order to break that covariance. Combining the long orbital period binary pulsar J1713$+$0747 with different short orbital period pulsar-WD systems, eventually led to \citep{lwj+09,zdw+19,ddf+20}
\begin{equation}
    \lvert\dot{G}/G\rvert \lesssim 10^{-12} \, {\rm yr}^{-1} = 0.014\,H_0 
    \quad \mbox{(95\% C.L.)} \,.
\label{eq:GdotLimPsr}
\end{equation}
It must be kept in mind that the exact value for that limit depends on the values of $c_\mathrm{p}$ (respectively $s_\mathrm{p} \approx c_\mathrm{p}/2$) for the different NS masses that enter into Eqs.~(\ref{eq:pbdotdipole}) and (\ref{eq:PbdotfromGdot2wd}), for which we can only give approximate numbers in generic tests, provided certain additional assumptions hold (see e.g. \citealt{fwe+12} for a discussion). 

When compared to limits from Solar System experiments (e.g. \citealt{gmg+18,hm18,bmt21}), the pulsar limit (\ref{eq:GdotLimPsr}) is more than an order of magnitude weaker. However, binary pulsar timing tests $\dot{G}$ in a very different gravity regime, and is therefore complementary to tests in weak-field environments like the Solar System. Just to give two examples. First, in theories like Barker's constant-$G$ theory \citep{bar78} there would still be a change to the NS mass if the background scalar field $\varphi$ changes with the expansion of the universe, and hence a corresponding change in the orbital period of a binary pulsar ($\dot{P}_\mathrm{b} \propto \dot\varphi$). Secondly, in above equations we have ignored that in the interaction between pulsar and companion there is actually an effective gravitational constant which has a body-dependent contribution. \cite{nor93} has written this effective gravitational constant as $G^{\rm eff}(t) = G(t) K_\mathrm{pc}(t)$, where $K_\mathrm{pc}(t)$ depends on the structure of the two bodies in the binary system ($K_{12}(t) \simeq 1$ for two weakly self-gravitating masses $m_1$ and $m_2$). As shown in \cite{wex14}, depending on the details of the gravity theory, i.e.\ the details of $K_\mathrm{pc}(t)$, the strong-field of a NS can lead to a very significant amplification of $\dot{G}_\mathrm{eff}$ so that $\lvert\dot{G}_\mathrm{eff}\rvert \gg \lvert\dot{G}\rvert$.


\subsubsection{Spatial variation of Newton's gravitational constant}
\label{sec:Gloc}

In addition to a temporal change in the locally measured gravitational constant $G_\mathrm{loc}$, beyond GR there can generally also be a spatially varying gravitational constant, which depends on the position relative to (external) masses. Even for theories that in the weak field only deviate in $\gamma_\mathrm{PPN}$ and/or $\beta_\mathrm{PPN}$, $G_\mathrm{loc}$ is generally expected to depend on the distance $r$ to the external mass $m$:
\begin{equation}\label{eq:GlocPPN}
    \frac{G_\mathrm{loc}(r)}{G_0} \simeq 
    1 - (4\beta_\mathrm{PPN} - \gamma_\mathrm{PPN} -3) \frac{G_0 m}{c^2 r} =
    1 - \eta_\mathrm{N} \,\frac{G_0 m}{c^2 r}
\end{equation}
where $G_0$ corresponds to the local gravitational constant at spatial infinity (see e.g.\ \citealt{wil18_book} for an expression that contains additional PPN parameters). For a strongly self-gravitating body, $\eta_\mathrm{N}$ in Eq.~(\ref{eq:GlocPPN}) gets replaced by a body dependent parameter, which in the literature is often denoted by $\eta^\ast$.  

As a consequence of a distance-dependent $G_\mathrm{loc}$, a pulsar moving in an eccentric orbit around a strongly self-gravitating companion will experience a periodic variation of $I_\mathrm{p}$ according to
\begin{equation}
    \frac{\Delta I_\mathrm{p}}{I_\mathrm{p}} \simeq 
    -\kappa_\mathrm{p} \, \frac{\Delta G_\mathrm{loc}(r)}{G_0} \simeq
     \kappa_\mathrm{p} \, \eta_\mathrm{c}^\ast \frac{G_0 m_\mathrm{c}}{c^2 r}
\end{equation}
which in turn leads to a variation in the pulsar's intrinsic rotation according to $\Delta\nu/\nu_0 = -\Delta I_\mathrm{p}/I_\mathrm{p}$ \citep{ear75,wil18_book}, due to the conservation of angular momentum. The (body-dependent) quantity $\kappa_\mathrm{p} \equiv (\partial\ln I_\mathrm{p}/\partial\ln G)_{N_b}$ is the \emph{sensitivity of the MoI} of the pulsar, describing how the MoI of a NS (with a specific mass) changes in response to a variation in the local gravitational constant. The variation in the (observed) spin-frequency $\nu$ caused by such an effect would become apparent as a modification of the PK parameter $\gamma$ of the Einstein delay (cf.\ Eq.~\ref{eq:deltaE} and e.g. \citealt{wil18_book}). Consequently, radio pulsars in eccentric DNS systems are ideal to test for a change of the local gravitational constant in the vicinity of a strongly self-gravitating body. The wealth of PK parameters measured in the Double Pulsar (see Sect.~\ref{sec:psrJ0737}) even allows a generic $\sim 10^{-3}$ constraint on such a strong-field violation of the SEP \citep{kw09,ksm+21}. Apart from that, this effect also plays an important role when constraining specific alternatives to GR, like scalar-tensor or TeVeS-like theories (see e.g.\ \citealt{de96_PSR,ksm+21}).

In the following we discuss constraints on preferred location effects related to a non-vanishing Whitehead parameter $\xi$. Such a deviation from GR is closely connected to a spatial anisotropy of the local gravitational constant.


\subsubsection{Limits on \texorpdfstring{$\hat\xi$}{xi}}
\label{sec:xi}

A non-vanishing Whitehead parameter $\xi$ leads to an anisotropy in the gravitational interaction of localised systems, induced by the mass distribution of our Galaxy. Such an anisotropy would lead to characteristic (non-GR) signatures in the dynamics of self-gravitating systems. While in general the best pulsar constraints on deviations from GR are obtained from binary systems or the pulsar J0337$+$1715 as part of a hierarchical triple system (Sect.~\ref{sec:Nordtvedt}), by far the best limits for (the strong-field generalisation of) $\xi$ come from fast spinning solitary MSPs. A violation of local position invariance related to a non-vanishing $\xi$ would lead to the precession of a spinning self-gravitating body around the direction to the Galactic centre (as the CM of the Milky Way) \citep{nor87}. The rate of precession is given by \citep{nor87,sw13}
\begin{equation}
  \Omega_\xi^{\rm prec} = \xi \,
  \frac{2\pi}{P} \left(\frac{v_\mathrm{G}}{c}\right)^2\cos\vartheta_\mathrm{G} \,,
\label{eq:OmegaPrec_xi}
\end{equation}
where $P$ denotes the rotational period of the body, $v_\mathrm{G}$ the rotational velocity of the Galaxy at the location of the pulsar (as a measure for the local Galactic potential $U_\mathrm{G}$), and $\vartheta_\mathrm{G}$ the angle between the direction to the Galactic centre and the spin of the body.\footnote{Note, the $\xi$ used in \cite{nor87} differs by a factor $-\frac{1}{2}$ from the $\xi$ of the ``standard'' PPN formalism.} 
\citet{nor87} used the alignment of the Sun with the planetary orbits to set a constraint on $\xi$ of the order of a few times $10^{-7}$. In the same publication, Nordtvedt already hinted at the potential of using fast rotating solitary pulsars to constrain $\xi$. Following this, \cite{sw13} have used 15 years of continuous observations of the two solitary MSPs B1937$+$21 ($P = 1.56$\,ms) and J1744$-$1134 ($P = 4.07$\,ms) with the 100-m Effelsberg radio telescope to infer a limit of 
\begin{equation}
    \lvert\hat\xi\rvert < 3.9 \times 10^{-9} \quad \mbox{(95\% confidence)}\,.
\label{eq:xilimit}
\end{equation}
where the hat indicates that pulsars are testing a strong-field generalisation of the Whitehead parameter $\xi$. Both of the pulsars are near the Galactic plane. The decisive factor for obtaining the limit (\ref{eq:xilimit}) was the stability of the pulse profiles over such a long period of time, which gave no indication of any precession of the spins of these two pulsars.

A non-vanishing $\xi$ also leads to a precession of the orbital angular momentum of a binary pulsar (in Eq.~(\ref{eq:OmegaPrec_alpha2}) the rotational period $P$ gets replaced by the orbital period $P_\mathrm{b}$). \cite{swk15} have used the binary pulsars J1012$+$5307 and J1738$+$0333 (both with a WD companion) to derive a (orbital dynamics related) limit of $\lvert\hat\xi\rvert < 3.1\times 10^{-4}$ (95\% confidence) which, however, is five orders of magnitude weaker than that from solitary pulsars.

The above limit on $\hat\xi$ can straightforwardly be converted into a limit on the anisotropy of the gravitational constant. If the local position invariance is violated, for a self-gravitating system falling freely in the gravitational potential of the Galaxy $U_\mathrm{G}$ one could have a directional dependence in the local gravitational constant. For a system with mass $m$, radius $R$ and MoI $I$ one then finds a variation in the gravitational constant $G$ of \citep{wil93}
\begin{equation}
    \frac{\Delta G}{G} \simeq \xi \left(1 - \frac{3I}{mR^2}\right) U_\mathrm{G}
     \cos^2\vartheta_{{\rm loc}} \,, 
\end{equation}
where $\vartheta_{{\rm loc}}$ denotes the angle between the direction to the Galactic centre and the direction to the location where $G$ is being measured, as seen from CM of the self-gravitating system. Such an anisotropy is generally expected to cause a precession of a rotating (self-gravitating) body like a pulsar. In case of a NS, one typically has $I/mR^2 \sim 0.4$. Using this number, from their limit on $\xi$ (Eq.~\ref{eq:xilimit}) \cite{sw13} then obtained
\begin{equation}
    \left\lvert \frac{\Delta G}{G} \right\rvert < 4 \times 10^{-16} \quad \mbox{(95\% confidence)}\,,
\end{equation}
which is the so far tightest limit on an anisotropy of $G$.


\subsection{Preferred-frame effects and violation of the conservation of momentum}
\label{sec:PFE}

So far, we have dealt with deviations from GR that can already occur in fully conservative theories of gravity. In the following we will summarise pulsar tests for PPN parameters that are linked to the violation of the Lorentz invariance of the gravitational interaction (preferred frame effects), leading to a violation of the conservation of angular momentum, as well as tests of parameters that result in a violation of the conservation of total momentum.


\subsubsection{Limits on \texorpdfstring{$\hat\alpha_1$}{alpha1}}
\label{sec:alpha1}

A non-vanishing $\hat {\alpha}_1$ implies that the (uniform) motion of a binary pulsar system with respect to a ``universal'' reference frame (defined by the rest frame of the Cosmic Microwave Background (CMB)) will affect its orbital evolution. Similar to the violation of the UFF (see Sect.~\ref{sec:Nordtvedt}), the time evolution of the observed eccentricity will depend on both a vector $\mathbf{e}_\mathrm{R}$ of constant length that rotates in the orbital plane with angular velocity $\dot \omega$ and a fixed vector $\mathbf{e}_\mathrm{F}$ as a result of the $\hat\alpha_1$-induced polarisation of the orbit. The ``forced eccentricity'' $\mathbf{e}_\mathrm{F}$ lies in the orbital plane and is perpendicular to $\mathbf{w}$, the velocity of the binary system with respect to the preferred frame. The magnitude of $\mathbf{e}_F$ (for $e \ll 1$) is written as \citep{de92_LLI,bcd96}:
\begin{equation}
  \lvert{\bf e}_\mathrm{F}\rvert \simeq \frac{\lvert\hat{\alpha}_1\rvert}{12} \,
  \left(\frac{P_\mathrm{b}}{2\pi}\right) \,
  \lvert X_\mathrm{p} - X_\mathrm{c}\rvert \,
  \frac{w_{\perp}}{a},
\label{eq:efalpha1}
\end{equation}
where $w_{\perp}$ denotes the length of the projection of the system's velocity $\mathbf{w}$ onto the orbital plane, and $a$ the semimajor axis of the relative motion. In the above equation we have neglected a factor $(\frac{2}{3}\hat\gamma - \frac{1}{3}\hat\beta + \frac{2}{3} + \frac{1}{3}\hat\alpha_1 X_\mathrm{p} X_\mathrm{c})^{-1}$, which is justified by the fact that the effective (strong-field) Eddington parameters $\hat\gamma$ and $\hat\beta$ are already sufficiently constrained (to their GR value, i.e.\ 1) by other pulsar experiments, and the $\hat{\alpha}_1$ term is small compared to one.

There are various binary pulsars with very small eccentricities ($e \lesssim 10^{-6}$). However, in principle a large $\mathbf{e}_\mathrm{F}$ could be hidden by an equally large $\mathbf{e}_\mathrm{R}$, since the observed eccentricity is the vector sum of the two. On the other hand, such a fortunate cancellation of $\mathbf{e}_\mathrm{F}$ and $\mathbf{e}_\mathrm{R}$ will eventually break down, as $\mathbf{e}_\mathrm{R}$ rotates with respect to $\mathbf{e}_\mathrm{F}$ at a rate of $\dot\omega$ (see Fig.~\ref{fig:eccrotate}). For that reason, small-eccentricity binary pulsar systems with a short orbital period, a large difference in the two masses, and a long observing time span should be the ideal systems for such a test.

Currently there are two systems that fit best the above criteria: PSRs J1738$+$0333 and J1909$-$3744. More details on these binary pulsars are given in Sects.~\ref{sec:psrJ1738} and \ref{sec:psrJ1909} respectively. For both systems, high resolution spectroscopy observations of their WD companions gave access to their systemic radial velocities, consequently---when combined with the proper motion from timing---allowing the determination of their 3D-velocity with respect to the Solar System and, furthermore, the determination of $\mathbf{w}$. For PSR~J1738$+$0333, 10 years of observation with the 305-m William E.\ Gordon Arecibo radio telescope lead to constraints of \citep{sw12}
\begin{equation}
    \hat\alpha_1 = -0.4^{+3.7}_{-3.1} \times 10^{-5} \quad \mbox{(95\% confidence)}\,.
\label{eq:alpha1limit1738}
\end{equation}
Using 15 years of observations of PSR~J1909$-$3744 with the Nan\c{c}ay Radio Telescope, \cite{lgi+20} were able to further improve this limit to
\begin{equation}
    \lvert\hat\alpha_1\rvert < 2.1 \times 10^{-5} \quad \mbox{(95\% confidence)}\,.
\label{eq:alpha1limit1909}
\end{equation}
The above limits are both better in magnitude than the weak-field results from LLR, and additionally also incorporate strong field effects related to the strong spacetime curvature inside and near the pulsars. 


\subsubsection{Limits on \texorpdfstring{$\hat\alpha_2$}{alpha2}}
\label{sec:alpha2}

As with the pulsar test of $\hat\xi$ in Sect.~\ref{sec:xi}, by far the best constraints on (the strong-field generalisation of) $\alpha_2$ also come from solitary MSPs. \cite{nor87} has shown that a non-zero $\alpha_2$ causes the spin of a self-gravitating body moving with velocity $\mathbf{w}$ (relative to a preferred frame) to precess about the direction of $\mathbf{w}$, with an angular velocity
\begin{equation}
  \Omega_{\alpha_2}^{\rm prec} = -\alpha_2 \,
  \frac{\pi}{P} \left(\frac{\lvert{\bf w}\rvert}{c}\right)^2\cos\vartheta_w \,,
\label{eq:OmegaPrec_alpha2}
\end{equation}
where $P$ denotes the rotational period of the body and $\vartheta_w$ the angle between $\mathbf{w}$ and the spin of the body.\footnote{Note, the $\alpha_2$ used in \cite{nor87} differs by a factor $\frac{1}{2}$ from the $\alpha_2$ of the ``standard'' PPN formalism.} While \cite{nor87} used the alignment of the Sun with the planetary orbits to set the tightest constraint on $\alpha_2$, he already identified pulsars as possible probes for an $\alpha_2$-related violation of Lorentz invariance. In a rough estimate he derived an early pulsar limit of order few times $10^{-6}$ from the (then) recently discovered MSP B1937+21. \cite{sck+13} has used 15 years of continuous observations of the two MSPs B1937$+$21 ($P = 1.56$\,ms) and J1744$-$1134 ($P = 4.07$\,ms) with the 100-m Effelsberg radio telescope to infer a limit of 
\begin{equation}
    \lvert\hat\alpha_2\rvert < 1.6\times 10^{-9}  \quad \mbox{(95\% confidence)}\,.
\label{eq:alpha2limit}
\end{equation}
Like for the $\xi$-test in Sect.~\ref{sec:xi}, the decisive factor for this analysis was again the stability of the pulse profiles over such a long period of time, which gave no indication of any precession of the spins of these two pulsars.

A non-vanishing $\alpha_2$ also leads to a precession of the orbital angular momentum of a binary pulsar (For a (nearly) circular orbit, the rotational period $P$ in Eq.~(\ref{eq:OmegaPrec_alpha2}) gets replaced by the orbital period $P_\mathrm{b}$). In \cite{sw12} the binary pulsars J1012$+$5307 and J1738$+$0333 were used to derive a (orbital dynamics related) limit of $\lvert\hat\alpha_2\rvert < 1.8\times 10^{-4}$ (95\% confidence) which, however, is five orders of magnitude weaker than that from solitary pulsars.


\subsubsection{Limits on \texorpdfstring{$\hat\alpha_3$}{alpha3}}
\label{sec:alpha3}

A non-vanishing $\hat\alpha_3$ implies both a violation of local Lorentz invariance and non-conservation of momentum in the gravitational sector. More specifically, it results in the self-acceleration of a rotating, gravitationally bound body that moves with respect to a preferred frame of reference. Within the first PN weak-field slow-motion approximation of the PPN formalism one finds \citep{nw72,wil93}:
\begin{equation}
  {\bf a}_A^{\rm self} = -\frac{\alpha_3}{3} \, 
    \frac{E_A^{\rm grav}}{m_A\,c^2} \, {\bf w} \times {\bf \Omega}.
\label{eq:alpha3accel}
\end{equation}
where $\mathbf{\Omega}$ is the rotational velocity vector of the body. As one can see from above equation, the self-acceleration is perpendicular to the body's spin and its velocity with respect to the preferred frame, i.e. $\mathbf{w}$. For a strongly self-gravitating body, one needs to replace the fractional gravitational binding energy by the sensitivity (i.e.\ $E_A^{\rm grav}/(m_A c^2) \rightarrow -s_A$) and the PPN parameter by its strong-field equivalent (i.e.\ $\alpha_3 \rightarrow \hat\alpha_3)$.

In the case of a binary system consisting of two spinning bodies, both components experience a self-acceleration according to Eq.~(\ref{eq:alpha3accel}). As a result, there is an acceleration of the CM of the whole binary system and a modification of the relative orbital motion of the two bodies \citep{bd96}. As it turns out, the second contribution is the key to constrain $\hat\alpha_3$ with the help of binary pulsars. In systems consisting of a MSP and a WD, both the sensitivity and rotational velocity of the pulsar completely dominate those of the WD and the self-acceleration of the WD therefore can be completely neglected. Furthermore, as a result of the recycling process, the spin of the pulsar is expected to be parallel to the orbital angular momentum of the system  (see e.g.\ \citealt{bv91}). In sum, in small-eccentricity pulsar-WD system the eccentricity vector $\mathbf{e}$ will experience a time evolution that is analogous to the one in Sects.~\ref{sec:Nordtvedt} and \ref{sec:alpha1}, with a forced eccentricity
given by \citep{bd96}:
\begin{equation}
  \lvert{\bf e}_\mathrm{F}\rvert \simeq \frac{\lvert\hat\alpha_3\rvert}{3} \,s_\mathrm{p}
    \left(\frac{P_\mathrm{b}}{2\pi}\right)^2 \frac{\pi\,\nu}{(\Tsun M)} \, 
    \left\lvert\frac{\mathbf{w}}{c}\right\rvert\sin\vartheta_w \,,
\label{eq:efalpha3}
\end{equation}
where $\vartheta_w$ is the (generally unknown) angle between $\mathbf{w}$ and ${\bf
\Omega}$, and $\nu = \lvert{\bf \Omega}\rvert/(2\pi)$ is the spin frequency of the pulsar.

The figure of merit for systems used to test $\hat\alpha_3$ is $\nu P_\mathrm{b}^2/e$, meaning fast spinning pulsars in wide orbits with low eccentricities. Based on probabilistic considerations, \cite{gsf+11} have used an ensemble of suitable binary pulsar systems to obtain a 95\% confidence limit of $\lvert\hat\alpha_3\rvert < 5.5\times 10^{-20}$. A similar limit could be obtained from utilising just a single binary system. Based on a combined data set from the North American Nanohertz Observatory for Gravitational Waves (NANOGrav) and the European Pulsar Timing Array (EPTA) for binary pulsar PSR~J1713$+$0747, \cite{zdw+19} obtained a 95\% confidence interval of 
\begin{equation}
    -3 \times 10^{-20} < \hat\alpha_3 < 4 \times 10^{-20} \,.
\label{eq:alpha3limit}
\end{equation}
This limit results from a direct constraint on a temporal variation in the eccentricity vector of the system, i.e.\ on $\dot{\bf e}$. For the PSR~J1713$+$0747 system, all parameters involved in the evaluation of $\dot{\bf e}$ are measurable, except for the radial velocity of the pulsar binary with respect to the Solar System, $v_r$, which enters the calculation of $\mathbf{w}$. For that reason, $v_r$ is kept as a free parameter and chosen such, that it gives the most conservative limits for $\hat\alpha_3$.


\subsubsection{Limits on \texorpdfstring{$\hat\zeta_2$}{zeta2}}
\label{sec:zeta2}

Another PPN parameter that is related to a violation of the conservation of momentum is $\zeta_2$. It will join $\alpha_3$ in accelerating the CM of a binary pulsar system \citep{wil92,wil93}:
\begin{equation}
  {\bf a}_\mathrm{CM} = (\hat\alpha_3 + \hat\zeta_2) 
    \left(\frac{2\pi}{P_\mathrm{b}}\right)^2 
    (\Tsun M) \,
    X_\mathrm{p}X_\mathrm{c}(X_\mathrm{p} - X_\mathrm{c}) \,
    \frac{e\,c}{2(1-e^2)^{3/2}} \,
    \hat{\bf n}_\mathrm{peri},
\label{eq:zeta2}
\end{equation}
where $\hat{\bf n}_\mathrm{peri}$ is a unit vector from the CM of the system to the periastron of the pulsar orbit. Again, the hat indicates that in the presence of NSs we have an effective strong-field generalisation of the PPN parameter. If not perpendicular to the line of sight, the acceleration $\mathbf{a}_\mathrm{CM}$ produces an extrinsic contribution to a binary pulsar's $\dot\nu$ as it changes the radial velocity $v_r$ of the binary system and therefore the corresponding Doppler effect. In general, this contribution would not be separable from the spin-down $\dot\nu$ intrinsic to the pulsar. However, in relativistic binary pulsar systems, like PSR~B1913+16 or the Double Pulsar, the large $\dot\omega$ has lead to a significant change of $\omega$ since the pulsar's discovery---for PSR~B1913+16, it has advanced by more than $200^{\circ}$, and for the Double Pulsar by nearly $360^\circ$. In such cases, the projection of $\mathbf{a}_\mathrm{CM}$ onto the line of sight would have changed considerably over the observing time span, producing apparent higher order derivatives of the pulse frequency, i.e.\ $\ddot\nu$, $\dddot\nu$, etc. For instance, the $\ddot\nu$ is given by \citep{wil92}:
\begin{equation}
  \frac{\ddot\nu}{\nu} = (\hat\alpha_3+\hat\zeta_2)  
    \left(\frac{2\pi}{P_\mathrm{b}}\right)^2 
    (\Tsun M) \,
    X_\mathrm{p}X_\mathrm{c}(X_\mathrm{p} - X_\mathrm{c}) \,
    \frac{e}{2(1-e^2)^{3/2}}\,
    \sin i\cos\omega\,\dot\omega \,.
\label{eq:pddotfromzeta2}
\end{equation}
With the extremely tight constraints on $\hat\alpha_3$ (see Sect.~\ref{sec:alpha3}), Eq.~(\ref{eq:pddotfromzeta2}) can be used to directly set a limit on $\hat\zeta_2$. A corresponding equation for $\dddot\nu$ can be used to derive additional constraints on $\hat\zeta_2$ which for some of the relativistic binary pulsars turned out to be even more constraining \citep{mzs+20}. A combination of four carefully selected short-orbital-period DNS systems (including the Hulse--Taylor pulsar) by \cite{mzs+20} lead to the so far best limit for a $\zeta_2$-related violation of the conservation of momentum:
\begin{equation}
    \lvert\hat\zeta_2\rvert < 1.3 \times 10^{-5} \quad \mbox{(95\% confidence)}.
\label{eq:zeta2limit}    
\end{equation}
The Double Pulsar is not included in the above result, since the analysis was done before the publication of \cite{ksm+21}.

With a limit such as (\ref{eq:zeta2limit}), which results from a combination of different pulsars with different masses, one must always bear in mind that the underlying assumption is that the body-dependent parameter $\hat\zeta_2$ has only a weak dependence on the mass of a NS.


\subsection{Alternative gravity theories}
\label{sec:altGrav}

The excellent agreement of pulsar experiments with GR and the tight generic constraints on deviations from GR in the presence of strongly self-gravitating masses, as discussed above, consequently also means tight constraints on numerous specific alternative theories of gravity. In this section we mention a few examples, with a particular focus on mono-scalar-tensor theories of gravity. In view of the large number of alternatives to GR, however, this overview must inevitably remain very incomplete.


\subsubsection{Damour--Esposito-Far{\`e}se gravity}
\label{sec:def_grav}

A particularly well studied alternative to GR, at least in the context of pulsar experiments, is the mono-scalar-tensor theory of Damour \& Esposito-Far{\`e}se with a quadratic coupling function (called \emph{DEF gravity} in this review; see~\cite{de92,de93,de96_PSR} for details). This two-parameter class of gravity theories shows various effects that quite generally illustrate how gravity could deviate from GR, in particular in the presence of strongly self-gravitating NSs. For this reason, DEF gravity is particularly suitable for a theory-space approach to interpret tests of GR with pulsars \citep{dam09}. JFBD gravity~\citep{jor55,fie56,bd61}, which for a long time was the most important competitor to GR, represents a one-parameter sub-class of DEF gravity. 

Besides a spacetime metric $g_{\mu\nu}$, DEF gravity contains a mass-less scalar field $\varphi$, with asymptotic value $\varphi_0$ at spatial infinity. The field equations of DEF gravity can be derived from the \emph{Einstein frame} action 
\begin{equation}
    \mathcal{S} = \frac{c^4}{16\pi G_\ast} \int \big(R[g_{\mu\nu}] 
          - 2g^{\mu\nu}\partial_\mu\varphi\partial_\nu\varphi\big)\,\sqrt{-g}\,d^4x + \mathcal{S}_\mathrm{mat}[\Psi_\mathrm{mat};\tilde{g}_{\mu\nu}] \,,
\end{equation}
where $G_\ast$ is the bare gravitational constant and $R$ the curvature scalar.\footnote{In principle the action $S$ may also contain a potential for the scalar field, $V(\varphi)$, for instance to add mass to the scalar field \citep{de92,abwz12}. We assume here that $V(\varphi)$ only changes on scales much larger than those associated with the pulsar systems (typically on cosmological scales) and therefore can be neglected in pulsar experiments.} 
All matter fields $\Psi_\mathrm{mat}$ couple universally to the physical (Jordan) metric \mbox{$\tilde{g}_{\mu\nu} \equiv g_{\mu\nu} \exp[2\alpha_0(\varphi - \varphi_0) + \beta_0(\varphi - \varphi_0)^2]$}, and hence DEF gravity is a metric theory of gravity and fulfills the EEP. Furthermore, DEF gravity is a fully conservative gravity theory where only $\gamma_\mathrm{PPN}$ and $\beta_\mathrm{PPN}$ differ from their GR values: 
\begin{equation}
  \gamma_\mathrm{PPN} = 1 - \frac{2\alpha_0^2}{1 + \alpha_0^2} \,, \quad
  \beta_\mathrm{PPN}  = 1 + \frac{\beta_0\alpha_0^2}{2(1 + \alpha_0^2)^2} \,.
\label{eq:PPN_DEF}
\end{equation}

The two parameters of DEF gravity, $\alpha_0$ and $\beta_0$, define the two-dimensional parameter space of this class of alternatives to GR, which contains JFBD gravity ($\beta_0 = 0$) and GR ($\alpha_0 = \beta_0 = 0$). The Newtonian gravitational constant, as measured in a Cavendish-type experiment, is related to the bare gravitational constant $G_\ast$ by $G = G_\ast(1 + \alpha_0^2)$. There is a tight limit for $\lvert\alpha_0\rvert$ of the order of a few times $10^{-3}$ from Solar-System experiments (through tests on $\gamma_\mathrm{PPN}$), while $\beta_0$ remains unconstrained in such weak-field experiments \citep{bit03,dam09,fm23}.

The quantities (``gravitational form factors'') of a body with mass $m_A$ and MoI $I_A$ that enter the PK parameters are
\begin{equation}
    \alpha_A\equiv \frac{\partial \ln m_A}{\partial\varphi_0} \,, \quad
    \beta_A    \equiv \frac{\partial \alpha_A}{\partial\varphi_0} \,, \quad
    \mathcal{K}_A \equiv -\frac{\partial\ln I_A}{\partial\varphi_0} \,,
\label{eq:grav_form_factors}
\end{equation}
where the number of baryons is kept fixed when taking the partial derivatives.\footnote{The sensitivity $s_A$ of Eq.~(\ref{eq:sensitivity}) and $\alpha_A$ are directly related by $\alpha_A = \alpha_0(1 - 2s_A)$ for $\alpha_0 \ne 0$ \citep{de92,mw13,wil18_book}. The relation between $\mathcal{K}_A$ and the quantity $\kappa_A$ in Sect.~\ref{sec:Gloc} can be found in \cite{de96_PSR}.}
The quantity $\alpha_A$ is the \emph{effective scalar coupling} of the body and gives its specific scalar charge. For weakly self-gravitating masses $\alpha_A$ approaches $\alpha_0$. In the parameter space where spontaneous scalarisation does occur for NSs ($\beta_0 \lesssim -4.5$), $\alpha_A$ can be of order unity even if $\alpha_0 = 0$. Hence, in the strong gravitational fields of NSs DEF gravity can deviate significantly from GR, even if it is very close (or even identical) to GR in the weak-field regime. The other two gravitational form factors, i.e. $\beta_A$ and $\mathcal{K}_A$ can show a similarly extreme, non-linear behaviour in the presence of strongly self-gravitating (material) bodies. For BHs, where there is a no-hair theorem they are identical to zero \citep{de92}.

\begin{figure}[ht]
\begin{center}
\includegraphics[width=10cm]{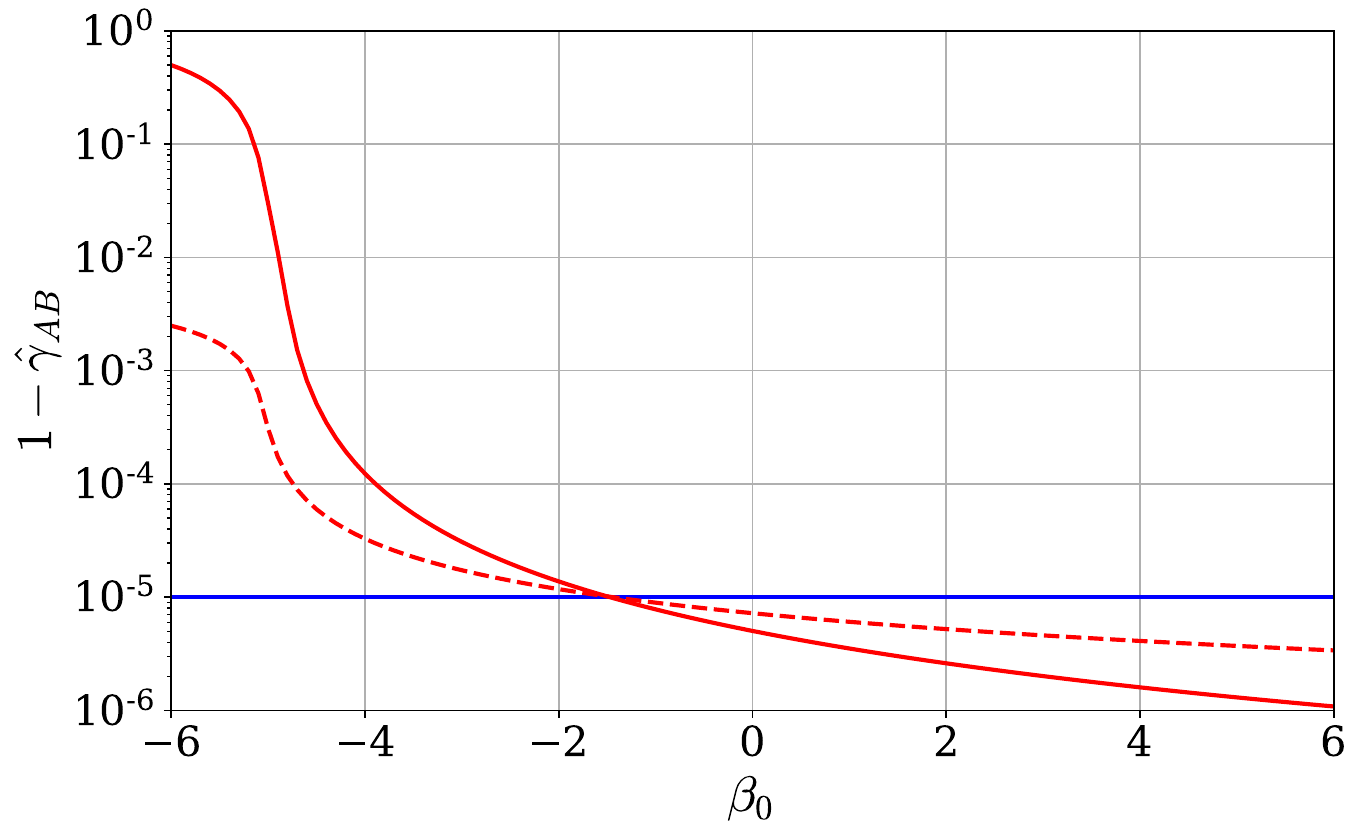}  
\caption{Strong-field Eddington parameter $\hat\gamma_{AB}$ in (quadratic) DEF gravity for 
  the Double Pulsar system. The plot shows the difference between GR and DEF, i.e.\ $1-\hat\gamma_{AB}$, as a function of $\beta_0$, for the interaction between the two NSs (red solid) and the interaction between pulsar B and a photon (red dashed). We have assumed $1 - \gamma_\mathrm{PPN} = 10^{-5}$ (i.e.\ $\lvert\alpha_0\rvert \simeq 0.00224$; blue horizontal line), which agrees well with current Solar-System experiments. To calculate the structure-dependent $\alpha_\mathrm{A}$ and $\alpha_\mathrm{B}$ we used the NS EoS ENG (see \citealt{lp01}).}
  \label{fig:gammaAB_DEF}
\end{center}
\end{figure}

In an N-body system, the effective gravitational constant for the interaction of two bodies is body dependent, and is given by 
\begin{equation}
  \hat{G}_{AB} = G_\ast(1 + \alpha_A\alpha_B) 
               = G\left(\frac{1 + \alpha_A\alpha_B}{1 + \alpha_0^2}\right) \,,
   \quad \mbox{with}~A\ne B \,.
\label{eq:Gab}
\end{equation}
Likewise, the PPN parameters become body dependent:
\begin{eqnarray}
  \hat\gamma_{AB}  &=& 1 - \frac{2\alpha_A\alpha_B}{1 + \alpha_A\alpha_B} \,,
  \quad \mbox{with}~A\ne B \,, \\
  \hat\beta^A_{BC} &=& 1 + \frac{\beta_A\alpha_B\alpha_C}{2(1 + \alpha_A\alpha_B)(1 + \alpha_A\alpha_C)} \,,
  \quad \mbox{with}~A\ne B, A\ne C  \,.
\end{eqnarray}
These PPN parameters can differ significantly from GR as well as from their weak field counterparts in Eq.~(\ref{eq:PPN_DEF}). Figure~\ref{fig:gammaAB_DEF} illustrates this for a specific case.

In a binary pulsar system, the PK parameters get modified by the gravitational form factors. For the quasi-stationary effects at the 1PN level one finds
\begin{eqnarray}\label{eq:PKst1}
  k &=& \frac{\hat\beta_\mathrm{O}^2}{1-e^2}
    \left(\frac{3  - \alpha_\mathrm{p}\alpha_\mathrm{c}}{1 + \alpha_\mathrm{p}\alpha_\mathrm{c}}
    - \frac{X_\mathrm{p}\alpha_\mathrm{p}^2\beta_\mathrm{c} + X_\mathrm{c}\alpha_\mathrm{c}^2\beta_\mathrm{p}}
    {2(1 + \alpha_\mathrm{p}\alpha_\mathrm{c})^2} \right) 
     \,,\label{eq:PK_k_DEF} \\
  \gamma &=& \frac{P_\mathrm{b}}{2\pi} \,\hat\beta_\mathrm{O}^2
    \left(\frac{1 + \mathcal{K}_\mathrm{p}\alpha_\mathrm{c}}{1 + \alpha_\mathrm{p}\alpha_\mathrm{c}} 
    + X_\mathrm{c}\right) X_\mathrm{c} \,e \,, \\
  s &=& x\left(\frac{P_\mathrm{b}}{2\pi}\right)^{-1} \hat\beta_\mathrm{O}^{-1} X_\mathrm{c}^{-1} \,,\\
  r &=& \frac{G_\ast m_\mathrm{c}}{c^3} \,,
\end{eqnarray}
where $\hat\beta_\mathrm{O} \equiv [2\pi\hat{G}_\mathrm{pc}(m_\mathrm{p} + m_\mathrm{c})/P_\mathrm{b}]^{1/3}/c$. Depending on the parameters $\alpha_0$ and $\beta_0$, for a given binary pulsar system some of these PK parameters can differ significantly from their GR values \citep{de96_PSR}. Above we have omitted PK parameters that so far have not played any role in constraining the DEF gravity parameter space, for instance, the rate of geodetic precession and the relativistic deformations of the orbit $\delta_\theta$ and $\delta_r$.

Concerning radiative aspects of gravity, in particular GW damping, one finds modifications already at the 1.5PN order in the orbital dynamics due to (scalar) dipolar GWs (see also Sect.~\ref{sec:dipolar} above). The corresponding change in the orbital period reads
\begin{equation}
    \dot{P}_\mathrm{b}^{\rm dipole} = -2\pi \, \hat\beta_\mathrm{O}^3 \,
     X_\mathrm{p} X_\mathrm{c} \,
    \frac{1 + e^2/2}{(1 - e^2)^{5/2}} \,
    \frac{(\alpha_\mathrm{p} - \alpha_\mathrm{c})^2}{1 + \alpha_\mathrm{p}\alpha_\mathrm{c}} \,
     + \mathcal{O}(\hat\beta_\mathrm{O}^5)\,.
\label{eq:PbdotDipoleDEF}
\end{equation}  
Since for $\lvert\alpha_\mathrm{p} - \alpha_\mathrm{c}\rvert \sim 1$ dipolar GW damping would be many orders of magnitude ($\sim (c/v)^2$) stronger than the GR GW damping, any confirmation of GR's quadrupole formula in the GW emission puts extremely tight constraints on $\lvert\alpha_\mathrm{p} - \alpha_\mathrm{c}\rvert$. However, only for sufficiently asymmetric binary systems, in terms of compactness of the two bodies, this converts into similarly stringent limits on DEF gravity. For that reason, pulsar-WD systems (see Sect.~\ref{sec:psr-wd}) are of particular interest here.

Apart from the dipole contribution, there are also monopole and quadrupole contributions related to the scalar field, which further enhance the orbital decay. However, both of them enter at the 2.5PN level, i.e.\ $\mathcal{O}(\hat\beta_\mathrm{O}^5)$, and are usually subdominant to $\dot{P}_\mathrm{b}^{\rm dipole}$. Detailed expressions can be found in \cite{de92}. 

Like for GR (previous sections), one can use different binary pulsar systems and their observed PK parameters to test the parameter space of DEF gravity. However, for DEF gravity there is no effacement of the internal structure of the bodies---a consequence of the violation of the SEP. Consequently, one has to assume an EoS for NS matter and for every pair $(\alpha_0,\beta_0)$ one needs to integrate the structure equations for slowly rotating NSs (see \citealt{de96_PSR}), certainly for the pulsar but also its companion if the latter is also a NS. For given central pressures one obtains the masses and the gravitational form factors for pulsar and companion, which then can be used to calculate the PK parameters (Eqs.~\ref{eq:PK_k_DEF}--\ref{eq:PbdotDipoleDEF}). By this, the PK parameters become (rather complicated) functions of the Keplerian parameters and the \emph{a priori} unknown masses of the binary system. A point $(\alpha_0,\beta_0)$ in the DEF gravity plane passes the test (for an assumed EoS) if there is a pair of masses, i.e.\ central pressures, where the corresponding PK parameters agree with the observations (see \citealt{de96_PSR,de98} for details of this approach).\footnote{Recently, a theory specific approach for testing DEF gravity has been developed, which does not require the intermediate step of fitting PK parameters \citep{bhw+23}} The procedure is somewhat simplified if the companion is a weakly self-gravitating body, since in this case one can assume $\alpha_\mathrm{c} \simeq \alpha_0$ and $\beta_\mathrm{c} \simeq \beta_0$. 

For the pulsar in the stellar triple system (see Sects.~\ref{sec:other-PSR-WD}, \ref{sec:Nordtvedt}) the situation is different. There we have no PK parameters to test the DEF gravity, but we can directly test the effective gravitational constant of Eq.~(\ref{eq:Gab}). A difference in the effective gravitation constant in the interaction between the pulsar and the outer WD and between the inner and the outer WDs is a strong effect that already enters at the Newtonian level in the equations of motion. To leading order the fractional difference $\Delta$ in the accelerations towards the outer WD reads (cf.~\citealt {dam09})
\begin{equation}
    \Delta \simeq \alpha_0 \, (\alpha_\mathrm{p} - \alpha_0)\,,
\label{eq:DeltaDEF}
\end{equation}
where for the weakly self-gravitating WDs $\alpha_A \simeq \alpha_0$ has been assumed. From the equation above, it can be seen that for very small $\alpha_0$ (or $\alpha_0 = 0$), where we could still have a scalarised pulsar (spontaneous scalarisation), the triple-system pulsar test does not give useful constraints on $\alpha_\mathrm{p}$.

\begin{figure}[ht]
\begin{center}
\includegraphics[width=\textwidth]{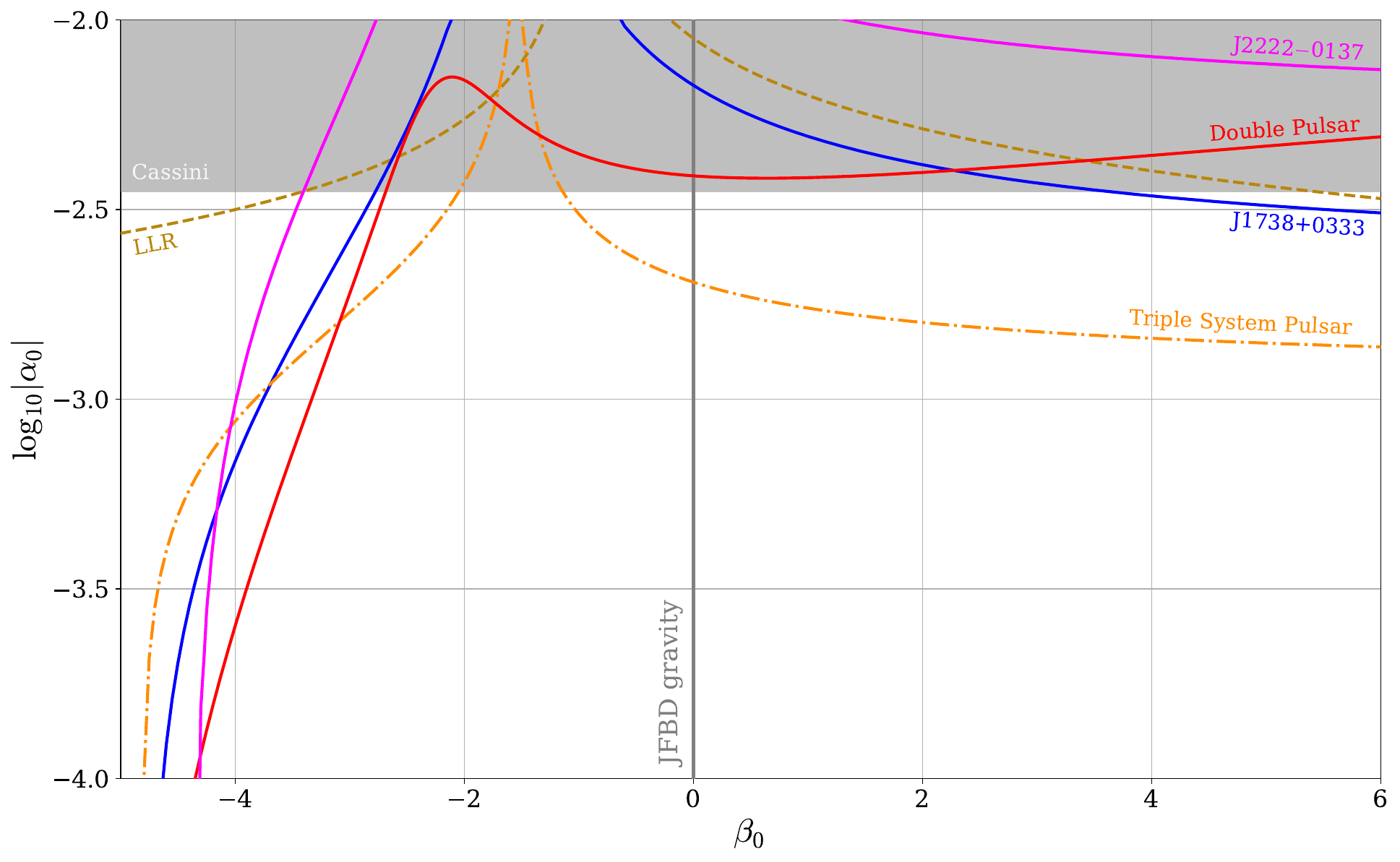}
\caption{Constraints on the DEF gravity parameter space from different experiments (95\% confidence): Shapiro delay with the Cassini spacecraft \citep{bit03}, dipolar radiation (J1738$+$0333, J2222$-$0137; Sections \ref{sec:psrJ1738}, \ref{sec:psrJ2222}), Nordtvedt effect (LLR \citealt{bmt21}, Triple System Pulsar; Sect.~\ref{sec:Nordtvedt}), and the Double Pulsar (Sect.~\ref{sec:psrJ0737}). Areas above a curve are excluded by the corresponding experiment (see \citealt{de96_PSR,de98} for details). Pulsar curves are computed with  a comparably stiff EoS (MPA1 in \citealt{lp01}), which for most of the parameter space gives conservative limits. GR corresponds to $\alpha_0 = \beta_0 = 0$, and JFBD theory is along the vertical $\beta_0 = 0$ line with Brans--Dicke parameter $\omega_\mathrm{BD} = (\alpha_0^{-2} - 3)/2$.}
\label{fig:DEFa0b0}
\end{center}
\end{figure}

Figure~\ref{fig:DEFa0b0} shows constraints in the DEF-gravity parameter space obtained from different Solar System and pulsar experiments. Concerning pulsar limits, one has to keep in mind that the gravitational form factors of Eq.~(\ref{eq:grav_form_factors}) depend on the structure of the NS and are therefore EoS dependent. Consequently, the pulsar limits in Fig.~\ref{fig:DEFa0b0} change to some extent if a different EoS is chosen to solve the NS structure equations. To obtain robust limits, one needs to follow an EoS-agnostic approach like in \cite{vcf+20}, where a point in the DEF gravity plane is only excluded if it is excluded for a whole range (from soft to stiff) of viable EoSs. In Fig.~\ref{fig:DEFa0b0} we have chosen only one (rather stiff) EoS to illustrate qualitatively the pulsar limits. In the highly non-linear regime of DEF gravity, the EoS dependence is particularly strong, and it requires a range of pulsars with a suitable distribution of masses in order to constrain DEF gravity \citep{ssb+17,zfk+22}.

\subsubsection{Various other alternatives to GR}

While DEF gravity is arguably the best studied class of alternatives to GR in the context of pulsar experiments, there are many other theories that have been significantly constrained or even ruled out using pulsar timing. A complete list is beyond the scope of this review, so in this subsection we will only give a few particularly informative examples. The biggest challenge in confronting an alternative theory of gravity with pulsar observations is to calculate the ``gravitational form factors'' for NSs, which goes far beyond a linear approximation, as it requires the full non-linearity of the theory. 

While in the previous section it was assumed that the potential of the scalar field $V(\varphi)$  only plays a role on cosmological scales, if at all, this assumption can certainly be relaxed, for example to have a massive scalar field with a Compton wavelength comparable to the length scales relevant in the pulsar experiment. Pulsar timing results have been used to exclude certain parts of the parameter space of such massive scalar-tensor theories (see e.g.\ \citealt{abwz12,rp16,ydp16,sy20a,sy20b}).

Binary pulsar observations have also been used to constrain or even exclude some MOND-like gravity theories, i.e.\ relativistic theories that have modified Newtonian dynamics as their non-relativistic limits and are an attempt to avoid the need of dark matter in the Universe, at least on certain scales. The most prominent example is Bekenstein's tensor-vector-scalar (TeVeS) theory \citep{bek04} which is practically excluded by the Double Pulsar \citep{fwe+12,ksm+21}.\footnote{Bekenstein's TeVeS (including generalised versions \citep{sei07,cww08}) is also excluded by the test for the speed of GWs with the GW170817 LIGO-Virgo event, which however tests a different aspect of the theory \citep{ghlp18}. Double Pulsar probes directly the scalar sector which is responsible for the MONDian behaviour in the (very) weak field (see e.g.\ \citealt{be07}). For that reason, the Double Pulsar test should remain of interest for modifications of TeVeS that are designed to bypass the GW170817 limit without modification of the coupling to the scalar field.}
The parameter space of a natural extension of Bekenstein's TeVeS with a quadratic coupling of matter to the scalar field, which by design satisfies Solar system tests, has been constrained with binary pulsars in \cite{fwe+12}.

Other examples of the application of pulsar observation for testing alternative gravity theories are tests of Mendes--Ortiz (MO) gravity \citep{mo16,afy19}, Einstein-Aether and khronometric gravity \citep{ybby14,ghb+21}, scalar-Gauss-Bonnet \citep{ddy22,ysyd24}, and cubic Galileon model \citep{swz20}, just to name a few.

On a final note, there are various alternatives to GR that naturally or by design pass all pulsar experiments the same way as GR does, because they either invoke screening mechanisms that make them indistinguishable in their so-called ``strong-field regime'' (which even includes the Solar System) or because they are sufficiently short range in their modifications to GR, so that effects do not show up in the orbital dynamics of binary pulsars.\footnote{If the short-range modification leads to a significant alteration of the MoI of a NS, such an effect could in principle be tested through the LT precession of a binary pulsar orbit (see the discussion in \citealt{hkw+20}).} Theories that predict deviations only in the context of BHs also have not been tested in binary pulsar experiments, simply because no binary pulsar with a BH companion was available until now, with the first strong candidate having been published only this year \citep{bdf+24}.


\section{Conclusions and future prospects}
\label{sec:conclusions}

\subsection{Summary}

As the first half century since the discovery of the first binary pulsar comes to a close, it is important to reflect on what has been achieved in terms of tests of gravity theories. This includes the first tests of gravity theories with compact, strongly self-gravitating objects and the first detection of GWs from the orbital decay of the first binary pulsar. These represent qualitatively new tests in comparison with all previous tests in the Solar System.

However, it is also important to realise that the most precise tests of gravity theories based on the timing of binary (and triple) systems have been published since the last \textit{Living Review in Relativity} on this topic \citep{sta03}:
\begin{itemize}
\item The measurement of the orbital decay in the Double Pulsar published in 2021 \citep{ksm+21} 
  improved the precision of tests of the radiative properties of gravity---especially the leading order quadrupolar term predicted by GR---by a factor of 25 over the best previous test. The results agree with GR within the relative 1-$\sigma$ uncertainty of $6.3 \times 10^{-5}$.
\item The same system allowed several other independent, high-precision tests of GR as well, 
  including the first pulsar tests of terms past the leading order.
\item These include the most precise pulsar tests of the Shapiro delay, carried out in a 
  spacetime with a curvature that is six orders of magnitude larger than the curvature probed with the Cassini-spacecraft test in the Solar system. More generally, it is the  photon propagation test with the highest spacetime curvature, exceeding the images of the supermassive BHs M87$^\ast$ \citep{EHT_2019} and Sgr~A$^\ast$ \citep{EHT_2022} by more than nine and three orders of magnitude respectively.
\item The MSP in a triple system, PSR~J0337+1715, has allowed an improvement in 
  our test of the UFF for NSs by three orders of magnitude \citep{agh+18,vcf+20}. This test of the SEP provides some of the tightest constraints for many alternatives to GR, including JFBD gravity and a large part of the DEF-gravity parameter space.
\item The latter parameter space was also constrained by tight constraints on the 
  possibility of dipolar GW emission in a set of pulsar-WD systems with a wide range of pulsar masses. 
\item Regarding other gravity theories, pulsar tests have not only provided 
  important limits on other types of scalar-tensor theories, scalar-Gauss-Bonnet gravity, Einstein-Aether (a tensor-vector theory which violates Lorentz invariance in the gravitational sector), etc., but also entirely ruled out others, like Bekenstein's TeVeS and some of its variations.
\end{itemize}

Impressively, GR still passes all these precise and diverse tests.
More generally, these experiments test some fundamental aspects and symmetries of gravitation and spacetime:
\begin{itemize}
\item The verification of the UFF for NSs via the non-detection of the Nordtvedt 
  effect and of dipolar GW emission.
\item The stringent limits on UFF violation and on preferred-location and preferred-frame 
  effects for the gravitational interaction further support the SEP. This is of fundamental importance, particularly in view of the conjecture that GR is the sole valid gravity theory that fully embodies the SEP.
\item Some of these radiative experiments are stringent probes into the nature of GWs, showing 
  that they are, to leading order, quadrupolar as predicted by GR. These GW pulsar tests nicely complement GW tests obtained from merger observations with ground-based GW observatories.
\item These radiative experiments have also excluded some strong-field highly nonlinear 
  deviations from GR, like the phenomenon of spontaneous scalarisation predicted by DEF gravity.
\item Finally, pulsars have also provided tight constraints for parameters of generic 
  frameworks, like strong-field generalizations of PPN parameters and parameters of the gravitational sector in the SME. Some of them are directly related to the aforementioned limits on violations of symmetries associated with the SEP, like the UFF, local position and local Lorentz invariance.
\end{itemize}
This flurry of recent results show that gravity experiments using radio pulsars are thriving. They, and many results from the Solar System, EHT, LIGO/Virgo/KAGRA, etc., demonstrate the continued interest in precision gravity experiments.


\subsection{Prospects}

The prospects for improvements in the precision of these tests for the near future appear to be excellent. First, the mere continuation of some timing experiments will greatly improve many of the tests done with these systems. As examples, two years of additional data on PSR~J0337+1715 allowed a (preliminary) doubling of the precision of the test of the UFF with this system \citep{vgc+22}. Furthermore, simulations showed that continued timing of the Double Pulsar might constrain the MoI of PSR~J0737$-$3037A to within 10\% until 2030 \citep{hkw+20}, apart from significantly improving the precision in the measurement of the orbital decay. Although such a determination of the MoI assumes GR to provide the correct description of the needed PK parameters (and will eventually help to constrain the EoS), interpreted as a LT test it still allows to probe for significant short-range deviations from GR that only affect pulsar A locally (see discussion in \citealt{hkw+20}).

A significantly improved radiative test and a LT test with the Double Pulsar rely on good independent constraints on the EoS of dense matter. These are provided by the measurement of large NS masses \citep{fcp+21}, the NICER constraints on the radius and mass of NSs (e.g., \citealt{mld+21,vsw+24}) and measurements of the NS tidal deformability by ground-based GW detectors \citep{LIGO2018} (see discussion in Sect.~\ref{sec:0737_periastron}); the latter are especially valuable as they can be translated directly to constraints on the MoI via the I-Love-Q relation \citep{yy13}. Improving these EoS constraints will not only lead to more precise radiative and LT tests with the Double Pulsar, but will also lead to more precise constraints on alternative theories of gravity, which as discussed in Sect.~\ref{sec:altGrav} are still subject to some uncertainties on this account.

The prospects for improvement of the pulsar gravity tests are further brightened by the higher sensitivity of telescopes like FAST, MeerKAT and in the future the SKA. FAST and MeerKAT are already improving the timing precision on all known pulsars, allowing new and more precise measurements of PK parameters. As an example, just three years of MeerKAT data yielded a photon propagation test in the Double Pulsar \citep{hkc+22} that is a factor of two better than the previous one based on 16 years of data from 6 different telescopes \citep{ksm+21}, which is significantly better than even the most optimistic simulations.

More importantly, it is clear that the pulsar field in general has been driven, from the start, by the discovery of new, better ``laboratories''. The rate of pulsar discoveries has recently increased significantly, with 1000 new pulsars having been found by FAST and MeerKAT already.\footnote{See e.g., GPPS (\url{http://zmtt.bao.ac.cn/GPPS}) and CRAFTS (\url{http://groups.bao.ac.cn/ism/CRAFTS/202203/t20220310_683697.html}) surveys with FAST and the TRAPUM survey (\url{https://trapum.org/discoveries/}) with MeerKAT.}
Furthermore, the rate of discovery of recycled pulsars---and especially recycled pulsars in very compact orbits---is increasing even faster, because of the much improved time and spectral resolution of the search data, the much improved computing capabilities and search algorithms.

All this will very likely lead to the discovery not only of more extreme versions of the currently known systems, which will allow new leaps in the precision for the types of tests described above, but also of completely new types of systems, such as pulsar---BH binaries, of which the first one might have already been found \citep{bdf+24}. Such systems will allow gravity tests that were until now beyond the testing power of pulsar timing \citep{wk99,lewk14,sy18}. In particular, the discovery of a pulsar in a relativistic orbit around the supermassive BH at the centre of our Galaxy would allow unprecedented tests of BH physics, in particular in combination with tests from other observations in this extreme gravity environment (see e.g.\ \citealt{pwk16} and references therein).

The prospect of detecting very compact binary pulsars, especially DNSs (or pulsar-BH systems), is very alluring for tests of gravity theories. A general reason is the attainable significance of the radiative test in the presence of contaminants, which improves as $P_\mathrm{b}^{-8/3}$. Such systems would also allow the measurement of the full precession cycle of relativistic spin-orbit coupling on reasonable timescales: for instance, a DNS with an orbital period of 30 minutes would have a geodetic precession period of about 5 years, which would then be measured precisely from the repeating changes in the pulse profile of the system. 
In fact, such a test of geodetic precession with repeating emission patterns could already be possible in the near future with double pulsar B, which is expected to precess back into our line of sight within the next few years \citep{Breton_2008,pmk+10,lks+23}.

Additionally, very compact binary pulsars will be detectable at good S/N by the Laser Interferometer Space Antenna (LISA) mission if they are not too distant from Earth \citep{tol20}. This mission will also find, independently, the most compact NS-NS, NS-WD or NS-BH systems of our Galaxy \citep{lmv+20}. Perhaps some of these NSs will be detectable as pulsars in targeted radio surveys. In either case, binary pulsar experiments would become ``multi-messenger'' experiments, allowing entirely new tests of gravity theories \citep{tol20,mxs+21}.


\bmhead{Acknowledgements}
\label{sec:acknowledgements}

We thank the referee Lijing Shao and the second anonymous referee for their many valuable comments which helped to improve the manuscript. We further thank Andrew Cameron, David Champion, Huanchen Hu, Michael Kramer, Jim Lattimer, Kuo Liu, Ingrid Stairs, Thomas Tauris, Guillaume Voisin, and Vivek Venkatraman Krishnan for valuable discussions. We thank the Max-Planck-Gesellschaft for continued support.


\phantomsection
\addcontentsline{toc}{section}{References}
\bibliography{pulsar_grav}

\end{document}